\documentclass[%
 reprint,
 amsmath,amssymb,
 aps, prd,preprintnumbers, superscriptaddress, nofootinbib
]{revtex4-2}

\usepackage{graphicx}	
\usepackage{amsmath}	
\usepackage{amssymb}	

\usepackage[english]{babel}
\usepackage{graphicx}
\usepackage{latexsym}
\usepackage{floatflt}
\usepackage{float}
\usepackage{units}
\usepackage{hyperref}
\usepackage{amsmath}
\usepackage{gensymb}
\usepackage{mathdots}
\usepackage{color}
\usepackage{xcolor}
\usepackage{mathtools}
\usepackage{lipsum}

\usepackage{dcolumn}
\usepackage{bm}

\newcommand{\id}{{\rm d}}

\begin{document}

\preprint{MIT-CTP/5861}

\title{Near-Peak Spectrum of Gravitational Waves from Collapsing Domain Walls}
\author{Bryce Cyr}
\email{brycecyr@mit.edu}
\affiliation{Center for Theoretical Physics, Massachusetts Institute of Technology, Cambridge, MA 02139,USA}

\author{Steven Cotterill}
\author{Richard Battye}
\affiliation{%
Jodrell Bank Centre for Astrophysics, School of Physics and Astronomy, The University of Manchester, Manchester M13 9PL, U.K.
}%

\begin{abstract}
Cosmological domain walls appear in many well-motivated extensions to the standard model of particle physics. If produced, they quickly enter into a self-similar scaling regime, where they are capable of efficiently sourcing a stochastic background of gravitational waves. In order to avoid a cosmological catastrophe, they must also decay before their enormous energy densities can have adverse effects on background dynamics. Here, we provide a suite of lattice simulations to comprehensively study the gravitational wave signatures of the domain wall network during this decay phase. The domain walls are initially formed through spontaneous breaking of a $\mathbb{Z}_2$ symmetry, and subsequently decay through the action of a small bias term which causes regions of false vacuum to collapse. We find that gravitational waves are produced in abundance throughout this collapsing phase, leading to a shift in the peak frequency and increase in the overall amplitude of the spectrum by an $\mathcal{O}(100)$ factor when compared against simple analytic arguments. Importantly, we also find that the characteristic frequency of emitted gravitational waves increases as the network decays, which leads to a softening of the high frequency spectral index. This high frequency spectrum therefore carries key information related to the dynamics of the collapsing phase, and can be used to discriminate between different domain wall scenarios using upcoming data.
\end{abstract}

\maketitle

\section{\label{sec:level1}Introduction}
\noindent Recent evidence for the detection of a stochastic gravitational wave background (SGWB) by the Pulsar Timing Array (PTA) consortium \cite{NANOGravDetection2023, EPTADetection2023, PPTADetection2023, CPTADetection2023} has sparked a renewed interest in domain wall scenarios. Evidence for a Hellings-Down correlation pattern \cite{Hellings1983} in the most recent data seems to point towards a gravitational wave origin for the common-spectrum process that was reported in earlier datasets (e.g. using the NANOGrav 12.5 year data \cite{NANOGrav2020}). While this result represents a major step forward in our understanding of gravitational wave backgrounds, a mild tension has developed between the data and the standard theory prediction \cite{NANOGravDetection2023,NANOGrav2023Exotic}. 

A known progenitor of nanohertz gravitational waves (the frequencies probed by PTAs) arises from a population of inspiraling supermassive black hole binaries (SMBHB) \cite{Begelman1980}. These systems are expected to form through galaxy mergers at early times, where dynamical friction drags both of the initial supermassive black holes to the centre of this new system. The amplitude of this signal is dependent on the dynamics of a complicated chain of hierarchical mergers occurring deep within the non-linear regime of structure formation, making it difficult to estimate. The spectral tilt, however, presents a more clean path for confirmation of this signal. PTA observations parameterize this information through the timing residual power spectral density, which is modeled as a power law with amplitude $A_{\rm GWB}$ and spectral index $\gamma_{\rm GWB}$. Under typical assumptions on the dark matter fluctuations, NANOGrav \cite{NANOGravDetection2023} reported maximum likelihood values of $A_{\rm GWB} \simeq 6.4 \times 10^{-15}$ and $\gamma_{\rm GWB} \simeq 3.2$, with error bars that put this result approximately $3\sigma$ away from the theory prediction from SMBHBs of $\gamma_{\rm BHB} = 13/3 \approx 4.33$ \cite{Phinney2001}\footnote{The exact significance of this tension can be relaxed using different modeling assumptions in the astrophysics.}.

While the background produced by SMBHBs is undoubtedly in the data at some level, this discrepancy in the spectral index motivates an investigation into  alternative explanations for the leading order source of a SGWB. Using their 15 year dataset, NANOGrav themselves have performed a preliminary analysis of other more exotic interpretations of this background \cite{NANOGrav2023Exotic}. In particular, they found that when the Bayes factor is used to compare various scenarios, certain classes of biased domain walls (DWs) are capable of offering some of the most convincing fits to the SGWB data. 

Domain walls are a class of topological defects that can form when the Universe undergoes a phase transition in which the true vacuum state of the theory consists of two (or more) degenerate, disconnected vacua \cite{Vilenkin2000}. In practice, the scalar field responsible for the phase transition populates different vacua on length scales separated by more than the correlation length of the field. 

As the Universe continues to expand, regions that were once causally disconnected from each other come into contact. By demanding continuity of the field profile across these regions of distinct vacua, sheet-like solitonic objects known as domain walls are produced in regions where the field is stabilized at the top of its potential. This effect is often referred to as the Kibble mechanism \cite{Kibble1976,Kibble1980}, and can be summarized by stating that if Nature permits a vacuum state that has the possibility of forming topological defects, they are undoubtedly formed in an expanding Universe.

Immediately after such a phase transition, a damping period occurs in which the initial field configurations settle into their respective minima. Following this, simulations and analytic arguments \cite{Press1989,Hindmarsh1996,Oliveira2004,Avelino2005,Leite2011,Martins2016} indicate that the network of domain walls enters into a so-called scaling regime, with $\mathcal{O}(1)$ Hubble-radius sized walls permeating our Universe at any given point in time. These objects possess significant energy density, which can have adverse effects on the standard cosmological history. Most notably, long-lived domain walls originating from the early Universe can be severely constrained by considering bounds from Cosmic Microwave Background (CMB) anisotropies and overclosure \cite{Zeldovich1974}. 

To evade these stringent limits, many authors have considered introducing a bias to the potential of the scalar field (which we call $\phi$). The purpose of this bias is to break the degeneracy of the vacuum, producing a pressure gradient on the walls that will induce the collapse of false vacuum regions \cite{Vilenkin1981,Widrow1989,Larsson1996,Gelmini1988,Correia2014,Correia2018}. Therefore, the life-cycle of a biased domain wall network can be described by three distinct phases: damping, scaling, and collapse.

As a domain wall network evolves in the scaling regime, copious amounts of gravitational radiation will be emitted through processes such as domain wall motion and collisions, the decay of substructure on walls, and the collapse of sub-horizon closed-surface walls under their own tension. Recently, a number of  groups have performed network simulations \cite{Hiramatsu2010, Kawasaki2011, Hiramatsu2013, Saikawa2017, Kitajima2023, Ferreira2024, Dankovsky2024} to assess the overall gravitational wave production. Results indicate that during the scaling regime, the gravitational wave spectrum roughly obeys a broken power law with peak frequency set by the scale of the largest domain wall in the simulation. The low frequency slope is typically well fit by a $k^3$ scaling\footnote{This is a enforced due to causality arguments as described in e.g. \cite{Caprini2009,Cai2019,Dankovsky2024}.}, while a precise characterization of the high frequency slope is one of the main aims of this work. 

Although gravitational wave production in the scaling regime has seen intense study, less is known about the final spectrum when one also includes the collapse phase. A commonly made assumption is that the network disappears instantaneously\footnote{We will extensively refer to this as the \textit{instantaneous decay approximation.}} once the pressure force induced by the bias overcomes the expansion rate of the wall. However, we know from scaling arguments that the maximum amplitude and peak value of the spectrum are sourced from the latest times that the network is active. It is then natural to expect that in realistic scenarios, the dynamics of this collapsing phase will play a crucial role in determining the near-peak spectrum of gravitational waves. If one hopes to use gravitational wave observations to probe the fundamental properties of domain walls, it is imperative to properly characterize the effects of this collapsing phase.

In this work, we study the spectrum of gravitational waves produced during the scaling and collapse phases for a network of (biased) $\mathbb{Z}_2$ symmetric domain walls. To this end, we perform lattice simulations consisting of $N=2048^3$ gridpoints in an expanding (radiation dominated) background, and compute the gravitational wave spectrum throughout. It has only recently become possible to run simulations which are capable of resolving both the scaling and decay phases.

To mediate the decay, we utilize two different forms for our bias term. First, we introduce a vacuum bias which is present at all times, but negligibly small at the start of the simulations. Afterwards, we consider a temperature-dependent bias which is initially absent in the simulation, before turning on at some critical redshift where it can in principle be large. This second form is motivated by a desire to study a toy model of non-perturbative effects that could be present between the scalar field and a thermal plasma (e.g. an axion during the Quantum ChromoDynamics (QCD)  phase transition), similar to what was considered recently in \cite{Kitajima2023}. In both cases we provide a parameterization for the lifetime of the network utilizing different metrics extracted from the simulation.

In the vacuum bias case, we also present a comprehensive analysis of the resultant gravitational wave spectrum, focusing on the near-peak region. Interestingly, we find that the spectral index of the high frequency slope can be used to extract information relating to the amplitude and time dependence of the bias term. We also agree with other groups\footnote{Similar to us, these works also compute the gravitational wave spectrum throughout a collapse phase.} \cite{Kitajima2023, Ferreira2024} in finding that the finite decay timescale causes the overall amplitude to rise, while the peak position shifts to lower frequencies when compared with the instantaneous decay approximation. This more complete characterization of the spectrum can be used alongside complementary information from other cosmic probes to help disentangle the sources of gravitational waves present in the early Universe \cite{Cyr2023}.

The remainder of this work is organized as follows. In Section \ref{sec:level2}, we briefly review key parts of the domain wall theory, describing in detail the two forms of bias we consider and the parameterizations that we apply to the numerical results. In Section \ref{sec:level4} we present our main results related to the lifetimes and near-peak spectra for various bias amplitudes, while we leave a dedicated description of our numerical setup (including details on the gravitational wave computation) to Appendix~\ref{sec:levelA1}. We discuss these results and conclude in Section \ref{sec:level5}. Throughout we use natural units with $c = \hbar = k_{\rm b} = 1$. Additionally, we will often use so-called numerical units, which can be distinguished from their physical counterparts by an overbar, e.g. $\bar{\tau} = \sqrt{\lambda} \eta \, \tau$ where $\bar{\tau}$ is a dimensionless conformal time. Appendix~\ref{sec:levelA1} also contains a useful numerical dictionary of these transformations. \\
%
\begin{figure*}
\centering 
\includegraphics[width=\columnwidth]{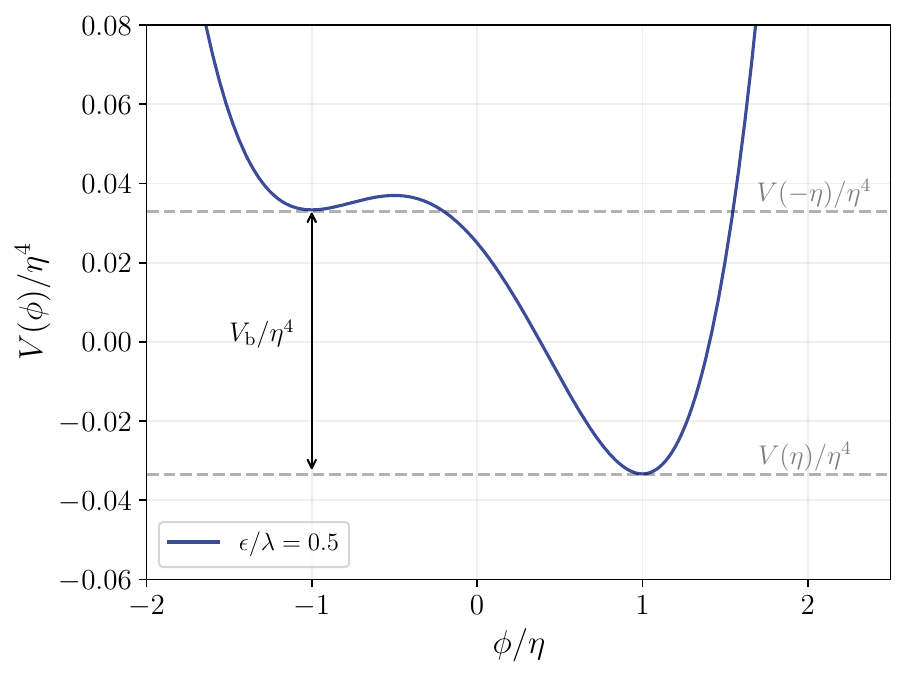}
\hspace{4mm}
\includegraphics[width=\columnwidth]{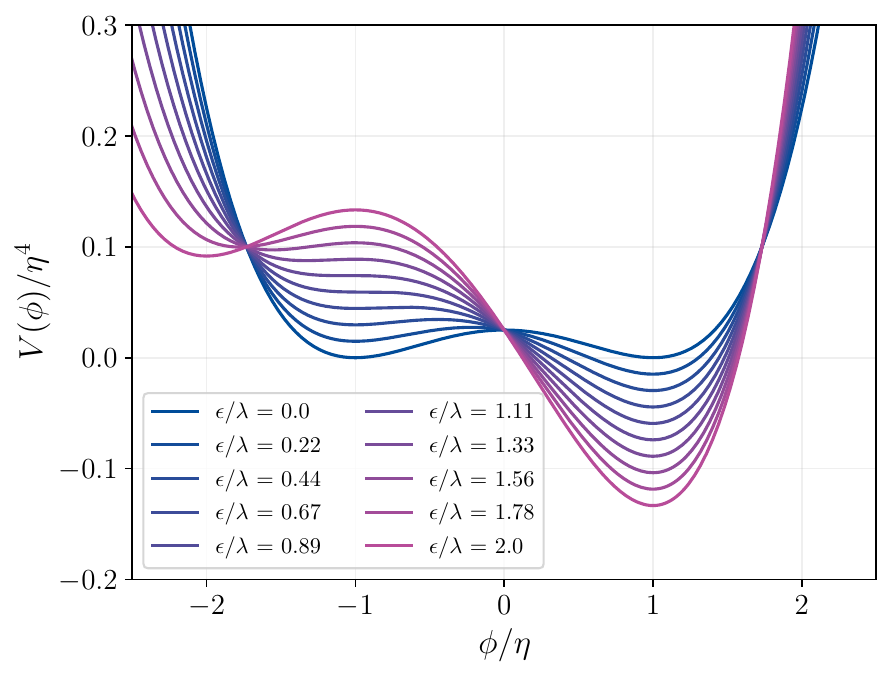}
\caption{Left: An (exaggerated) example of a biased potential. The potential difference induces a pressure force on the domain walls, causing regions of false vacuum to collapse. In principle, tunneling or thermal fluctuations can expedite the decay process, though their contributions are typically subdominant to the pressure force. Right: An illustration of an evolving potential as the (temperature dependent) bias term is turned on smoothly from $\epsilon/\lambda = 0 \rightarrow 2$. At the critical point $\epsilon = \lambda$ the minimum at $\phi = -\eta$ flips to a local maximum, and the false vacuum regions will undergo a rapid and violent collapse.}
\label{Fig:Potential}
\end{figure*}
%

\section{Decaying Domain Walls} \label{sec:level2}
\noindent In this work we choose to study the simplest scenario which permits domain wall formation, namely that of a single real scalar field in a $\mathbb{Z}_2$ symmetric double well potential centered at the origin, expressed by
%
\begin{align} \label{eq:bare-potential}
   V_0 = \frac{\lambda}{4}(\phi^2-\eta^2)^2.
\end{align}
%
At early times this symmetry is spontaneously broken and the scalar field settles into one of the two vacua at $\phi = \pm \eta$ in different regions of space. Continuity of the field profile across these distinct regions necessitates the formation of domain walls. In the Minkowski limit, the equations of motion can be analytically solved \cite{Vilenkin2000}, yielding solitonic objects which can be characterized by an effective width, $\delta \simeq 2/m$ (where $m = \sqrt{2\lambda}\eta$ is the mass of $\phi$ in the vacuum) and surface tension $\sigma = \int T_0^0 \id z = 2 \sqrt{2\lambda} \eta^3/3$.

If such a phase transition takes place in the early Universe, it is expected that a tangled network of domains initially forms, before eventually evolving towards a scaling solution in which the network achieves self-similarity.
At this time, one expects $\mathcal{O}(1)$ Hubble-sized domain walls in any given casual patch \cite{Garagounis2002}, with sub-Hubble closed walls collapsing under their own tension. In the scaling regime, the (physical) energy density of the network is therefore dominated by the large walls and can be estimated as
%
\begin{align} \label{eq:scaling-Energy}
    \rho_{\rm dw}^{\rm s}(z) \simeq 2\mathcal{A} \frac{\sigma}{H(z)^{-1}},
\end{align}
%
where $\mathcal{A}$ is a roughly constant area parameter that we find to be $\mathcal{A} = 0.78 \pm 0.03$ from our simulations (see the following section for details). A value of $\mathcal{A} = 1/2$ would imply one perfectly flat wall stretching through the Hubble patch\footnote{While this may seem like an odd choice, we write Eq.~\eqref{eq:scaling-Energy} in such a way as to match the definition of $\mathcal{A}$ given in the oft cited Ref. \cite{Hiramatsu2013} for ease of comparison.}. Early simulations by Hiramatsu \textit{et al.} \cite{Hiramatsu2013} found this area parameter to be $\mathcal{A} = 0.8 \pm 0.1$. More recent work by Ferreira \textit{et al.} \cite{Ferreira2024} find $\mathcal{A}\simeq 0.9$, while Dankovsky \textit{et al.} \cite{Dankovsky2024} determine $\mathcal{A} = 0.85 \pm 0.04$ from their work with vacuum initial conditions. Generally speaking, the constancy of $\mathcal{A}$ is how one can assess the time at which scaling has been achieved. From a numerical standpoint, it is essential that the network persists into the scaling regime in order to mitigate the influence of initial conditions on our final results. 

Eq.~\eqref{eq:scaling-Energy} makes evident an obvious challenge for domain wall networks: during radiation domination, the energy in the network scales as $\rho_{\rm dw}^{\rm s} \sim (1+z)^2$, while the critical density goes like $\rho_{\rm c} \sim (1+z)^4$. A simple bound on the wall tension can be set by demanding that the relative energy density in domain walls remains a subdominant component of the Universe at any point during their existence
%
\begin{align}
    \sigma \lesssim \frac{3H(z)}{16\pi G} \frac{1}{\mathcal{A}}, \hspace{15mm} {\rm (overclosure)}.
\end{align}
%
These overclosure bounds have been used to set useful constraints on specific domain wall models, such as the $\mathbb{Z}_2$ symmetric realization of the two Higgs doublet model \cite{Battye2020}.

If one insists on having a domain wall network survive until the present time, a more stringent bound can be placed by noting a lack of large scale density fluctuations imprinted on the CMB. This constraint, known as the Zel'dovich-Kobzrev-Okun bound \cite{Zeldovich1974}, restricts the wall tension to be $\sigma^{1/3} \lesssim \mathcal{O}(1 \, {\rm MeV})$ (setting $\mathcal{A} = 1$). In other words, for walls to be cosmologically long-lived, the symmetry breaking transition must take place such that $\eta \lesssim $ MeV\footnote{Note that for thermal phase transitions, $\eta \simeq T_{\rm SSB}$, while for vacuum phase transitions $\eta \simeq T_{\rm SSB}^2/m_{\rm pl}$ where $m_{\rm pl}$ is the Planck mass. The majority of simulations performed in the literature (including ours) take place assuming a $\phi$-field in vacuum and so the latter expression is most relevant.}.

\subsection{\label{sec:level2.1}Potential Biasing}
The constraints on long-lived domain wall configurations are quite stringent, thus different techniques been developed to evade them \cite{Vilenkin1981,Gelmini1988,Larsson1996}. One such method that has enjoyed the spotlight recently is by supplementing the potential given in Eq.~\eqref{eq:bare-potential} with a small bias term, $V = V_0+\Delta V$. Depending on the form of $\Delta V$, this can break the degeneracy of the vacuum and will signal an impending epoch of domain wall collapse. An example of this type of potential is shown in the left panel of Fig.~\ref{Fig:Potential}.

For the purposes of this work, we facilitate the collapse by introducing a modification to the potential, namely 
%
\begin{align}
    V = V_0 + \epsilon \eta \phi \left( \frac{\phi^2}{3} - \eta^2\right),
\end{align}
%
where $\epsilon$ is the dimensionless amplitude of the bias. The functional form is chosen due to the convenient property that the location of the vacua remain at $\phi = \pm \eta$, while a potential offset is introduced with 
%
\begin{align} \label{eq:Vbias}
    V_{\rm b} &= V(-\eta) - V(\eta) \nonumber\\
    &= \frac{4 \epsilon \eta^4}{3}.
\end{align}
%
It is this potential offset that induces a pressure force on the walls. The inclusion of a bias directly proportional to $\phi^3$ has been studied by other authors \cite{Kitajima2023, Ferreira2024} and we find roughly consistent results for the network evolution\footnote{We argue that this bias term is quite general in Appendix~\ref{sec:levelA1}.}. The inclusion of a linear term in $\Delta V$ does shift the maximum of the potential from the origin to $\phi \rightarrow -\eta \epsilon/\lambda$, though the energy density in the domain walls only receives a correction $V_{\rm max} \simeq V(\phi=0) + \mathcal{O}(\epsilon^2/\lambda^2)$ which we now argue must be small.

We wish to consider two distinct cases, one where this additional term is present at all times (a so-called vacuum bias), and one in which it switches on at a particular redshift. In the vacuum bias scenario, a large potential difference between the minima can preclude initial domain wall formation at the phase transition, as the lower energy state will be disproportionately populated \cite{Gelmini1988, Saikawa2017}, and any regions of false vacuum that are populated will rapidly collapse during the damping phase of the network. To avoid this uninteresting situation, the amplitude of the bias must not be too large, or more precisely, the ratio $\epsilon/\lambda$ must satisfy
%
\begin{align} \label{eq:vac_bias_limit}
    \frac{\epsilon}{\lambda} \leq \frac{3}{16} \zeta \simeq 0.15, \hspace{15mm} (\textrm{vacuum bias}),
\end{align}
%
where $\zeta \simeq 0.795$ is a dimensionless factor derived from percolation theory \cite{Stauffer1978, Saikawa2017}. Thus, corrections to the domain wall energy density are typically negligible in the vacuum bias scenario. Domain walls can only form through the breaking of a discrete global symmetry. Swampland conjectures \cite{Vafa2005,Banks2010,Harlow2018} argue that there should be no fundamental global symmetries, so it is possible to argue that an effective vacuum bias should always exist, being sourced by higher dimensional operators present in a full quantum gravity theory \cite{King2023,Gouttenoire2025}.

Another option to consider for the bias term is one that switches on at some critical redshift. In this case the phase transition takes place in a fully $\mathbb{Z}_2$ symmetric background, and Eq.~\eqref{eq:vac_bias_limit} is not straightforwardly applicable ($\epsilon \simeq 0$ at time of domain wall formation, with both vacua being equally populated). For the purposes of illustration, imagine a bias that turns on smoothly at some conformal time $\tau$, such as
%
\begin{align} \label{eq:eps_Temp}
    \epsilon(\tau) = \frac{\epsilon_0}{2}\left[1+ \tanh{\left(\frac{\tau - \tau_{\rm crit}}{\Delta \tau}\right)}\right].
\end{align}
%
Here, $\epsilon_0$ is the late-time value of the bias, while $\tau_{\rm crit}$ and $\Delta \tau$ are the central time and duration of the dynamics governing the activation of a bias. We stress that we are not attempting to model any specific scenario here, but instead explore a toy model which could in principle be identified with e.g. non-perturbative or non-trivial effects in other sectors which couple to $\phi$.

In the right hand panel of Fig.~\ref{Fig:Potential}, we illustrate the case where $\epsilon_0 = 2\lambda$. Qualitatively, for $\epsilon/\lambda < 1$, the situation proceeds identically to the case of a vacuum bias. That is, a pressure force is developed as $V_{\rm b}$ increases, which leads to the collapse of the domain wall network. This pressure force gains an additional time dependence which could produce interesting effects. As $\epsilon/\lambda \simeq 1$, we observe an inflection point at $\phi = -\eta$, in which this false vacuum becomes unstable, flipping to be a local maximum. For $\epsilon > \lambda$, a new false vacuum develops at $\phi = -\eta \, \epsilon/\lambda$. 

At this point one might be tempted to conclude that at this inflection point, a second phase transition takes place in the false vacuum regions leading to a new network of domain walls. This is however not the case, as at the onset of this second symmetry breaking, one must once again apply the percolation bound given in Eq.~\eqref{eq:vac_bias_limit}. Application of this bound implies that the old false vacuum located at $\phi = -\eta$ will preferentially fall into the true vacuum state, marking a rapid and violent end to the domain wall network. 

A popular model invoking a temperature dependent bias has been studied recently \cite{Kitajima2023} where it was argued that non-perturbative dynamics during the QCD phase transition could induce wall collapse. Interestingly, the authors found that the production of gravitational waves from this model could potentially explain the PTA signal. The stability of domain wall networks in more general classes of potentials has also been studied \cite{Krajewski2021,Kitajima2023b}.

The presence of a bias creates a pressure between the true and false vacuum given by $p_{\rm b} = V_{\rm b}$, which will eventually overwhelm the contribution coming from the wall tension, $p_{\rm T} \simeq  \sigma/H^{-1}$. Equating these two gives the condition $H(\bar{\tau}_{\rm b}) = V_{\rm b}/\sigma$, which can be recast into the well-known estimate for the timescale of domain wall collapse (assuming radiation domination) \cite{Saikawa2017,Roshan2024}, 
%
\begin{align} \label{eq:TVb}
    T_{\rm b} = 3.41 &\times 10^{-2}  \, C_{\rm b}^{-1/2}  \left(\frac{g_*(T_{\rm b})}{10} \right)^{-1/4} \nonumber\\
    &\times \left( \frac{\sigma}{{\rm TeV}^3}\right)^{-1/2} \left( \frac{V_{\rm b}}{{\rm MeV}^4}\right)^{1/2}\, {\rm GeV},
\end{align}
%
where $C_{\rm b} \simeq 2$ for a $\mathbb{Z}_2$ symmetry, and $g_*$ is the number of relativistic degrees of freedom in the radiation bath at the decay time\footnote{Note that previous works \cite{Saikawa2017,Roshan2024} have claimed a dependence on $\mathcal{A}^{-1/2}$ in Eq.~\eqref{eq:TVb}. This comes from a slightly different definition of the tension-induced pressure, namely $p_{\rm T} = \sigma \mathcal{A}/H^{-1}$. The area parameter encodes information related more to the dynamics of the network as a whole, and not necessarily the radius of curvature of any single domain wall, thus we neglect it. This is also implicitly assumed in other recent work \cite{Kitajima2023b, Ferreira2024}.}. Most work done on domain wall signatures make the assumption that the decay occurs instantaneously, i.e. that the network disappears at precisely $T = T_{\rm b}$. This temperature is then used to fix the peak position and overall amplitude of the gravitational waves. However, as pointed out in our simulations and by other recent work \cite{Kitajima2023,Ferreira2024}, the network is still efficient at radiating gravitational waves for $T \lesssim T_{\rm b}$. As we will show, this leads to an appreciable increase in the amplitude (by roughly two orders of magnitude), a shifting of the peak position to lower frequencies, and a softening of the high frequency spectral index, dependent on the details of the bias parameter. 

Gravitational waves notwithstanding, the collapse of domain wall networks may also be constrained by their decay products. Recently \cite{Ferreira2022}, it was shown that for networks decaying into dark radiation around the time of the QCD phase transition ($T\simeq{\rm MeV}$), the change in the number of relativistic species ($\Delta N_{\rm eff}$) may be sufficient to be detected by the Simons Observatory \cite{SimonsObservatory2018}. Decay into standard model particles could introduce additional signatures, both cosmologically and in colliders, though this scenario contains more model dependence. We do not speculate further on the decay products of such a network, and instead choose to focus strictly on the gravitational wave signatures.

\subsection{\label{sec:level2.2}Gravitational Wave Estimation}
As domain walls enter into the scaling regime, the network becomes dominated by a small number of large walls that run through the Hubble patch. If we imagine for simplicity that we have horizon sized domain walls with a total area parameter $\mathcal{A}$, it is possible to estimate the power emitted in gravitational radiation using the quadrupole formula \cite{Carroll2004}
%
\begin{align}
    \dot{E}_{\rm GW} = \frac{G}{5} \left( \frac{\id^3 J_{ij}}{\id t^3} \, \frac{\id^3 J^{ij}}{\id t^3} \right),
\end{align}
%
where $I_{ij} = \int \id^3 y \,\, y_i \, y_j \, \rho_{\rm dw}(\Vec{y},z)$ is the quadrupole moment of the mass distribution, related to its traceless part by $J_{ij} = I_{ij} - \frac{1}{3} \delta_{ij}\delta^{kl} I_{kl}$. Without loss of generality, we can make the assumption that the domain wall is localized to a region of size $\delta$ about one of the axes (for concreteness, the $z$ axis here). It is then straightforward to show that the dominant contribution to the gravitational wave emission comes from the $J_{xy} \sim \frac{\lambda}{4} \mathcal{A} \eta^4 \delta H^{-4}$ component, yielding
%
\begin{align}
    \dot{E}_{\rm GW} \simeq \mathcal{C} G \mathcal{A}^2 \sigma^2 H^{-2}.
\end{align}
%
Here $\mathcal{C} \sim \mathcal{O}(10)$ is a numerical coefficient determined by simulations. The total energy density in gravitational waves from the network can then be estimated by 
%
\begin{align}
    \rho_{\rm GW} &\simeq \dot{E}_{\rm GW} \frac{H^{-1}}{H^{-3}} \nonumber\\
    &\simeq \mathcal{C} G \mathcal{A}^2 \sigma^2.
\end{align}
%
From this it is clear that the total comoving energy density in gravitational waves remains constant in time during the scaling regime. The full spectrum is typically expressed by 
%
\begin{align} \label{eq:GW-spectrum}
    \Omega_{\rm GW}(k) = \frac{1}{\rho_{\rm c}} \frac{\id \rho_{\rm GW}}{\id \ln k},
\end{align}
%
and has been analytically estimated recently both in the context of non-biased domain walls \cite{Dankovsky2024}, as well as biased ones under the assumption of a velocity-dependent one scale model for the network evolution \cite{Gruber2024}. In this work, we numerically determine the spectrum for a variety of bias parameters by fitting our results to a spectral function 
%
\begin{align} \label{eq:Sk-to-GW}
    S_{\rm k}(t) = \frac{2\pi^2 V a^4(t)}{G} \frac{\id \rho_{\rm GW}}{\id \ln k},
\end{align}
%
where $V$ is the comoving volume of the Universe at time $t$. This expression was initially derived in the context of gravitational waves from preheating \cite{Dufaux2007}, and has since been applied to domain walls \cite{Hiramatsu2013,Saikawa2017}. We remind the reader that in what follows we will use dimensionless variables, denoted with tildes and bars, and we elaborate more on this function in Appendix \ref{sec:levelA1}.

%
\begin{figure*}
\centering 
\includegraphics[width=\columnwidth]{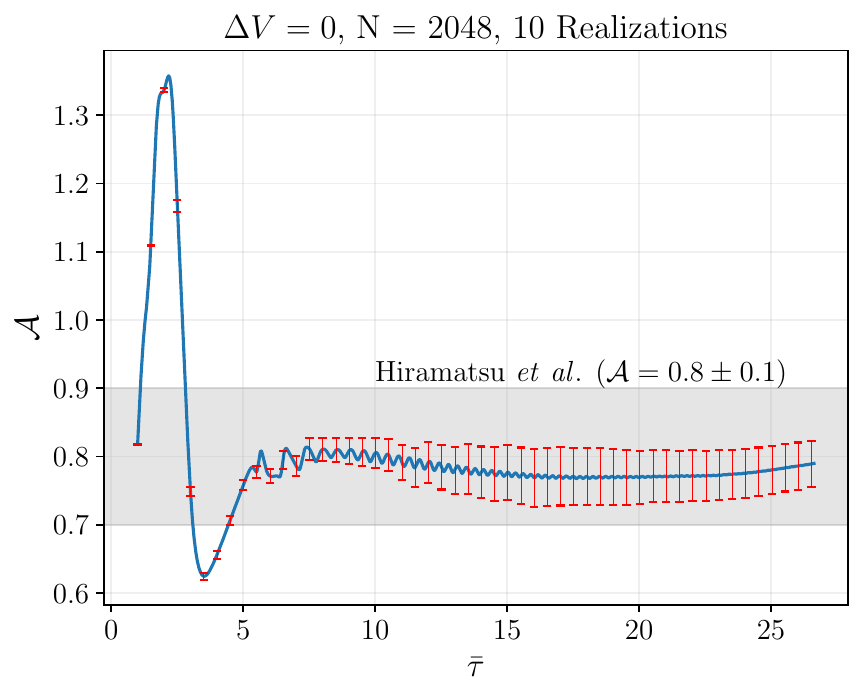}
\hspace{4mm}
\includegraphics[width=\columnwidth]{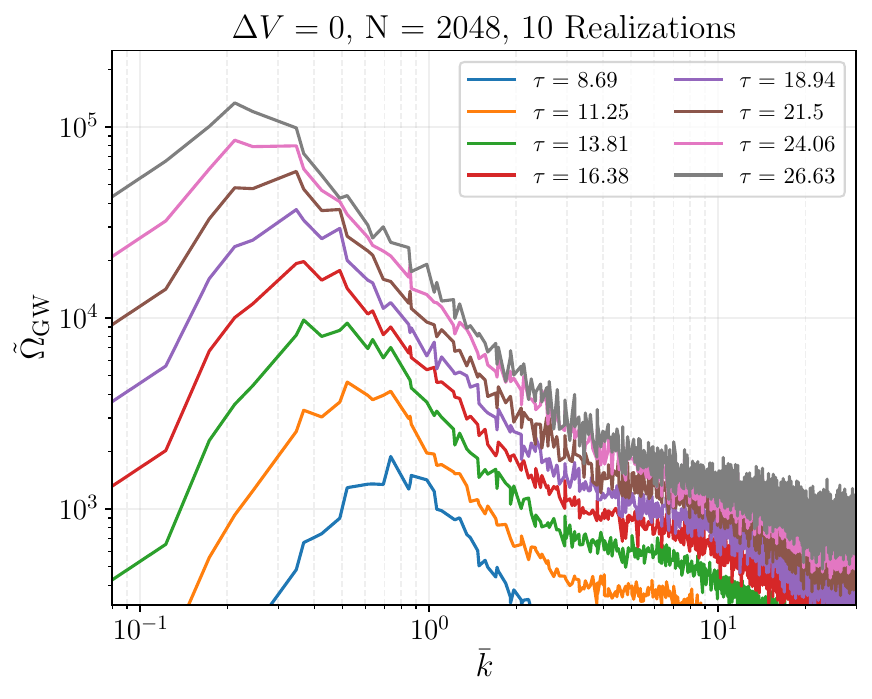}
\caption{Left: The area parameter $\mathcal{A}$ as determined by our simulation for the case of unbiased ($\Delta V = 0$) domain walls. Scaling is achieved at $\bar{\tau} \simeq 7$. The error bars show the $1\sigma$ spread about the mean from ten different realizations of the network. We only plot the errors for $1\%$ of our datapoints to aid visualization. For comparison, we show the uncertainty on $\mathcal{A}$ from previous work \cite{Hiramatsu2013} in the gray shaded region, and following a similar procedure we find an average value from our simulations of $\mathcal{A} = 0.78 \pm 0.03$. Right: The growth of the scaled gravitational wave spectrum at various timeslices in the simulation.}
\label{Fig:No-bias-summary}
\end{figure*}
%

\section{\label{sec:level4}Results}
\noindent Here, we discuss results obtained by performing a suite of lattice simulations to study the evolution of a real scalar field, in both cases of a biased and unbiased $\mathbb{Z}_2$ symmetry. These simulations occur on a comoving cubic grid consisting of $N = 2048^3$ points, and evolve the scalar field in a radiation dominated background. This implies that the physical spacing between gridpoints grows over the course of the simulation. 

We use periodic boundary conditions at the edges of the box and optimize the runtime of the simulation such that light-crossing occurs when the domain walls are only just resolved, with thickness $\delta$ roughly spanning two gridpoints. This provides a maximal dynamic range, allowing us to resolve the damping, scaling, and decay phase of the network for a variety of bias parameters. For each set of parameters, we perform ten independent realizations and average the results.

Throughout the simulation suite, we capture two main quantities. First, the total area of the domain wall network is determined using the well known PRS linking formula \cite{Press1989} at each timestep. This information allows us to reconstruct the dimensionless area parameter $\mathcal{A}$ from the simulation, and provides us with a way to determine whether the network is in the damping, scaling, or decaying regime as a function of time. 

Note that while we will consider the symmetry breaking phase transition to take place during radiation domination, the $\phi$ field itself remains in vacuum (i.e. it is not coupled to the radiation bath). The relationship between the symmetry breaking temperature ($T_{\rm SSB}$) as measured by the photons and the scale $\eta$ is $T_{\rm SSB} \propto \sqrt{\eta \, m_{\rm pl}}$, where $m_{\rm pl} = G_{\rm N}^{1/2}$ is the Planck mass. This is in contrast to a thermal phase transition in which one expects $T_{\rm SSB} \simeq \eta$. Physically, vacuum phase transition are triggered by the time at which Hubble friction releases the scalar field, e.g. when $H_{\rm SSB} \simeq m_{\phi}$. By design, our simulations start at a value of $H_{\rm i} \simeq H_{\rm SSB}$. 

We extract the area parameter for the unbiased case ($\Delta V = 0$) in the left hand plot of Fig.~\ref{Fig:No-bias-summary}. Averaging over our realizations, we find $\mathcal{A} = 0.78 \pm 0.03$, consistent with simulations performed\footnote{Note that area parameters extracted from these works utilized box sizes ranging from $N = 1024^3$  to $N = 3240^3$.} by other authors \cite{Hiramatsu2013, Dankovsky2024, Ferreira2024}. As expected, at early times the network undergoes a transient damping regime where many small domains form, causing an initial rise in $\mathcal{A}$. Scaling in this case is achieved by around $\bar{\tau} \simeq 7$, as indicated by the near-constancy of $\mathcal{A}$ from that point onwards. Error bars represent the standard deviation over the ten realizations. They are computed at every timestep, however we only show $1\%$ of them to avoid overcrowding the plot. 

The second quantity we capture is the amplitude and frequency dependence of $\bar{S}_{\rm k}$, which is related to the gravitational wave spectrum through Eqs.~\eqref{eq:GW-spectrum} and \eqref{eq:Sk-to-GW}. Similar to the area parameter, this is continuously computed, however we choose to only output the spectrum at 11 different timesteps, linearly spaced over the duration of the simulation. For $N=2048^3$ simulations, the initial and final (conformal) times are $\bar{\tau}_{\rm i} = 1$ and $\bar{\tau}_{\rm f} = 26.63$. Details of the numerical calculation of $\bar{S}_{\rm k}$ can be found in Appendix~\ref{sec:levelA1}.

The right hand panel of Fig.~\ref{Fig:No-bias-summary} showcases the evolution of the gravitational wave spectrum in the unbiased case over the course of the simulation. In order to make maximal use of our data, for each realization we compute $\bar{S}_{\rm k}$ along 13 different lines of sight. This allows us to not only reduce uncertainty on our measurements with increased statistics, but also to sample a more dense range of $k$-modes as these 13 projections probe three independent effective grid spacings. Doing this relies on an assumption that the gravitational wave signal is isotropic, which we justify in Fig.~\ref{Fig:Anisotropies} located in the App.~\ref{sec:levelA1} . Not surprisingly, we find that the peak frequency redshifts and the amplitude of the signal increases as the largest domain walls in the simulation radiate gravitational waves at progressively later times. At high frequencies ($\bar{k} \gtrsim 5$ for the final timestep), deviations from a simple broken power law appear. We choose not to speculate on their origin here, instead focusing on performing a comprehensive study of the near-peak region.

\subsection{\label{sec:level3.1}Biased Networks: Lifetimes}

%
\begin{figure}
\centering 
\includegraphics[width=\columnwidth]{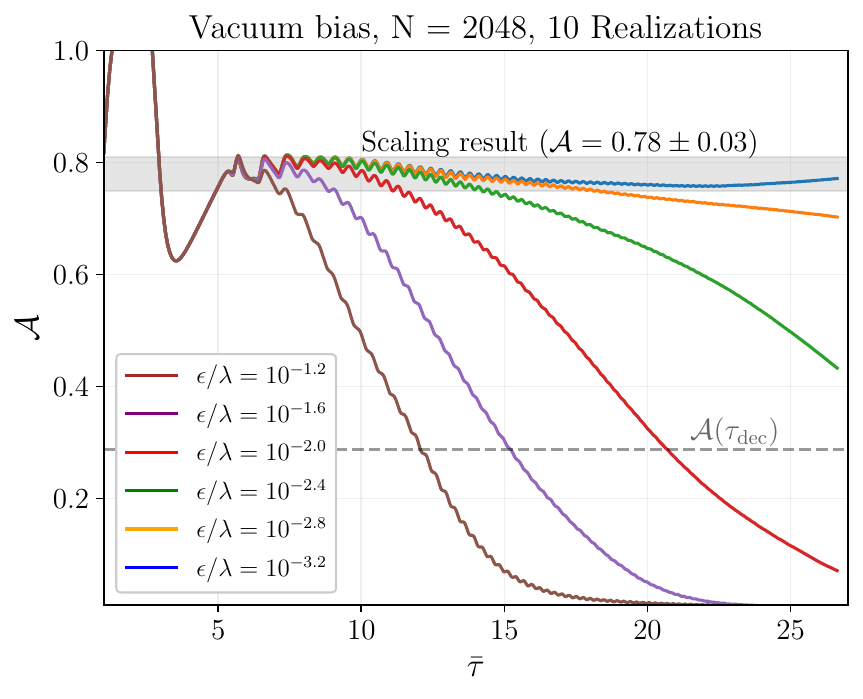}\\
\includegraphics[width=\columnwidth]{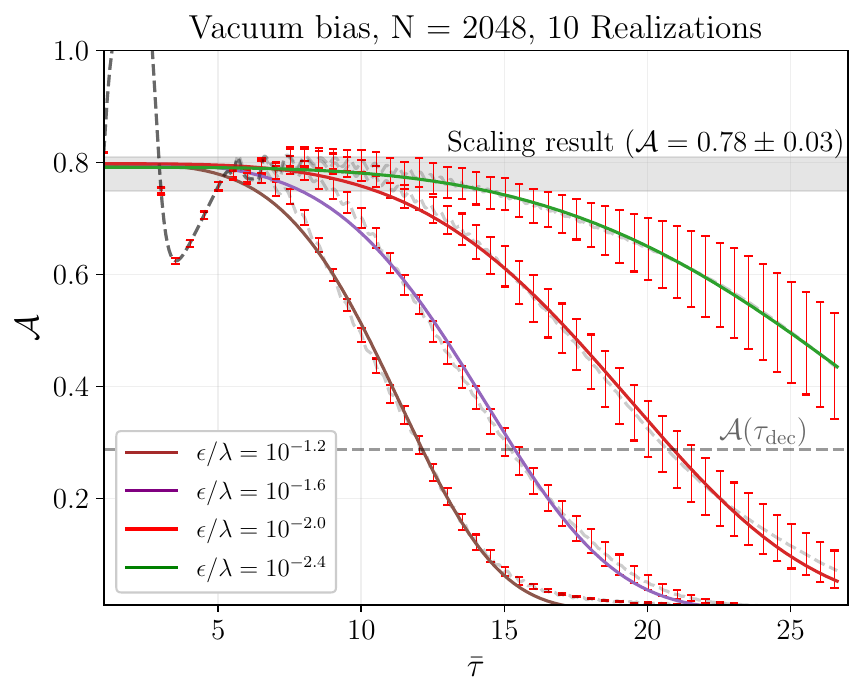}
\caption{Top: Evolution of the area parameter for different choices of the vacuum bias amplitude ($\epsilon/\lambda$). The networks are marginally in the scaling regime for the largest bias considered (brown line, $\epsilon/\lambda = 10^{-1.2}$), while once the bias drops to $\epsilon/\lambda \lesssim 10^{-3}$ the network remains in scaling for the entirety of the simulation. Bottom: Numerical fits to the area parameter following Eq.~\eqref{eq:A-fit}. The solid lines represent the fits, while the error bars encode the standard deviation on the data (dashed lines) over all ten realizations. The mean and standard deviation for the fitting parameters can be found in Table~\ref{tab:A-fits_VB}.}
\label{Fig:Vac-bias-comp}
\end{figure}
%

\noindent We are now in a position to evaluate how the lifetime of a given domain wall network scales with different bias parameters. As discussed earlier, the finite decay time of a domain wall network is expected to significantly change both the peak frequency and amplitude of gravitational waves relative to what is commonly computed using the instantaneous decay approximation. Other attempts have been made to determine this offset \cite{Hindmarsh1996, Larsson1996, Correia2014, Correia2018, Pujolas2022}, though only recently \cite{Kitajima2023b, Ferreira2024} has it become possible to perform simulations with the dynamic range necessary to reliable extract this information.

Recent work \cite{Ferreira2024} has performed a similar analysis by fitting the decay rate of the false volume fraction in their simulations. While we do not track the fraction of gridpoints in the false vacuum, the dimensionless area parameter $\mathcal{A}$ can act as a reasonable proxy for this and hence our results can be easily compared with them. This is due to the fact that the decay will largely be mediated by the collapse of the $\mathcal{O}(1)$ large domain walls present during scaling, thus as $\mathcal{A}$ decreases, so does the false volume fraction. Quantitatively, we fit the area parameter to the following expression
%
\begin{align} \label{eq:A-fit}
    \mathcal{A}(\bar{\tau}) = \mathcal{A}_{\rm scale} \, \, {\rm exp}\left[ - \left(\frac{\bar{\tau}}{\bar{\tau}_{\rm dec}} \right)^p\right],
\end{align}
%
in order to numerically determine the parameters $p$ and $\bar{\tau}_{\rm dec}$. Here, we allow the amplitude to vary between $\mathcal{A}_{\rm scale} = 0.78 \pm 0.03$ as inferred from our unbiased results. We define $\bar{\tau}_{\rm dec}$ as the true decay time of the network, which we then compare with the value of $\bar{\tau}_{ \rm b}$ as computed using the instantaneous decay approximation, $H(\bar{\tau}_{\rm b}) = V_{\rm b}/\sigma$ with $V_{\rm b}$ given in Eq.~\eqref{eq:Vbias}. Note that significant gravitational waves can still be produced for $\bar{\tau} \geq \bar{\tau}_{\rm dec}$ as we will see in the following subsection.

%
\begin{table}[H]
    \centering
    \renewcommand{\arraystretch}{1.5} 
    \begin{tabular}{|c|c|c|c|c|}
        \hline
        $\epsilon/\lambda$ & $\mathcal{A}_{\rm scale}$ & $p$ & $\bar{\tau}_{\rm dec}$ & $\bar{\tau}_{\rm b}$ \rule{0pt}{12pt} \\
        \hline
        $10^{-1.2}$ & $0.797\pm0.003$ & $4.33\pm0.21$ & $12.08\pm0.15$ & 3.35  \\
        $10^{-1.6}$ & $0.798\pm0.005$ & $4.22\pm0.30$ & $15.29\pm0.31$ & 5.31  \\
        $10^{-2.0}$ & $0.798\pm0.007$ & $4.15\pm0.48$ & $20.82\pm0.97$ & 8.41  \\
        $10^{-2.4}$ & $0.793\pm0.009$ & $3.97\pm0.96$ & $31.71\pm5.26$ & 13.33 \\
        \hline
    \end{tabular}
    \caption{Parameters determined by fitting the $\mathcal{A}(\bar{\tau})$ data according to Eq.~\eqref{eq:A-fit} for four choices of vacuum amplitude bias. Uncertainties represent the standard deviation determined by computing the fit to each realization individually.}
    \label{tab:A-fits_VB}
\end{table}
%

First, we consider the case of a vacuum bias, i.e. an $\epsilon$ which is turned on from the beginning of the simulation. The decay of the area parameter for various $\epsilon$ values is shown in Fig.~\ref{Fig:Vac-bias-comp}. After a brief period of damping, the network achieves scaling, before eventually succumbing to the vacuum pressure at some later time. We restrict ourselves to relatively small biases such that the percolation bound given in Eq.~\eqref{eq:vac_bias_limit} is always satisfied. We find that $\epsilon/\lambda = 10^{-1.2}$ is the largest amplitude bias that apparently decays after the network has achieved scaling, with larger values decaying earlier and therefore carrying undesirable information about the initial conditions. In contrast, it appears that for $\epsilon/\lambda \lesssim 10^{-3}$, the network remains in scaling on timescales accessible to our simulation. 

We restrict our fitting procedure to scenarios where the area parameter drops significantly over the course of the simulation. We plot these cases in the bottom panel of Fig.~\ref{Fig:Vac-bias-comp}. In this panel, we also show error bars representing the standard deviation\footnote{As before, we only show error bars on $1\%$ of the datapoints to avoid overcrowding.} determined when combining our ten realizations. Once the network has undergone significant decay, the area fitting algorithm becomes less reliable. As a result, discrepancies may appear between the fitting function and the data for $\bar{\tau} > \bar{\tau}_{\rm dec}$.

The results of the fits are shown in Table~\ref{tab:A-fits_TB}. The amplitude parameter is remarkably stable throughout the different realizations. Interestingly, the central value of the spectral index $p$ shifts to lower values for decreasing $\epsilon/\lambda$, implying that the decay is less sharp. Unfortunately, the large error bars prevent us from making a statistically rigorous statement, though it would be interesting to reassess this result with more simulation data. 

Of significant interest is the comparison between $\bar{\tau}_{\rm dec}$ and $\bar{\tau}_{\rm b}$, where we find that the instantaneous decay timescale predicts that the network will collapse much earlier than it actually does. This is expected, as $\bar{\tau}_{\rm b}$ only defines the moment when the pressure force begins to dominate, and doesn't take into account the fact that a horizon scale structure should take at least one Hubble time to fully collapse.

While we find that the ratio $\bar{\tau}_{\rm dec}/\bar{\tau}_{\rm b}$ decreases as $\epsilon/\lambda$ is lowered, we provide a word of warning. While clearly $\bar{\tau}_{\rm b}$ does not do a good job of defining the lifetime of the network, it can be useful as a tool to diagnose when deviations away from scaling should be expected. Therefore, when $\bar{\tau}_{\rm b}$ provides a value corresponding to a time when our simulation is still in the damping phase, caution should be exercised. This is certainly the case for $\epsilon/\lambda = 10^{-1.2}$, which implies that pressure forces may have played a non-trivial role in the approach of the network to scaling. 

With this in mind, perhaps one of the more informative fits to gleam qualitative information from is the $\epsilon/\lambda = 10^{-2.4}$ case. Indeed, here we can see that at $\bar{\tau}_{\rm b} \simeq 13$, the area parameter has just started to drop by more than $1\sigma$ away from its scaling value. From the $\bar{\tau}_{\rm dec}/\bar{\tau}_{\rm b}$ ratio it is clear that the decay of the network is a rather slow process, drawn out by more than a Hubble time. As a rough estimate, the delay appears to provide an offset of $\bar{\tau}_{\rm dec}/\bar{\tau}_{\rm b} \simeq 3$, or in other words, the true decay temperature of the network is set by $T_{\rm dec} \simeq T_{\rm b}/3$ where $T_{\rm b}$ is given in Eq.~\eqref{eq:TVb}. As we will show, this has a large effect on the inferred amplitude and peak position of the resultant gravitational wave spectrum.

%
\begin{figure}
\centering 
\includegraphics[width=\columnwidth]{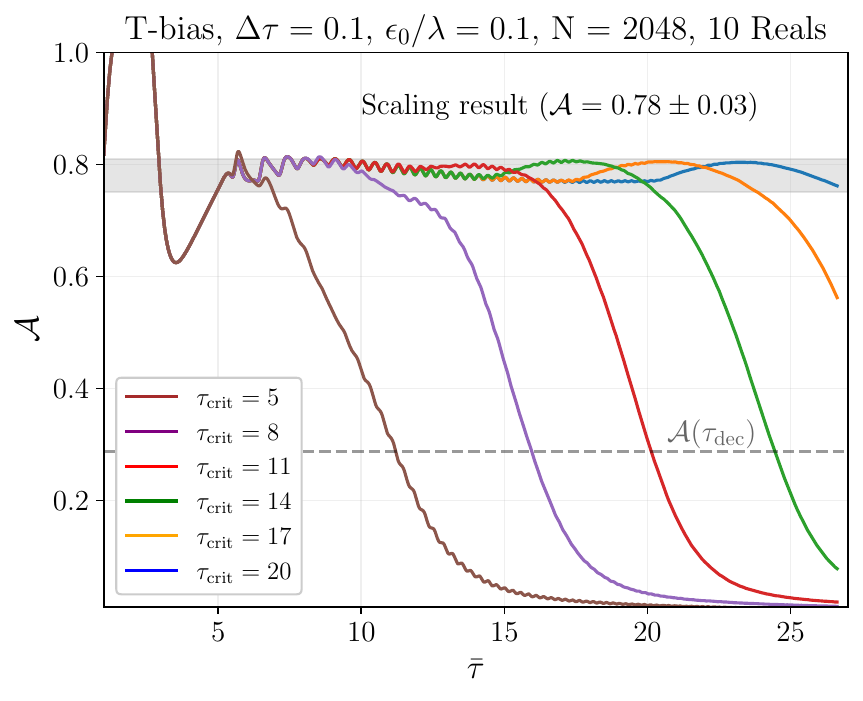}\\
\includegraphics[width=\columnwidth]{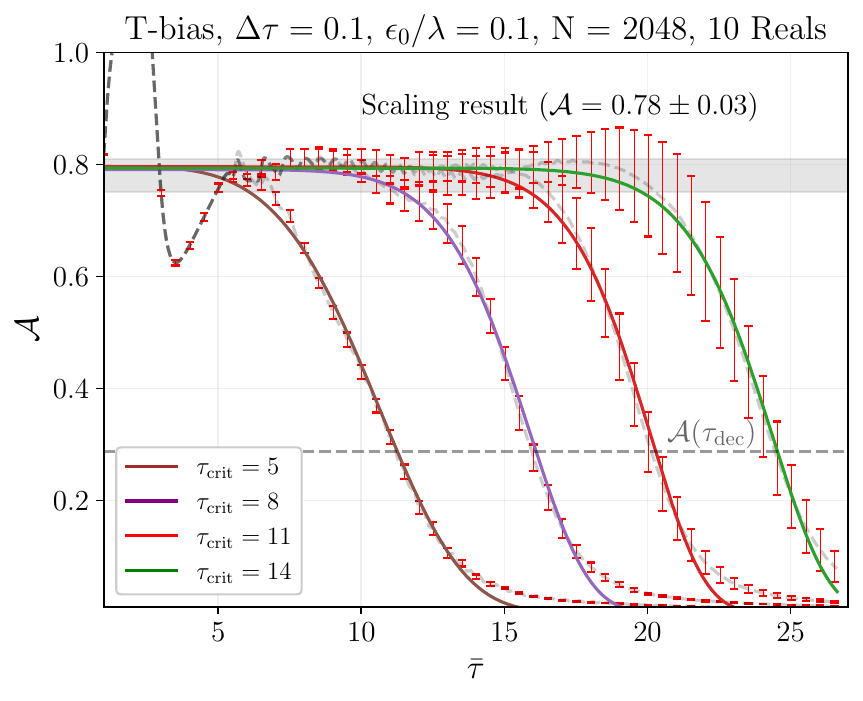}
\caption{Same as Fig.~\ref{Fig:Vac-bias-comp}, but for the temperature dependent bias. Unlike the vacuum bias case, external energy is injected into the system as $\epsilon(\tau)$ increases, which for certain parameter values can give rise to numerical instabilities. Therefore, We choose to fix $\Delta \bar{\tau}$ and $\epsilon_0/\lambda$ while only varying $\bar{\tau}_{\rm crit}$.}
\label{Fig:T-bias-comp}
\end{figure}
%
%
\begin{figure*}
\centering 
\includegraphics[width=\columnwidth]{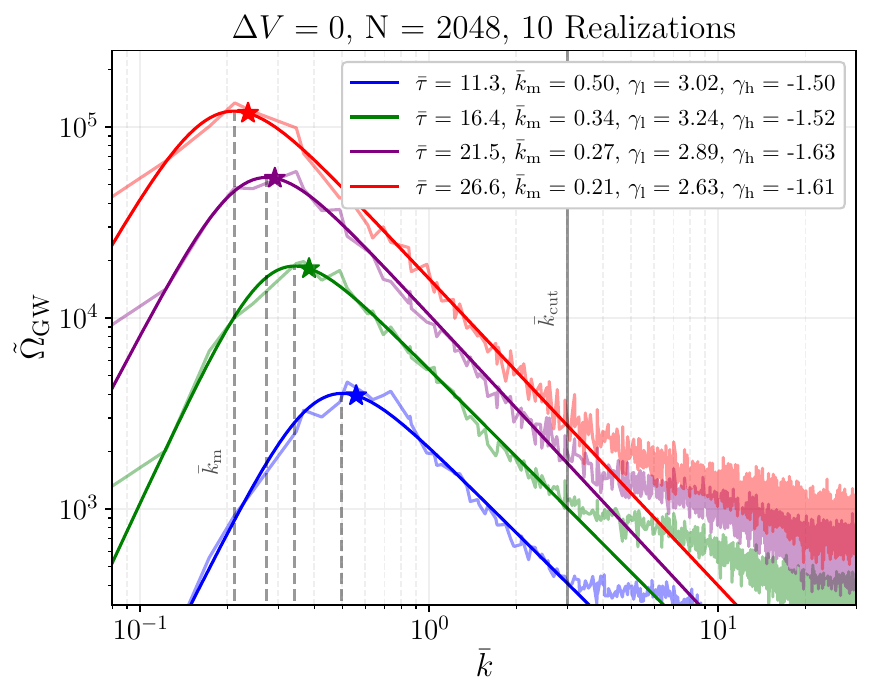}
\hspace{4mm}
\includegraphics[width=\columnwidth]{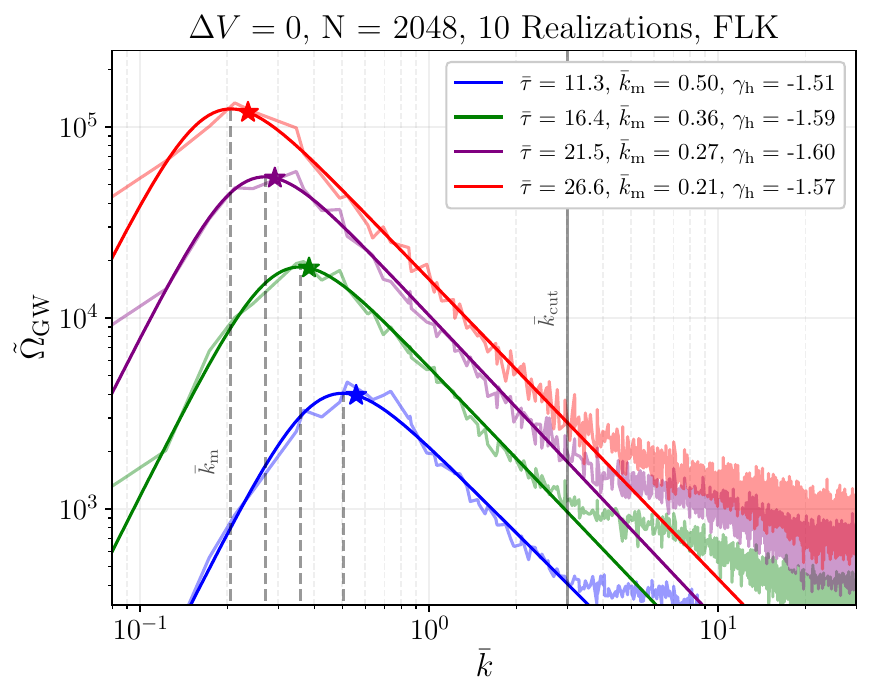}
\caption{The gravitational wave spectrum at different timesteps for an unbiased domain wall network. Left: We allow the causal (low-$k$) slope to vary as a free parameter. Right: We fix $\gamma_{\rm l} = 3$ to match theoretical expectations. Note that $\bar{k}$ is the comoving momentum mode in units of $\sqrt{\lambda} \eta$, and $\Omega_{\rm GW} = \tilde{\Omega}_{\rm GW} (\eta/m_{\rm pl})^4$ is the relationship between the physical spectrum (at time $\bar{\tau}$) and the rescaled version. Stars represent the comoving Hubble scale in numerical units, and we find $\bar{\mathcal{H}}/\bar{k} _{\rm m}= 0.91 \pm 0.02$ for the unbiased fixed lower$-k$ (FLK) scenario, in which we set $\gamma_{\rm l} = 3$, its causally expected value.}
\label{Fig:No-bias-timeslice-fits}
\end{figure*}
%

Now we turn to the results from the temperature dependent bias, as described by Eq.~\eqref{eq:eps_Temp}. In this setup, we once again choose a value for the amplitude of the bias (given here by $\epsilon_0/\lambda$), which is then activated at some value $\bar{\tau}_{\rm crit}$ and over some period $\Delta \bar{\tau}$. This is meant to serve as a toy model to describe a scenario in which non-trivial couplings between $\phi$ and an additional field are ``turned on" due to dynamics in the auxiliary sector. In practice, we fix the overall bias amplitude to be $\epsilon_0/\lambda = 0.1$ and take the activation timescale to be $\Delta \bar{\tau} = 0.1$, while examining different critical times $\bar{\tau}_{\rm crit}$. The data for this is presented in the top panel of Fig.~\ref{Fig:T-bias-comp}. 

%
\begin{table}[H]
    \centering
    \renewcommand{\arraystretch}{1.5} 
    \begin{tabular}{|c|c|c|c|c|}
        \hline
        $\bar{\tau}_{\rm crit}$ & $\mathcal{A}_{\rm scale}$ & $p$ & $\bar{\tau}_{\rm dec}$ & $\bar{\tau}_{\rm b}$ \rule{0pt}{12pt} \\
        \hline
        $5$ & $0.796\pm0.002$ & $4.64\pm0.17$ & $11.21\pm0.11$ & 4.95\\
        $8$ & $0.791\pm0.008$ & $8.74\pm0.94$ & $16.07\pm0.17$ & 7.90 \\
        $11$ & $0.796\pm0.010$ & $11.98\pm1.92$ & $20.22\pm0.37$ & 10.86 \\
        $14$ & $0.795\pm0.014$ & $14.28\pm3.70$ & $24.42\pm0.58$ & 13.84 \\
        \hline
    \end{tabular}
    \caption{Temperature dependent bias fit results, all performed at $\epsilon_0/\lambda = 0.1$ and $\Delta \bar{\tau} = 0.1$. As expected, the instantaneous decay approximation finds $\bar{\tau}_{\rm b} \simeq \bar{\tau}_{\rm crit} < \bar{\tau}_{\rm dec}$. The spectral index $p$ grows with later decays, indicating a more violent decay due to the large pressure buildup associated with these cases. }
    \label{tab:A-fits_TB}
\end{table}
%

Recall that deviations of $\mathcal{A}$ away from scaling can only begin after the instantaneous decay timescale, $\bar{\tau}_{\rm b}$. For large values of $\epsilon_0/\lambda$, this will typically occur at $\bar{\tau} \simeq \bar{\tau}_{\rm crit}$. However, as one lowers the bias amplitude, the vacuum equivalent decay timescale\footnote{We define this timescale as the one computed when considering a vacuum bias with amplitude $\epsilon_0/\lambda$.} ($\bar{\tau}^0_{\rm b}$) increases, and the true value of $\bar{\tau}_{\rm b}$ can be found through the expression $\bar{\tau}_{\rm b} = {\rm Max}(\bar{\tau}_{\rm crit}, \bar{\tau}^0_{\rm b})$. 

To probe the effects of a time-dependent bias, we should therefore only consider scenarios in which the instantaneous decay timescale is set by $\bar{\tau}_{\rm crit}$. We achieve this by fixing $\epsilon_0/\lambda = 0.1$ (corresponding to a $\bar{\tau}_{\rm b}^0 \simeq 2.66$) throughout these simulations. In order to further amplify the effects of a temperature-dependent bias, we also choose a relatively short activation timescale, $\Delta \bar{\tau} = 0.1$. Smaller values (shorter activation times) led to some numerical instabilities, thus we don't discuss them further. 

With these two parameters fixed, we vary $\bar{\tau}_{\rm crit} = (5,8,11,14,17,20)$. Two main differences arise when compared against the vacuum case. First, the activation of the bias acts as a type of energy injection into the $\phi$ field, which seems to briefly raise the amplitude of the area parameter (though not by more than $1\sigma$) before decay takes place. Second, the decay rate itself appears to be much sharper than the vacuum case, a statement which is justified by examining the $p$ values in Table~\ref{tab:A-fits_TB}. The reason for this is rather intuitive. In the vacuum case, the onset of decay was always set by the gradual and monotonic rise of the bias pressure. In contrast, for $\bar{\tau} \leq \bar{\tau}_{\rm crit}$, this same pressure is held behind a floodgate far past when it would have become dominant. When the floodgate finally bursts at $\bar{\tau} \approx \bar{\tau}_{\rm crit}$, this excess buildup of pressure mediates a much more rapid and violent collapse of the network. As one would expect, the larger one sets $\bar{\tau}_{\rm crit}$, the steeper the decay.

\subsection{Gravitational Waves from Biased Networks}
%
\begin{figure*}
\centering 
\includegraphics[width=\columnwidth]{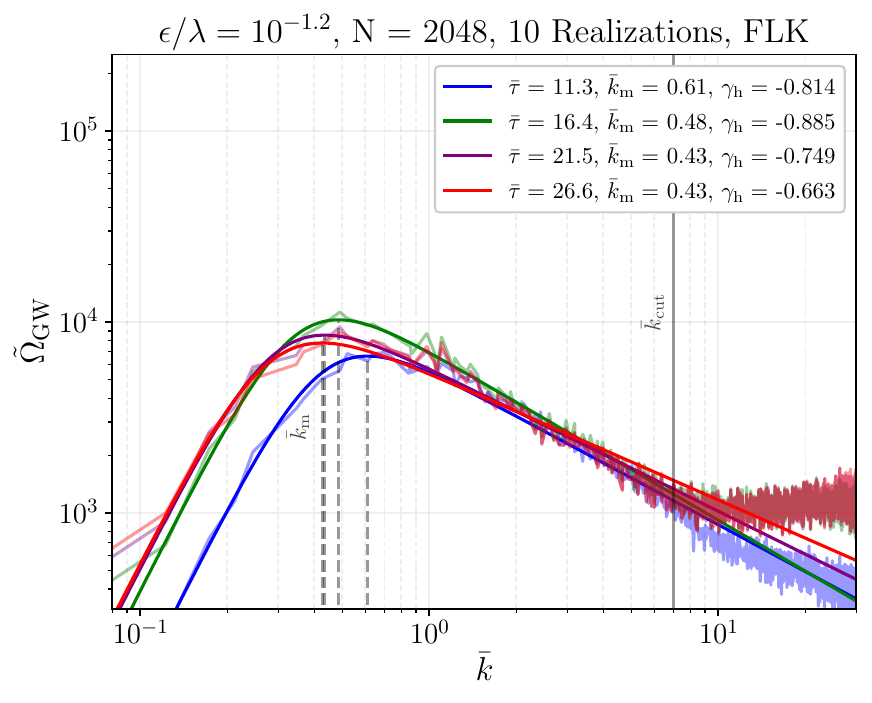}
\hspace{4mm}
\includegraphics[width=\columnwidth]{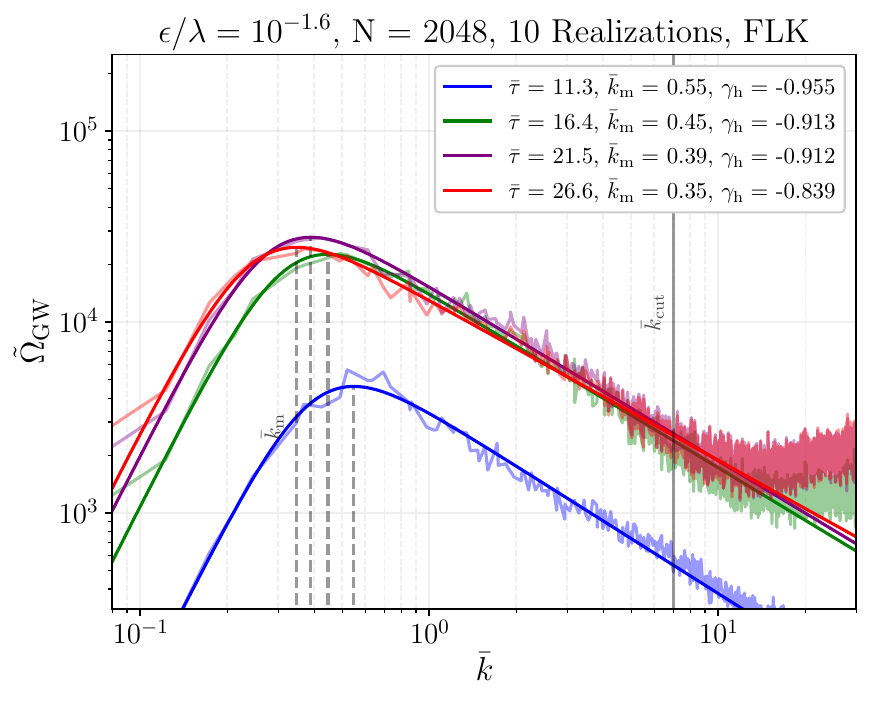}\\
\includegraphics[width=\columnwidth]{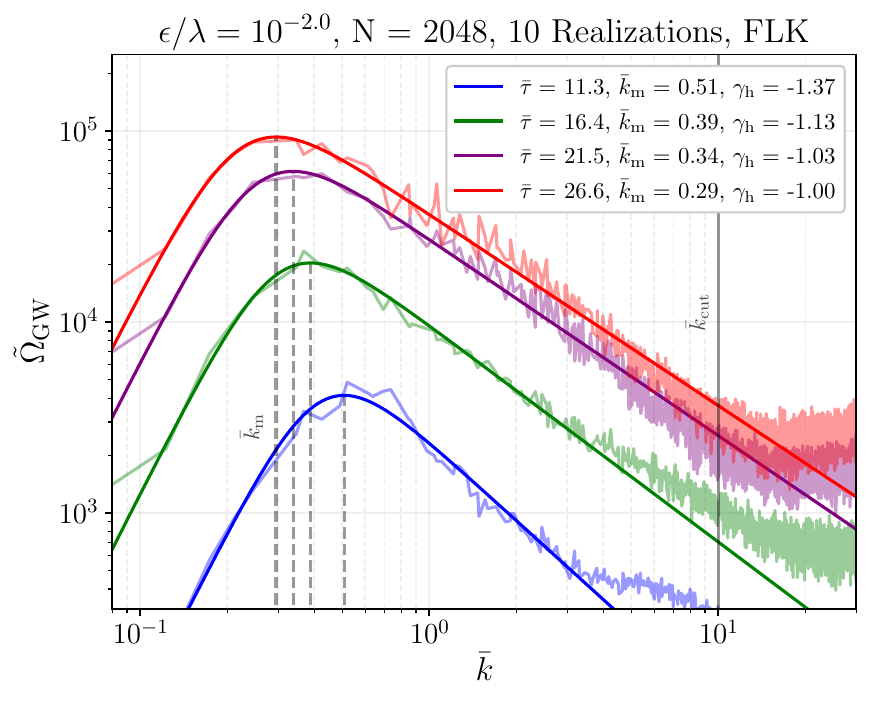}
\hspace{4mm}
\includegraphics[width=\columnwidth]{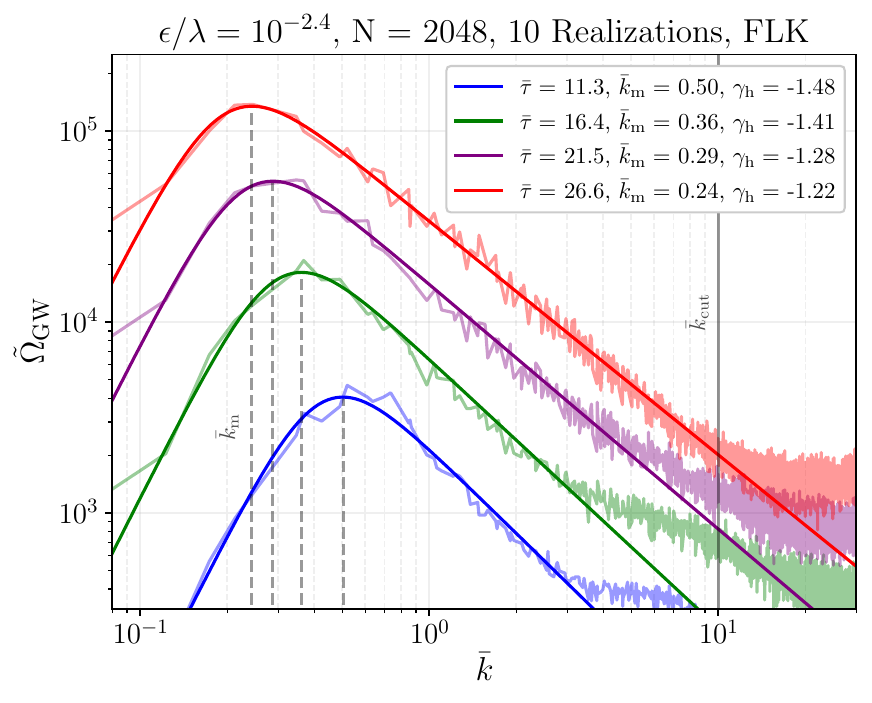}\\
\includegraphics[width=\columnwidth]{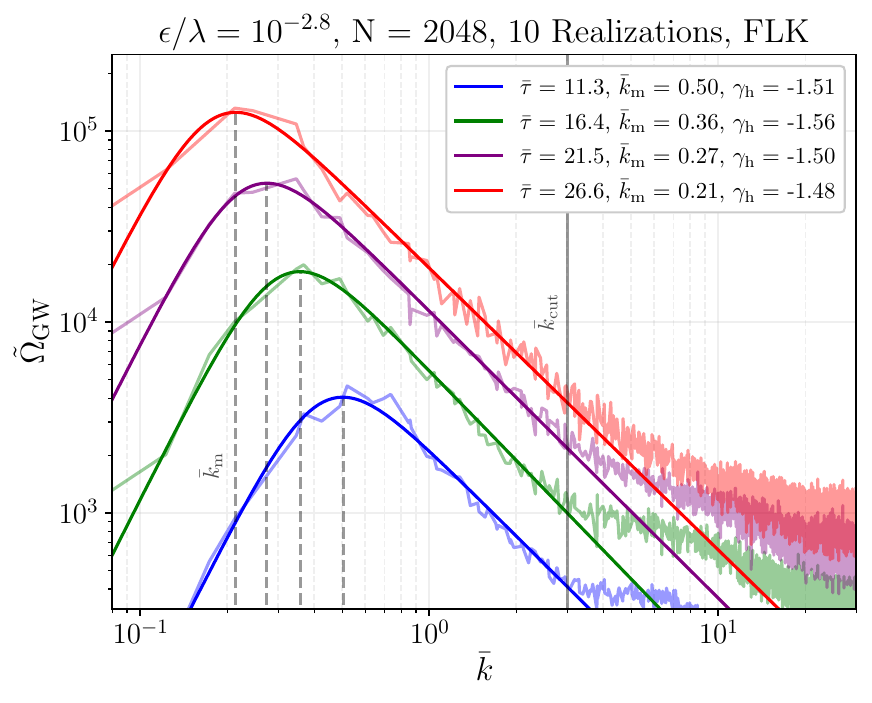}
\hspace{4mm}
\includegraphics[width=\columnwidth]{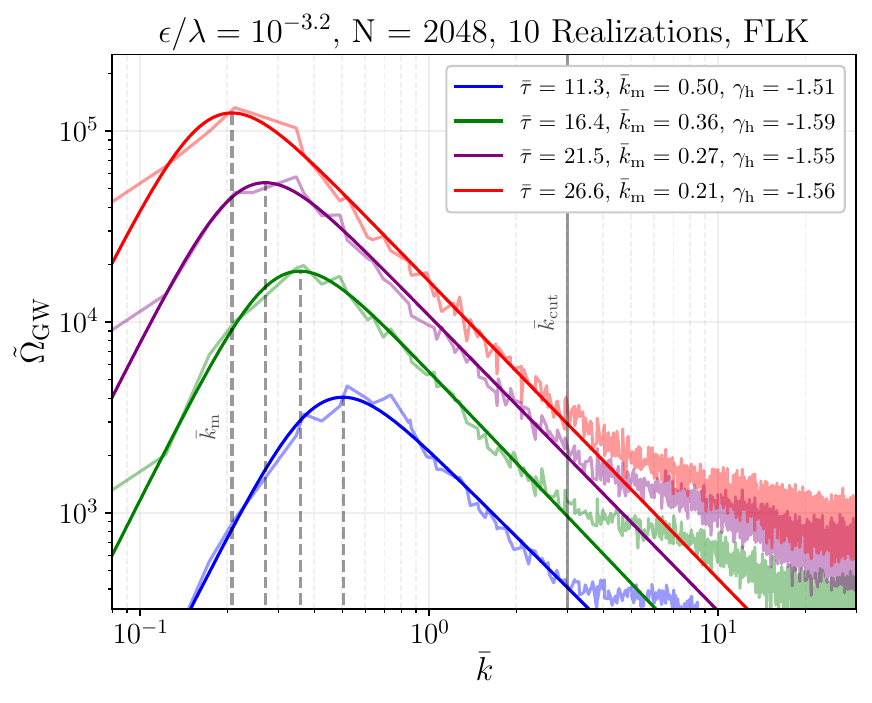}
\caption{A comparison of the gravitational wave spectra for various vacuum bias parameters at different times during the simulation. For all plots we fix the lower-$k$ (FLK) slope to its expected causal value of $\gamma_{\rm l} = 3$. The position of $\bar{k}_{\rm cut}$ represents the highest $\bar{k}$-mode fit for the final ($\bar{\tau} \simeq 26$) timestep. Values of this cutoff for earlier timesteps can be found in Appendix~\ref{sec:levelA4}.}
\label{Fig:Vac-bias-timeslice-fits-comparison}
\end{figure*}
%

\setlength{\tabcolsep}{10pt}
%
\begin{table*}[t]
    \centering
    \renewcommand{\arraystretch}{1.5} 
    \begin{tabular}{|c|c|c|c|c|c|c|}
        \hline
        $\epsilon/\lambda$ & $\bar{k}_{\rm cut}$ & $\bar{k}_{\rm m}$ & $\tilde{\Omega}(\bar{k}_{\rm m})$ & $\bar{k}_{\rm width}$ & $\gamma_{\rm h}$ & $\bar{\tau}_{\rm dec}$ \rule{0pt}{12pt} \\
        \hline
        $10^{-1.2}$ & $7$ & $0.436\pm0.035$ & $(7.41\pm0.85)\times 10^{3}$ & $1.313\pm0.256$ & $-0.65 \pm 0.095$ & $12.08 \pm 0.15$\\
        $10^{-1.6}$ & $6$ & $0.349\pm0.020$ & $(2.39\pm0.37)\times 10^{4}$ & $0.764\pm0.152$ & $-0.82 \pm 0.100$ & $15.29 \pm 0.31$\\
        $10^{-2.0}$ & $10$ & $0.297\pm0.028$ & $(9.10\pm1.52)\times 10^{4}$ & $0.494\pm0.066$ & $-1.01 \pm 0.081$ & $20.82 \pm 0.97$\\
        $10^{-2.4}$ & $10$ & $0.241\pm0.020$ & $(1.30\pm0.23)\times 10^{5}$ & $0.323\pm0.021$ & $-1.22 \pm 0.122$ & $31.71^* \pm 5.26$\\
        $10^{-2.8}$ & $3$ & $0.212\pm0.025$ & $(1.20\pm0.22)\times 10^{5}$ & $0.232\pm0.016$ & $-1.47 \pm 0.167$ & N/A\\
        $10^{-3.2}$ & $3$ & $0.206\pm0.026$ & $(1.20\pm0.21)\times 10^{5}$ & $0.213\pm0.014$ & $-1.55 \pm 0.198$ & N/A\\
        $0$ & $3$ & $0.205\pm0.026$ & $(1.20\pm0.21)\times 10^{5}$ & $0.211\pm0.015$ & $-1.56 \pm 0.201$ & N/A\\
        \hline
    \end{tabular}
    \caption{Gravitational wave spectral parameters extracted from the final timestep in the simulation ($\bar{\tau}_{\rm f} \simeq 26$). Error bars represent the standard deviation over the ten realizations of each $\epsilon/\lambda$ values considered. Here, $\bar{k}_{\rm cut}$ refers to the maximum value of $\bar{k}$ that we fit using Eq.~\eqref{eq:Sk_fit} in order to avoid contamination from the high frequency modes. We also restate the $\bar{\tau}_{\rm dec}$ derived in Table~\ref{tab:A-fits_VB} from fitting the area parameter, where an asterisk represents an decay time beyond the dynamic range of our simulation.}
    \label{tab:GWs-comp}
\end{table*}
%
\setlength{\tabcolsep}{6pt}
Let us now turn to the spectrum of gravitational waves produced by networks of domain walls. During the simulation, we compute the spectral function $\bar{S}_{\rm k}$ throughout the evolution and decay of the network\footnote{Note the transformation $S_{\rm k} = \bar{S}_{\rm k} \cdot  (\eta^3/\sqrt{\lambda})$ which allows us to compute the physically relevant quantity $\Omega_{\rm GW}$.} following the procedure laid out in Ref. \cite{Dufaux2007}. As expected from previous results, this spectrum is well fit by a broken power law whose peak frequency is given by roughly the scale of the Hubble horizon at the time of network decay. To more quantitatively determine this scale, we fit the data to an $\bar{S}_{\rm k}$ of the form
%
\begin{align} \label{eq:Sk_fit}
    \bar{S}_{\rm k} = \alpha\left( \frac{\bar{k}}{\bar{k}_{\rm p}} \right)^{\gamma_l} \left[1+\beta \left(\frac{\bar{k}}{\bar{k}_{\rm p}} \right)^{\gamma_l - \gamma_h} \right]^{-1}.
\end{align}
%
Here, $\bar{k}_{\rm p}$ is the pivot scale, $\alpha$ and $\beta$ are free and dimensionless fitting parameters which encode the overall amplitude of $\bar{S}_{\rm k}$, while $\gamma_{\rm l}$ and $\gamma_{\rm h}$ are the low- and high-$k$ slopes respectively. Note that the pivot scale ($\bar{k}_{\rm p}$) and the peak scale ($\bar{k}_{\rm m}$) are not identical in general, but are related by the expression
%
\begin{align}
    \bar{k}_{\rm m} = \bar{k}_{\rm p} \, \left|\frac{1}{\beta}\frac{\gamma_l}{ \gamma_{\rm h}} \right|^{\frac{1}{\gamma_{\rm l} - \gamma_{\rm h}}}. \nonumber
\end{align}
%
We show the raw data and the results of this fitting procedure for the unbiased case in Fig.~\ref{Fig:No-bias-timeslice-fits}, highlighting four specific timeslices. The left hand panel shows the evolution of the fit to all five free parameters ($\alpha,\beta,\bar{k}_{\rm p},\gamma_{\rm l},\gamma_{\rm h}$), while in the right hand panel we simply fix $\gamma_{\rm l} = 3$, which is expected to be a universal feature \cite{Dankovsky2024,Caprini2009,Cai2019}. Generally speaking, the quality of the fits do not suffer much degradation by fixing this parameter, so in what follows we eliminate $\gamma_{\rm l}$ as a free parameter. We also restrict our fitting procedure to values of $\bar{k} \leq \bar{k}_{\rm cut}$ as labeled in the plots in order to avoid contamination from the high frequency part of the spectrum, whose origin we have chosen not to investigate. 

We express the amplitude of the gravitational wave signal by $\tilde{\Omega}_{\rm GW}$ derived through purely numerical quantities. The physical gravitational wave amplitude (at time $\bar{\tau}$) can be recovered by the simple scaling relation $\Omega_{\rm GW} = \tilde{\Omega}_{\rm GW} (\eta/m_{\rm pl})^4$. Note that the simulations take place in vacuum, which means that the scale $\eta$ is related to a physical spontaneous symmetry breaking temperature by $\eta \simeq T_{\rm SSB}^2/m_{\rm pl}$.

We found that for the fixed lower-$k$ (FLK) scenario, the peak of the gravitational wave spectrum lies at $k_{\rm m} = H/f_{\rm H}$, where $f_{\rm H} = 0.91 \pm 0.02$. Importantly, we also found the spectral index to the right of the peak to be $\gamma_{\rm h} = -1.56 \pm 0.201$ at the end of the simulation. This is in some disagreement with the often cited value of $\gamma_{\rm h} = -1$ \cite{Hiramatsu2010,Hiramatsu2013}, though it is in agreement with other more contemporary studies \cite{Ferreira2022,Ferreira2024, Dankovsky2024} who appear to find $\gamma_{\rm h} \in (-1.7,-1.5)$.

In Fig.~\ref{Fig:Vac-bias-timeslice-fits-comparison}, we show similar plots for each of the vacuum bias parameters considered in our simulation suite. Table~\ref{tab:GWs-comp} provides supplemental information such as uncertainties, decay timescales computed from the area parameter data (see Sec.~\ref{sec:level3.1}), and choices for the fitting cutoff scale $\bar{k}_{\rm cut}$. Note that the central values presented in Table~\ref{tab:GWs-comp} differ marginally from those presented in Fig.~\ref{Fig:Vac-bias-timeslice-fits-comparison} due to the fact that the curves computed in the figure are fit to the combined datasets, as opposed to an average over fitting each dataset independently. The data presented in this table are for the final ($\bar{\tau} \simeq 26$) timestep in the simulation, while similar information for earlier timesteps can be found in Appendix~\ref{sec:levelA4} for all of cases considered.

We clearly observe the expected behaviour of the amplitude and peak position of the gravitational wave spectrum as the $\epsilon/\lambda$ parameter is increased. In scenarios where we can reliably resolve all three phases (damping, scaling, and decay) of network evolution, larger $\epsilon/\lambda$ values lead to lower overall $\Omega_{\rm GW}$ as the network fully decays at some $\bar{\tau} < \bar{\tau}_{\rm f}$. They also exhibit a relative blueshifting of the peak position as the characteristic $k$-mode emitted by the network corresponds roughly to $k \simeq 0.9 \, H_{\rm dec}$. 

\begin{figure}
\centering 
\includegraphics[width=\columnwidth]{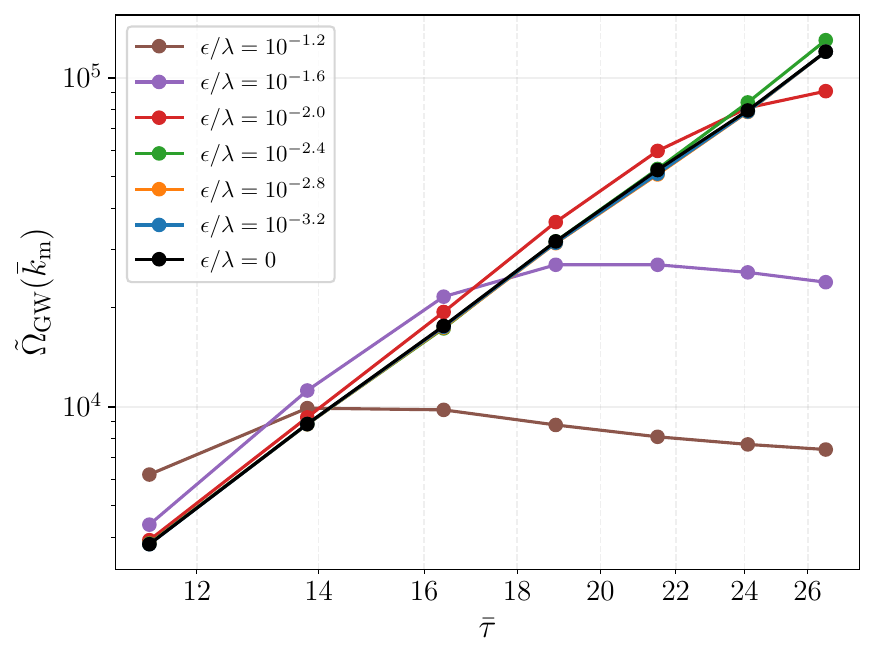}\\
\includegraphics[width=\columnwidth]{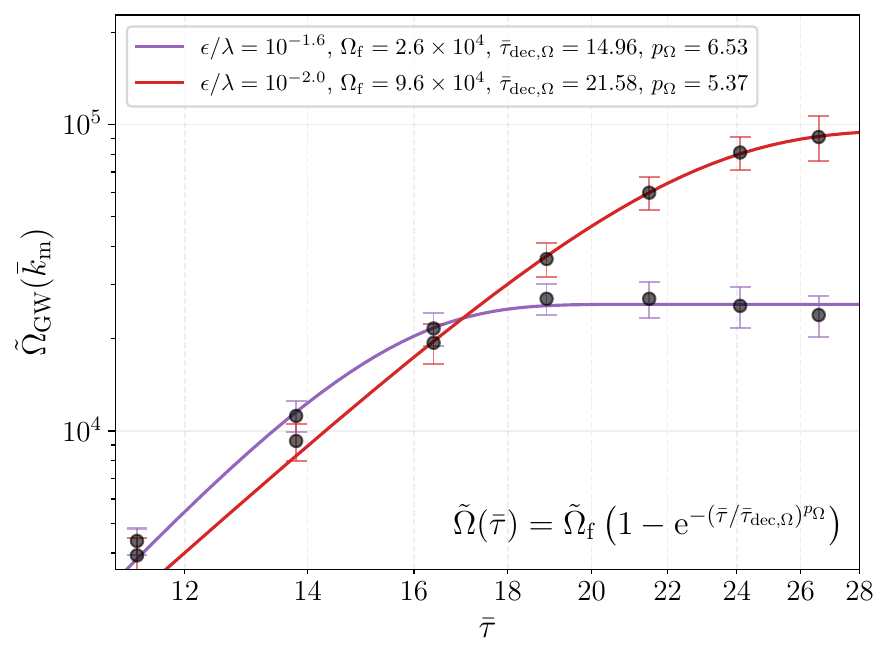}
\caption{Top: Data at seven different timesteps throughout the simulation for each choice of vacuum bias parameter. Note that both $\epsilon/\lambda = 10^{-1.2}, 10^{-1.6}$ have already begun to deviate from scaling at the earliest timestep. Bottom: Fitting the data for our two most robust simulations, one finds that $\bar{\tau}_{\rm dec, \Omega}$ is within $1\sigma$ of the $\bar{\tau}_{\rm dec}$ inferred from fitting the area parameter (see Table~\ref{tab:A-fits_VB} or \ref{tab:GWs-comp}).}
\label{Fig:OGW-summary}
\end{figure}

To gain a better understanding on the evolution of $\Omega_{\rm GW}$, in Fig.~\ref{Fig:OGW-summary} we show the comparative growth of this quantity over seven timesteps in the simulation, running from roughly $\bar{\tau} \simeq 11$ until $\bar{\tau} = \bar{\tau}_{\rm f} \simeq 26$. In the top panel, the amplitude of the spectrum grows monotonically and near-identically for the no bias, $\epsilon/\lambda = 10^{-3.2}$, and $\epsilon=10^{-2.8}$ cases. Small deviations near the end of the simulation begin to emerge for $\epsilon/\lambda  = 10^{-2.4}$, as the network enters into its collapse stage and energy is more abundantly liberated into gravitational waves. 

As we go to larger values for the vacuum bias, more of the collapse phase enters into our dynamic range and we observe interesting differences from monotonic growth. For example, the $\epsilon/\lambda = 10^{-1.6},10^{-2.0}$ cases both exhibit a period of initial growth over the $\Delta V = 0$ (no-bias) scenario. Interestingly, this deviation seems to appear at roughly the timescale indicated by the instantaneous decay approximation, namely $H(\tau_{\rm b}) = V_{\rm b}/\sigma$. This is perhaps unsurprising, as it is precisely this time that the bias begins to have an $\mathcal{O}(1)$ effect on the global evolution of the walls which causes an enhancement to the amplitude. 
%
%

In the bottom panel of Fig.~\ref{Fig:OGW-summary}, we perform a fit to the amplitude for these two cases, similar in spirit to what was done for the area parameter. The decay timescale inferred from this fitting procedure was found to be $\bar{\tau}_{\rm dec,\Omega} = 14.96$ for $\epsilon/\lambda = 10^{-1.6}$, and  $\bar{\tau}_{\rm dec,\Omega} = 21.58$ for $\epsilon/\lambda = 10^{-2.0}$, both within $1\sigma$ of their respective $\bar{\tau}_{\rm dec}$ values computed in Sec.~\ref{sec:level3.1}. Recall also that use of the instantaneous decay approximation would have wrongly concluded that the network ceases to emit gravitational waves after $\bar{\tau}_{\rm b} = 5.31$ and $8.41$ respectively. Fig.~\ref{Fig:OGW-summary} shows that neglecting gravitational waves from the collapse phase would have resulted in a prediction for the gravitational wave amplitude roughly two orders of magnitude lower than what we find in our simulations\footnote{It should also be noted that $\Omega_{\rm GW}$ continues to grow slightly even for $\bar{\tau} \gtrsim \bar{\tau}_{\rm dec}$, e.g. in the $\epsilon/\lambda = 10^{-1.6}$ we see that we reach our asymptotic value at roughly $\bar{\tau} \simeq 18$}. 

The astute reader will notice that we have chosen not to fit the $\epsilon/\lambda = 10^{-1.2}$ case. To echo arguments made in Sec.~\ref{sec:level3.1}, a level of theoretical uncertainty exists that is difficult to quantify. In this case, $\mathcal{O}(1)$ corrections to the potential become important around $\bar{\tau}_{\rm b} = 3.35$, at which time it is clear that the domain wall network has not yet achieved scaling (see the top panel of Fig.~\ref{Fig:Vac-bias-comp}). Thus, while by eye the amplitude of $\Omega_{\rm GW}$ and its time dependence appear reasonable, we view this data with caution as undesirable dynamics during the damping stage of network evolution may have had an impact on the overall normalization of the gravitational wave amplitude.

\begin{figure}
\centering 
\includegraphics[width=\columnwidth]{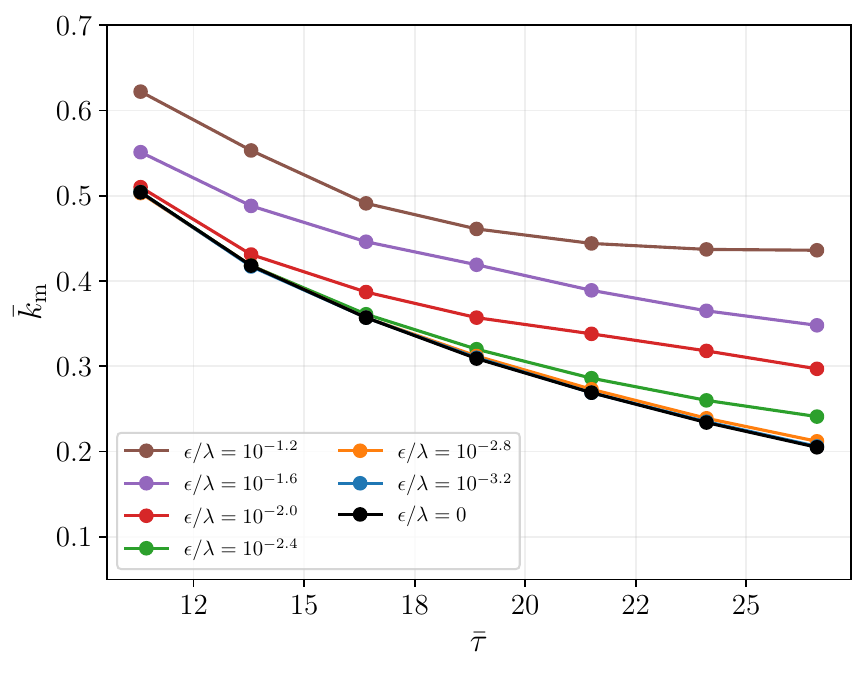}\\
\includegraphics[width=\columnwidth]{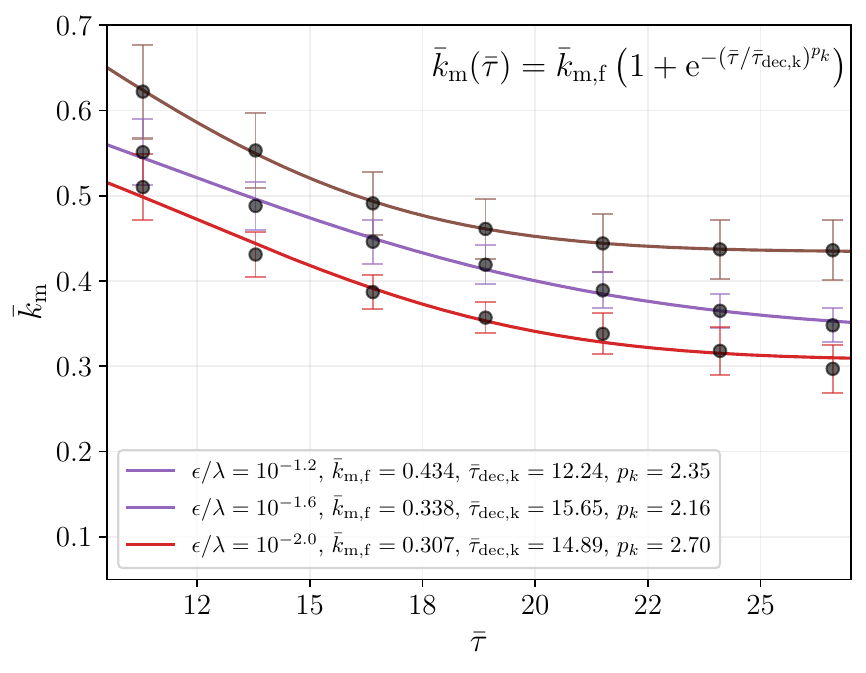}
\caption{Top: Evolution of the peak frequency across the various bias parameters and timesteps of the simulation. Bottom: Fits to the largest bias cases, which illustrate the freezing in of $\bar{k}_{\rm m}$ after the network has fully decayed.}
\label{Fig:km-summary}
\end{figure}

Next, we plot the evolution of the peak position of the spectrum in Fig.~\ref{Fig:km-summary}. The monotonic, near identical behaviour of this quantity is once again observed for the low ($\epsilon/\lambda \lesssim 10^{-2.4}$) scenarios. Similar to the trends found in $\Omega_{\rm GW}$, deviations start to occur for larger values of the bias. In these cases, the (comoving) peak frequency begins to freeze in to a specific value for $\bar{\tau} \gtrsim \bar{\tau}_{\rm dec}$, where it will remain until arbitrarily late times. Strictly speaking, only the $\epsilon/\lambda = 10^{-1.2}$ fully exhibits this freeze in over the length of our simulations, though from the top panel of Fig.~\ref{Fig:km-summary} it is evident we would observe smaller bias parameters following the same trajectory given a larger simulation size. 

In the bottom panel of Fig.~\ref{Fig:km-summary}, we again perform a fitting procedure much akin to the functional forms considered earlier. For the $\epsilon/\lambda = 10^{-1.2},10^{-1.6}$ fits, we find agreement within $1\sigma$ between the decay timescale $\bar{\tau}_{\rm dec,k}$, and those inferred from earlier fitting procedures. Unfortunately, the fit is less good for the $\epsilon/\lambda = 10^{-2.0}$ case, likely due to the fact that the data has not sufficiently flattened by the end of the simulation. This has also lead to a marginally higher value of $p_{\rm k}$, which describes the steepness of the transition around $\bar{\tau}_{\rm dec,k}$. Larger simulations would allow us to probe higher $\bar{\tau}$ values, in which case we would expect better agreement between the decay timescale determined here, and the earlier fits. 

One point of note is that the value of $\bar{k}_{\rm m}$ for $\epsilon/\lambda = 10^{-1.2},10^{-1.6}$ has already undergone significant blueshifting relative to smaller bias values at $\bar{\tau}\simeq 11$. This is because for these larger values, network decay has already begun at $\bar{\tau} \lesssim 11$, as is evident from Fig.~\ref{Fig:Vac-bias-comp}. Additionally, we note that even though the overall decay of the $\epsilon/\lambda = 10^{-1.2}$ network may begin prematurely (that is, during the damping phase), horizon sized objects are still expected to have formed. Therefore, the evolution of $\bar{k}_{\rm m}$ is expected to be less sensitive to the initial conditions than $\Omega_{\rm GW}$, which is the reason why we choose to fit that case here.
%
%
%
\begin{figure}
\centering 
\includegraphics[width=\columnwidth]{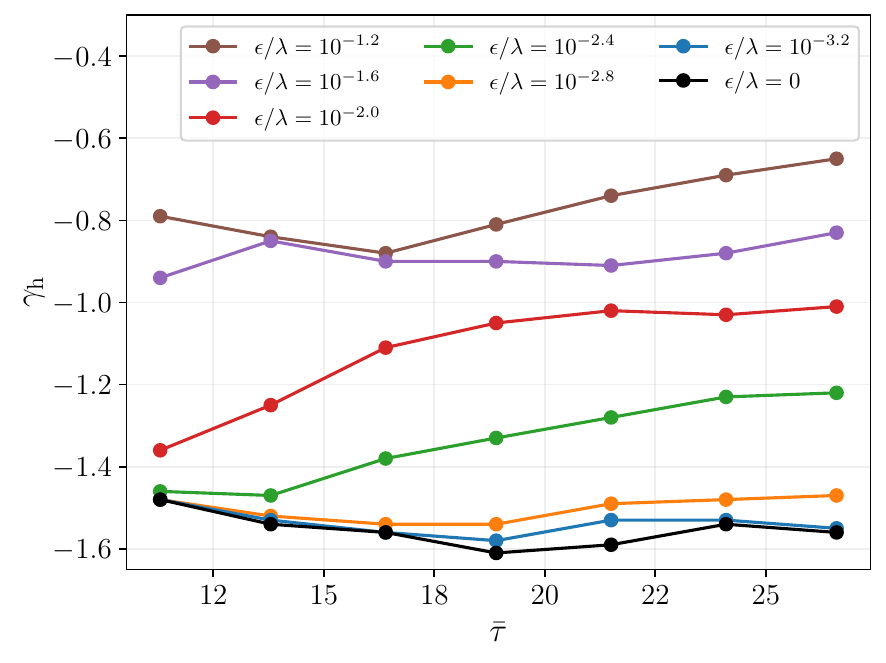}
\caption{Spectral indices for the high frequency slope of the gravitational wave spectrum, for various vacuum bias parameters as a function of time. While not shown, the standard deviation of these curves typically lies between $\pm 0.1$ and $\pm 0.2$. Exact values for this can be found in the supplemental tables.}
\label{Fig:gh-summary}
\end{figure}
%

%
\begin{figure*}
\centering 
\includegraphics[width=2\columnwidth]{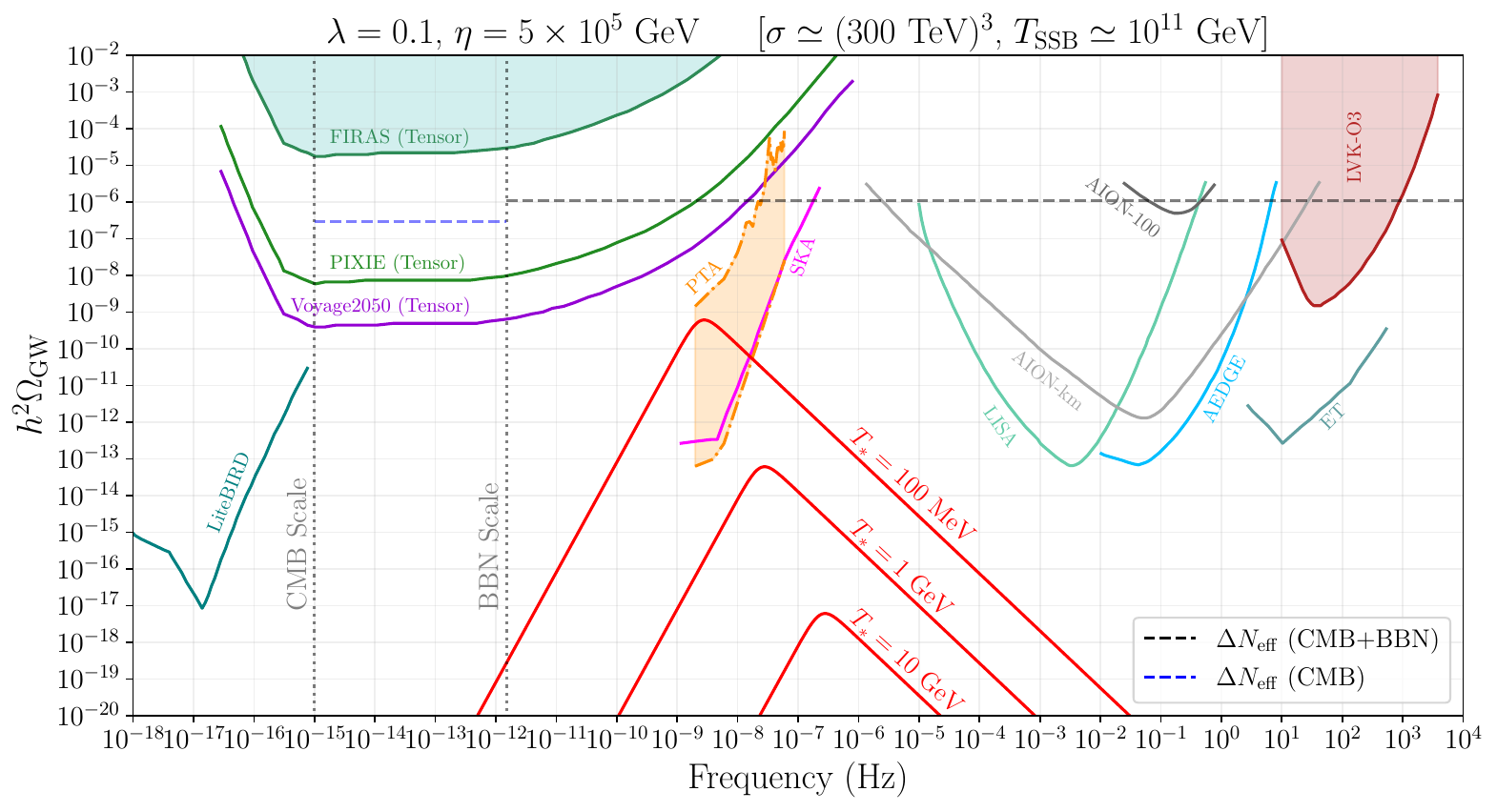}
\caption{The domain wall gravitational wave signal for a set of parameters ($\lambda,\eta$) as seen today, assuming instantaneous decay at some temperature $T_*$. Sensitivity curves for a selection of gravitational wave observatories are also plotted, see \cite{Campeti2020,Ellis2023,Cyr2023, Aggarwal2025} and references therein for additional details. Filled regions are currently constrained, while unfilled are forecasted sensitivities. Recall that $\sigma$ and $T_{\rm SSB}$ are derived parameters, and as such are uniquely determined once one specifies $\lambda$ and $\eta$.}
\label{Fig:GW-constraints-expanded}
\end{figure*}
%

Finally, in Fig.~\ref{Fig:gh-summary}, we show the evolution of the high frequency slope ($\bar{k} > \bar{k}_{\rm m}$) of the gravitational wave spectrum. Similar to the other plots in this section, the larger bias values are offset relative to the no bias case due to evolution at $\bar{\tau} \lesssim 11$. Standard deviations for these slopes can be found in Appendix \ref{sec:levelA4}, though typical values range between $\sigma_{\gamma_{\rm h}}\simeq 0.1-0.2$. With this in mind, the $\epsilon/\lambda \leq 10^{-2.8}$ cases are all consistent with the unbiased value of $\gamma_{\rm h} \simeq -1.56 \pm 0.201$. For $\epsilon/\lambda = 10^{-2.0},10^{-2.4}$, a softening of the spectral index is observed over the duration of the simulation. This is due to the fact that as the largest domain walls in the simulation begin to shrink, the characteristic frequency of gravitational waves produced by the network begins moving back towards higher frequencies. This provides a frequency dependent boost to the amplitude of $\Omega_{\rm GW}$ (with larger boosts seen at high frequencies) as walls shrink and eventually disappear from the network. The $\epsilon/\lambda = 10^{-2.0}$ case specifically appears to asymptote to a stable value at around $\bar{\tau}\simeq 21$. Similarly, for $\epsilon/\lambda = 10^{-1.6}$ the slope appears to already be frozen in.

The largest bias case, however, appears to exhibit some growth at late-times. This late time softening appears to be a symptom of a different shortcoming in the simulation, namely the leakage of energy at late times. By looking at the brown curve in Fig.~\ref{Fig:OGW-summary}, we can see that the peak amplitude of gravitational waves slowly decreases after network decay. This energy leakage is unphysical, as once the network fully collapses one expects that $\Omega_{\rm GW}$ should remain constant (during radiation domination). This reduction in amplitude manifests itself as a softening of $\gamma_{\rm h}$ at late times, giving the observed increase seen in Fig.~\ref{Fig:gh-summary}. While we have not fully investigated the source of this leakage, it does not appear to affect results for other bias parameters in any meaningful way.

The final metric we keep track of in Table~\ref{tab:GWs-comp} is the half-width half-max (HWHM) of the spectrum, $\bar{k}_{\rm wid}$. This quantity can be derived once $\bar{k}_{\rm m}$, $\gamma_{\rm h}$, and $\Omega_{\rm GW}$ are known for a given $\epsilon/\lambda$, and simply serves as an intuitive proxy for the relative size of the ``peaked'' region of the spectrum. As expected, its value grows with larger $\epsilon/\lambda$. 

In contrast to our discussions regarding the area parameter, where we considered both vacuum and temperature dependent bias, here we limit our analysis solely to the vacuum bias scenario. The vacuum bias case has two desirable qualities that our temperature dependent simulations do not. First, in all cases (except perhaps $\epsilon/\lambda = 10^{-1.2}$), the pressure force due to the bias is initially negligible compared to the tension of the walls. As time goes on, this force slowly becomes relevant, and a somewhat smooth transition occurs between the scaling and decaying regimes for the network. For the temperature dependent case, this smooth transition does not exist, and the walls violently begin their collapse in response to a highly non-perturbative change in the pressure force at $\bar{\tau}_{\rm crit}$, which can induce some numerical instabilities in the computation of the gravitational wave spectrum.

Secondly, the potential energy (and therefore the energy density in the core) of the domain walls remains constant in the vacuum bias case. When $\epsilon = \epsilon(\bar{\tau})$ however, a large energy injection to the cores of these objects is activated at $\bar{\tau}_{\rm crit}$. That is, the local maximum of the domain wall potential shifts from $\phi = 0$ to $\phi = 0.1 \, \eta$ in a highly non-adiabatic way. In addition to numerical instabilities, it is unclear whether such an energy injection to the walls is representative of a physically realistic scenario. As discussed in Sec.~\ref{sec:level3.1}, calculating the lifetime of the network can be done reliably, however, the procedure to compute the gravitational wave spectrum is more sophisticated, and is therefore more susceptible to errors from this non-physical energy injection. For these reasons we choose only to analyze the vacuum bias gravitational wave spectra, leaving a more comprehensive investigation into the temperature dependent case for future work. 

In Fig.~\ref{Fig:GW-constraints-expanded}, we make contact with observations by presenting a non-exhaustive list of constraints (filled contours, with the exception of PTAs) and forecasted sensitivities (unfilled contours) on stochastic gravitational wave backgrounds. Recent PTA results present strong evidence for the existence of a SGWB in the nanohertz frequency range, whose signal appears to lie in the shaded orange region of this Figure. The (Tensor) sensitivity curves are computed from tensor-induced scalar fluctuations, which are known to source CMB spectral distortions \cite{Kite2020,Cyr2023}. The horizontal dashed lines represent recent constraints \cite{Yeh2022,Aggarwal2025} on $\Delta N_{\rm eff}$ from Big-Bang nucleosynthesis (BBN) and the CMB. Note that these constraints assume that gravitational waves are the sole source of relativistic degrees of freedom emitted by the network\footnote{One should also not forget that $\Delta N_{\rm eff}$ bounds require an integration over the full spectrum of gravitational waves. What we (and others) have plotted assume a sharply peaked feature, and should be corrected if one wishes to study a particular model in detail}. Supplemental constraints have been derived recently by Ferreira \textit{et al.} \cite{Ferreira2022} when considering the production of additional dark and standard model radiation.

In this Figure, we highlight a specific case where the domain wall parameters are set to be $(\lambda,\eta) = (0.1, \, 5 \times 10^5 \, {\rm GeV})$. Concretely, this implies a vacuum phase at $T_{\rm SSB} \simeq 10^{11}$ GeV, giving rise to domain walls with surface tension $\sigma \simeq (300 \, {\rm TeV})^3$. The red solid lines show the spectrum under the (incorrect) assumption of an instantaneous network decay taking place at $T_* = (0.1,\, 1, \, 10)$ GeV, as inferred from an extrapolation of our \textit{no bias} results. Heuristically, the location of the peak frequency is set by $T_{*}$ while the amplitude of the signal depends on $\lambda$ and $\eta$. Thus, by simply tuning $T_*$, $\eta$, and $\lambda$ one can in principle get this spectrum to peak over an enormous range of frequencies and amplitudes. As a suggestive example, our parameter choice provides a peak in the pulsar timing band, at a temperature near the expected QCD phase transition ($T_{\rm QCD} \simeq 100$ MeV).

%
\begin{figure}
\centering 
\includegraphics[width=\columnwidth]{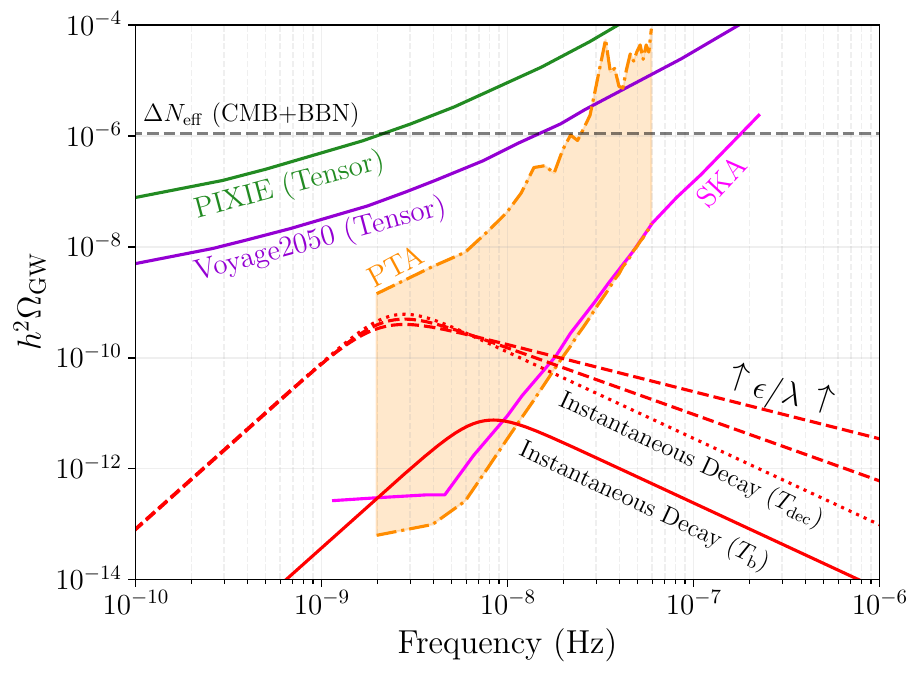}
\caption{A zoom-in of the PTA region of parameter space in Fig.~\ref{Fig:GW-constraints-expanded}, now assuming a non-instantaneous decay. For a particular scenario, as $\epsilon/\lambda$ is increased, the high frequency spectral index softens. A precise measurement of this high frequency slope could be used to determine the amplitude of the bias term for collapsing domain walls.}
\label{Fig:GW-constraints-zoom-in}
\end{figure}
%

The spectra in Fig.~\ref{Fig:GW-constraints-expanded} are fundamentally unphysical, as they assume instantaneous decay, thereby neglecting network dynamics during the collapse phase. In Fig.~\ref{Fig:GW-constraints-zoom-in}, we provide a zoom-in of the peak region of the spectrum, now showing the qualitative differences when one includes the gravitational waves produced during network collapse. 

First, note that from Table~\ref{tab:A-fits_TB}, the most well resolved simulations (namely, $\epsilon/\lambda = 10^{-1.6},10^{-2.0}$) exhibit an approximate delay between the instantaneous and observed decay time of $\bar{\tau}_{\rm dec} \approx 3 \bar{\tau}_{\rm b}$. Assuming radiation domination, this translates into a delay on the inferred decay temperature of $T_{\rm dec} \approx T_{\rm b}/3$, where $T_{\rm b}$ was given in Eq.~\eqref{eq:TVb}. Recalling that $\Omega_{\rm GW} \propto T_{\rm dec}^{-4}$, we see that this delay has the effect of both enhancing the amplitude of the inferred $h^2 \Omega_{\rm GW}$ by roughly two orders of magnitude, and shifting the peak position to lower frequencies. This enhancement can be seen in Fig.~\ref{Fig:GW-constraints-zoom-in} by comparing the red solid and dotted lines. 

This red dotted line still remains unphysical, as all we have done is shift the temperature of an instantaneous network decay. Once gravitational waves from the collapsing network are included, one sees that the high frequency slope becomes less steep, as indicated by the red dashed lines. The exact degree of softening depends upon the ratio of $\epsilon/\lambda$, and so we merely illustrate the direction of the trend in Fig.~\ref{Fig:GW-constraints-zoom-in}. The limited dynamic range of our simulations ($T_{\rm ini}/T_{\rm f} \simeq 26$) precludes us from being able to study both a sizeable $\Omega_{\rm GW}$ as well as a $T_{\rm dec}$ leading to a peak in the pulsar timing band from a purely numerical standpoint. Larger simulations would allow us to confidently extrapolate the high frequency spectral index for the very small $\epsilon/\lambda$ ratios necessary to produce the spectra shown in Figs.~\ref{Fig:GW-constraints-expanded} and \ref{Fig:GW-constraints-zoom-in}, and is an aim of future work. For now, the high frequency slopes provided remain a qualitative extrapolation based on the $\epsilon/\lambda = 10^{-1.6}$ and $10^{-2.0}$ simulations.
%

\section{\label{sec:level5}Discussion and Conclusions}
In this work, we have performed a suite of lattice simulations to investigate the detailed dynamics of a domain wall network formed via a real scalar field in a $\mathbb{Z}_2$ symmetric potential. We have extracted observables both in the case that the $\mathbb{Z}_2$ remains exact (our so-called \textit{no bias} scenario), as well as when it is broken by a perturbative bias term, whose strength is parameterized by $\epsilon/\lambda$ and form can be seen in Eq.~\eqref{eq:Vbias}. In a subset of the biased scenarios, our simulations have sufficient dynamic range to resolve all three stages of the network lifecycle: damping, scaling, and decay. Observing these three phases is critical in quantifying how measurable properties of the network, such as the effective lifetime and gravitational wave spectrum, respond to the presence of a bias.

The first quantity we extracted from our unbiased simulations was the so-called area parameter, $\mathcal{A}$, defined in Eq.~\eqref{eq:scaling-Energy}. As the name suggests, the area parameter is a dimensionless quantity that measures the relative size of the domain walls in the simulation. Physically, $\mathcal{A}=1/2$ can correspond to one domain wall with area $A = H^{-2}$ at any point in the simulation. Most importantly though, is the fact that $\mathcal{A}$ approaches a constant value when the network achieves scaling. From the left hand plot of Fig.~\ref{Fig:No-bias-summary}, we see that in our simulations this happens around $\bar{\tau} \simeq 7$, where we find $\mathcal{A} \simeq 0.78 \pm 0.03$, in agreement with other classical \cite{Hiramatsu2013} and contemporary \cite{Ferreira2024, Dankovsky2024} simulations performed under similar conditions.

While $\mathcal{A}$ remains constant throughout the simulation for $\Delta V = 0$, introducing a bias causes the network to collapse, leading to time evolution of $\mathcal{A = \mathcal{A}(\bar\tau)}$ which is well fit by the generalized exponential function given in Eq.~\eqref{eq:A-fit}. Using this, we computed the decay timescale ($\bar{\tau}_{\rm dec}$) of the network for different values of vacuum ($\epsilon/\lambda$) and temperature-dependent ($\epsilon(\bar{\tau})/\lambda$) bias parameters. The results of this fitting procedure can be seen in Fig.~\ref{Fig:Vac-bias-comp}/Table~\ref{tab:A-fits_VB}, and Fig.~\ref{Fig:T-bias-comp}/Table~\ref{tab:A-fits_TB} for the vacuum and temperature dependent cases respectively.

Much of the previous work on biased domain walls have performed what is known as the \textit{instantaneous decay approximation}, in which the network ceases to exist once the pressure force from the potential bias overcomes the intrinsic tension of the walls. Quantitatively, this happens when $H(T_{\rm b}) = V_{\rm b}/\sigma$, more explicitly calculated in Eq.~\eqref{eq:TVb}. Converting this into $\bar{\tau}_{\rm b}$, we compare this quantity with $\bar{\tau}_{\rm dec}$ and find that the network persists long past the time indicated by the instantaneous decay approximation. We find that $\bar{\tau}_{\rm dec}/\bar{\tau}_{\rm b} \approx 3$, which implies that in actuality, the true temperature that the network loses much of its cosmological importance is $T_{\rm dec} \simeq T_{\rm b}/3$.

The second main focus of our work has been to investigate the exact response of the gravitational wave spectrum to the collapse of the network. For this part of the analysis, we restricted ourselves to the unbiased and vacuum bias scenarios only, leaving the temperature dependent case for future work. As discussed above, the temperature dependent bias represents a qualitatively different picture in which significant energy injection into $\phi$ occurs at some $\bar{\tau}_{\rm crit}$. This non-perturbative injection disrupts the structure of the potential for $\phi$, which can lead to some numerical instabilities when attempting to extract the gravitational waves.

The growth of the GW spectrum for the unbiased scenario was presented in Fig.~\ref{Fig:No-bias-timeslice-fits}, where we extract the numerical quantity $\bar{S}_{\rm k}$ by way of the broken power law fitting function given in Eq.~\eqref{eq:Sk_fit}. In this figure, we show both the fits where low frequency slope ($\gamma_{\rm l}$) is allowed to vary, as well as when we fix it to its expected causal value of $\gamma_{\rm l} = 3$ (FLK). The quality of the fits is not degraded by any appreciable amount when fixing $\gamma_{\rm l}$, thus we eliminate it as a free parameter in all further analysis. The high frequency slope that we extract from our fits can also be used to calibrate models of the GW spectrum and we discuss one such model in Appendix \ref{sec:levelA3}. Note that the quantity $\bar{S}_{\rm k}$ can be easily recast into the more familiar $\Omega_{\rm GW}$ using Eqs.\eqref{eq:GW-spectrum} and \eqref{eq:Sk-to-GW} (see also our numerical dictionary in Appendix \ref{sec:A-numerical-dictionary}).

By fitting the spectra of gravitational waves from unbiased domain walls at different timesteps, we are able to reproduce various expected features. It is well known that domain walls produce gravitational waves with peak frequency at roughly the Hubble scale at any given time. In our simulations, we find this ratio to be $H/k_{\rm m} = 0.91 \pm 0.02$ at all times after scaling is reached, in agreement with other recent works. While the network persists, we also find that $\Omega_{\rm GW} \propto (T_{\rm i}/T)^4$ throughout the simulation, again matching theoretical expectations\footnote{Numerically what we find is $\bar{S}_{\rm k} \propto (\bar{\tau}/\bar{\tau}_i)^4$ which in radiation domination translates into $\bar{S}_{\rm k} \propto (T_{\rm i}/T)^4$.}. Importantly, we find the high frequency spectral index (at the end of the simulation) to be $\gamma_{\rm h} = -1.56 \pm 0.201$. This is in conflict with the oft-cited value of $\gamma_{\rm h} = -1$ (e.g. in Hiramatsu \textit{et al.} \cite{Hiramatsu2013}), but consistent with more contemporary simulations \cite{Ferreira2024, Kitajima2023, Dankovsky2024}.

The bulk of our numerical results are presented in Fig.~\ref{Fig:Vac-bias-timeslice-fits-comparison}, where the spectra for various bias parameters ranging from $\epsilon/\lambda = 10^{-3.2} - 10^{-1.2}$ were considered. The smallest bias parameters chosen ($\epsilon/\lambda = 10^{-2.8},10^{-3.2}$) were relatively uninteresting, as our dynamic range was not large enough to observe them experience any significant departure away from scaling, as can be seen by the area parameter plot, Fig.~\ref{Fig:Vac-bias-comp}. As such, their spectra essentially match that of the unbiased scenario. For $\epsilon/\lambda = 10^{-2.4}$ various quantities begin to show deviations, such as the high frequency slope near the end of the simulation. We consider the $\epsilon/\lambda = 10^{-1.6},10^{-2.0}$ to be particularly good benchmark cases, as they clearly exhibit an unperturbed approach to scaling ($\bar{\tau} \lesssim 7$), followed by production of significant gravitational waves during the scaling regime, finally culminating in a nearly complete annihilation by the end of the simulation at $\bar{\tau}_{\rm f} \simeq 26$. Though the $\epsilon/\lambda = 10^{-1.2}$ case also contains useful information, its approach to scaling was likely perturbed by the large bias pressure, which adds some uncertainty to the level of accuracy of those gravitational wave results.

In Figs.~\ref{Fig:OGW-summary} and \ref{Fig:km-summary}, we performed additional fits to the time evolution of the amplitude of the spectrum at its maximum ($\tilde{\Omega}_{\rm GW}(\bar{k}_{\rm m})$), and the peak frequency ($\bar{k}_{\rm m}$) for some benchmark cases. We recover the expected behaviour for a collapsing network, namely, once the network has decayed the amplitude and peak frequency freeze in to their respective values around $\bar{\tau}_{\rm dec}$. This independent fitting procedure allowed us to derive decay timescales based on the $\tilde{\Omega}_{\rm GW}$ and $\bar{k}_{\rm m}$ data, which were consistent\footnote{With the exception of the $\epsilon/\lambda = 10^{-2.0}$ case from the $\bar{k}_{\rm m}$ data. Details on this can be found near Fig.~\ref{Fig:km-summary}.} with the $\bar{\tau}_{\rm dec}$ derived in Sec.~\ref{sec:level3.1}. 

The high frequency slope of the gravitational wave spectrum is perhaps the cleanest way to probe the amplitude of the bias term, as indicated in Fig.\ref{Fig:gh-summary}. From a physical standpoint, the softening of $\gamma_{\rm h}$ is somewhat intuitive. As the network decays, the characteristic length scale of domain walls decreases, and thus higher frequency gravitational waves are emitted in greater abundance. The death throes of the network should thus lead to a generic boost of the high frequency spectrum. The exact interplay between the amplitude of $\epsilon/\lambda$ and the asymptotic spectral index ($\gamma_{\rm h} (\bar{\tau}\rightarrow \infty)$) will require larger simulations in order to resolve a larger range of vacuum bias amplitudes, and is an aim of future work.

To highlight the importance of this relationship, we have illustrated the physical gravitational wave spectrum for suggestive values of $\lambda$ and $\eta$ in Figs.~\ref{Fig:GW-constraints-expanded} and \ref{Fig:GW-constraints-zoom-in}, where we also overlay various constraints and forecasts on the GW parameter space. For this model, the spectrum peaks in the nano-Hertz frequency range where pulsar timing arrays are actively searching for stochastic backgrounds. Fig.~\ref{Fig:GW-constraints-zoom-in} shows two important features. First, utilizing the instantaneous decay approximation will lead to an underestimation of the peak amplitude (by a factor of $\mathcal{O}(100)$), as well as a mischaracterization of the peak frequency by a factor of roughly $1/3$. Second, the dynamics of the collapse phase (characterized through $\epsilon/\lambda$) will change $\gamma_{\rm h}$ away from its unbiased value of $\gamma_{\rm h}^{\rm UB} \simeq -1.56 \pm 0.201$. As data from the pulsar timing consortium improves, it may be possible to resolve these differences, and thus put constraints on the exact form of the bias which would have led to the collapse of the domain wall network. This high frequency behaviour underscores the importance of making measurements of any stochastic background away from its peak position.

Domain walls remain an interesting and well-motivated extension to the standard models of cosmology and particle physics. Among its observational signatures, a stochastic background of gravitational waves is perhaps one of the more promising avenues to explore in light of the recent successes of the LIGO collaboration and pulsar timing array consortium. Due to their enormous energy density, domain walls produced in the very early universe must not persist until today. Therefore, their existence precipitates the need for a subsequent decay, which is most often mediated by introducing a small bias between the two vacuum states. As we have shown in this work, the decay of the network induces large deviations to the overall spectrum of gravitational waves when compared against the spectrum produced during the scaling regime. This decay phase is ubiquitous, and thus these effects must be taken into account when reconciling domain wall scenarios with data.

\begin{acknowledgments}
\noindent BC would like to acknowledge support from both an NSERC Banting fellowship, as well as the Simons Foundation (Grant Number 929255). SC would like to acknowledge the assistance given by Research IT and the use of the Computational Shared Facility at The University of Manchester as well as resources provided by the Cambridge Service for Data Driven Discovery (CSD3) operated by the University of Cambridge Research Computing Service (www.csd3.cam.ac.uk), provided by Dell EMC and Intel using Tier-2 funding from the Engineering and Physical Sciences Research Council (capital grant EP/T022159/1), and DiRAC funding from the Science and Technology Facilities Council (www.dirac.ac.uk).
\end{acknowledgments}

\appendix
\section{\label{sec:levelA1}Numerical Setup}

\subsection{\label{sec:generality of bias}Generality of the bias term}

It may seem like the bias term that we have chosen to use in this paper has been chosen somewhat arbitrarily and that our results are dependent upon this choice. In fact, the bias term that we are using is general in the sense that all other polynomial potentials up to quartic powers of $\phi$ can be related to this one via field redefinitions. We can show this explicitly by writing down the full potential,
\begin{align}
    V = \frac{1}{4}c_4\phi^4 + \frac{1}{3}c_3\phi^3 + \frac{1}{2}c_2\phi^2 + c_1\phi \,,
\end{align}
and then a redefinition of the field $\phi = \Tilde{\phi} + \alpha$ leaves the kinetic term unchanged and the potential with the same functional form (neglecting constants) under the transformations, $\Tilde{c}_4 = c_4$, $\Tilde{c}_3 = c_3 + 3c_4\alpha$, $\Tilde{c}_2 = c_2 + 2c_3\alpha + 3c_4\alpha^2$ and $\Tilde{c}_1 = c_1 + c_2\alpha + c_3\alpha^2 + c_4\alpha^3$. The mapping to our potential can then be derived by setting $\Tilde{c}_4 = \lambda$, $\Tilde{c}_3 = \epsilon\eta$, $\Tilde{c}_2 = -\lambda\eta^2$ and $\Tilde{c}_1 = -\epsilon\eta^3$, which results in a cubic equation in $\alpha$, for which there will always be at least one real solution.

\subsection{\label{sec:rescaling}Rescaled parameters for simulation}

It is possible to perform additional scalings and field redefinitions to show that some of the otherwise free parameters do not change the physics being described, only the scales at which it occurs. Consider the action of a $\mathbb{Z}_2$ symmetric model consisting of a single real scalar field,
%
\begin{align}
    S_0 = \int \sqrt{-g}\id^4 x \left[ \frac{1}{2}\partial_\mu\phi\partial^\mu\phi - \frac{\lambda}{4}(\phi^2 - \eta^2)^2 \right] \,,
\end{align}
%
where $g$ is the determinant of the metric and $\sqrt{-g} = a^4$ if we work with comoving Cartesian coordinates and conformal time. One can perform a field redefinition $\bar{\phi} = \phi/\eta$ as well as a rescaling of the coordinates $\bar{x} = x\sqrt{\lambda}\eta$ so that the action becomes
%
\begin{align} \label{eq:rescaled-L}
    S_0 = \frac{1}{\lambda} \int \sqrt{-\bar{g}}\id^4 \bar{x} \left[ \frac{1}{2}\bar{\partial}_\mu\bar{\phi}\bar{\partial}^\mu\bar{\phi} - \frac{1}{4}(\bar{\phi}^2 - 1)^2 \right] \,.
\end{align}
%
Therefore, in simulations one can evolve the dimensionless system using the equation of motions derived from Eq.~\eqref{eq:rescaled-L}, yielding generic results that can be scaled to any choice of the arbitrary parameters $\eta$ and $\lambda$ (in other words, one does not need to set a value for these parameters at the onset of the simulation). In this situation, the rescaled mass of the field becomes $\bar{m} = \sqrt{2} = m/\sqrt{\lambda}\eta$. 

With the introduction of a bias, it becomes impossible to scale out all of the parameters. In terms of the same rescaling we have
%
\begin{align} \label{eq:rescaled-L-vacBias}
    S = S_0 + \frac{1}{\lambda} \int \bar{\epsilon}\bar{\phi}\bigg(\frac{\bar{\phi}^2}{3}-1\bigg)\sqrt{-\bar{g}} \id^4 \bar{x} \,,
\end{align}
%
where, $\bar{\epsilon} = \epsilon/\lambda$. This ratio is a remaining free parameter that cannot be eliminated and will affect the behaviour of the system.


\subsection{\label{sec:evolution}Evolution}

We simulate this system by implementing a Cartesian (comoving) grid of points, with periodic boundary conditions and where each grid point has a value associated with it that represents $\phi$ at that location. At every location, the value can be updated according to the equation (with the above rescalings implicitly performed)
%
\begin{align}
    \Ddot{\phi} = \nabla^2\phi - a^{\beta}(\phi^2-1)(\phi+\epsilon) - F\dot{\phi} \,,
\end{align}
%
where $H = a^\prime(\Bar{\tau})/a$ while $\beta$ and $F$ are parameters that have been introduced in order to be able to perform non-standard types of evolution, e.g. the PRS algorithm or an initial period of damping. For standard evolution, they take the values $\beta=2$ and $F = 2\mathcal{H}$.

The spatial derivatives are approximated with second-order finite difference operators,
%
\begin{align}
    \frac{\partial^2\phi(x)}{\partial x^2} \approx \frac{\phi(x+\Delta x) - 2\phi(x) + \phi(x-\Delta x)}{\Delta x^2} \,,
\end{align}
%
and the system is evolved using the leapfrog algorithm. We set the initial conditions in the same manner as ref. \cite{Kawasaki2011}, with $\bar{k}_{\rm cut} = 2\pi$, which corresponds to the length scale $\mathcal{H}^{-1}$ at the beginning of the simulation.

As the Universe expands in these simulations, the width of the domain walls remains the same in physical coordinates and therefore it contracts in co-moving coordinates. This presents a problem for our simulations because we need to ensure that we are resolving the domain walls.  The claim of the PRS algorithm is that the dynamics of the domain wall networks is not significantly changed if the equations of motion are modified by setting $\beta=0$ and $F = 3\mathcal{H}$, while it has the beneficial effect that the widths of the domain walls grows proportionally to the expansion, and the resolution issue is never encountered. We avoid this approach throughout this paper since it is unclear how this modified evolution will affect the the spectrum of gravitational waves emitted from the network, although it would be an interesting avenue for future research. 

Instead, we must ensure that we are still resolving the cores of the domain walls by the end of the simulation. We parameterize this by setting $p = [a_f\Delta x]^{-1}$, where $a_f$ is the scale factor at the end of the simulation, which can be interpreted as the statement that, at the very end of the simulation, there are $p$ grid points per each physical length scale of size $(\sqrt{\lambda}\eta)^{-1}$, which has been scaled to one. The other constraint that must be satisfied is that we do not exceed the light-crossing time, which we saturate by setting $\bar{\tau}_f - \bar{\tau}_i = \frac{1}{2}N_x\Delta x$, which is approximately the same as the statement that the horizon is equal to the size of the box. Satisfying both of these equations fixes the lattice spacing to be
%
\begin{align}
    \Delta \bar{x} = \frac{1}{N_x}\left[ -\bar{\tau}_i + \sqrt{\bar{\tau}_i^2 + \frac{2N_x}{p}} \right] \,.
    \label{eq: lattice spacing}
\end{align}
%
All of our main results are computed using simulations with $N_x = 2048$, in all three spatial directions, and we set $p=1.5$ --- a choice which is justified in Appendix \ref{sec:convergence}. We evolve all of our simulations with timesteps of $\Delta\bar{\tau} = \Delta\bar{x}/5$.

\subsection{\label{sec: GW calc}Gravitational Wave Calculation}

We calculate the spectrum of gravitational waves produced from the network using the method developed in \cite{Dufaux2007}. The numerical procedure is to first calculate the energy momentum tensor at all grid points,
%
\begin{align}
    \bar{T}_{ij}({\mathbf{\bar{x}}},\bar{\tau}) = \bar{\partial}_i\bar{\phi}\bar{\partial}_j\bar{\phi} - \bar{g}_{ij}\bar{\mathcal{L}} \,.
\end{align}
%
In practice, it is only necessary to calculate the first term because the next step is to project out the transverse, traceless part, and the second term is pure trace. The projection is done by first taking the Fourier transform and then calculating
%
\begin{align}
    \bar{T}^{\text{TT}}_{ij}(\mathbf{\bar{k}},\bar{\tau}) &= \bigg[P_{ik}P_{jl} - \frac{1}{2}P_{ij}P_{kl}\bigg]\bar{T}_{kl}(\mathbf{\bar{k}},\bar{\tau})\,, \\
    P_{ij}(\mathbf{\hat{\bar{k}}}) &= \delta_{ij} - \hat{\bar{k}}_k\hat{\bar{k}}_l \,.
\end{align}
%
The spectrum of gravitational waves is then given by
%
\begin{align}
    \frac{\id\bar{\rho}_{\rm GW}}{\id\log\bar{k}} &= \frac{G}{2\pi^2\bar{V}a^4}S_k \,, \\
    \bar{S}_k &= \bar{k}\int d\Omega_k \sum_{ija}|\Bar{C}^{(a)}_{ij}|^2 \,,
\end{align}
with the surface element given by $\id\Omega_k = \sin\theta_k \id\theta_k \id\phi_k$ --- expressed in terms of spherical polar coordinates in Fourier space, $\bar{k}$, $\theta_k$ and $\phi_k$ --- and $\bar{C}_{ij}^{(a)}$ is defined by
%
\begin{align}
    \Bar{C}^{(1)}_{ij} &= -\bar{k}\int_{\bar{\tau}_i}^{\bar{\tau}} \id\bar{\tau}^\prime a(\bar{\tau}^\prime)\sin \bar{k}\bar{\tau}^\prime \bar{T}^{\text{TT}}_{ij}(\mathbf{\bar{k}},\bar{\tau}^\prime) \,, \nonumber \\
    \Bar{C}^{(2)}_{ij} &= \bar{k}\int_{\bar{\tau}_i}^{\bar{\tau}} \id\bar{\tau}^\prime a(\bar{\tau}^\prime)\cos \bar{k}\bar{\tau}^\prime \bar{T}^{\text{TT}}_{ij}(\mathbf{\bar{k}},\bar{\tau}^\prime) \,.
\end{align}
%
From a numerical perspective, it is inconvenient that these calculations are performed in a spherical polar coordinate system, because the data that we extract from the simulations is arranged in a Cartesian coordinate system, and the calculation of the spectrum is also quite numerically costly. We avoid the coordinate issue and save computational resources by only calculating $\bar{C}_{ij}^{(a)}$ along a few different axes and assuming that they are approximately isotropic.

This entire procedure is performed every $8$ timesteps throughout our simulations and the values of $\bar{C}_{ij}^{(a)}$ along the $13$ axes of interest are stored and continually updated. Using this data, we can compute $\bar{S}_k$ at any time of our choosing, but in practice we choose to output it at $11$ linearly spaced time intervals.

\subsection{\label{sec:convergence}Convergence testing and Validation}

One of the main challenges present in predicting the gravitational wave signature from decaying domain walls is that the ratio of the energy scales at which domain walls form and later decay in our simulations is much smaller than is expected in most realistic scenarios. Ideally, we would like the network to be confidently within the scaling regime for a significant period of time before the bias causes the network to decay. This desire is in conflict with the requirement that the width of a domain wall must remain resolvable throughout the simulation. There is, therefore, a trade-off to be made between numerical accuracy and the dynamic range of our simulations.

In order to understand the dependence of our results on the lattice spacing of our simulations, we have performed additional simulations on a comoving grid with side-length $L \approx 31$, which all start from the same initial conditions but have different lattice spacings. In order to keep the size of the box the same across these simulations, $p$ must increase proportionally with $N_x$, where the lattice spacing is given by Eq. (\ref{eq: lattice spacing}). 

\begin{figure*}
\centering 
\includegraphics[trim={0.5cm 0cm 1.4cm 1.4cm},clip,width=\columnwidth,]{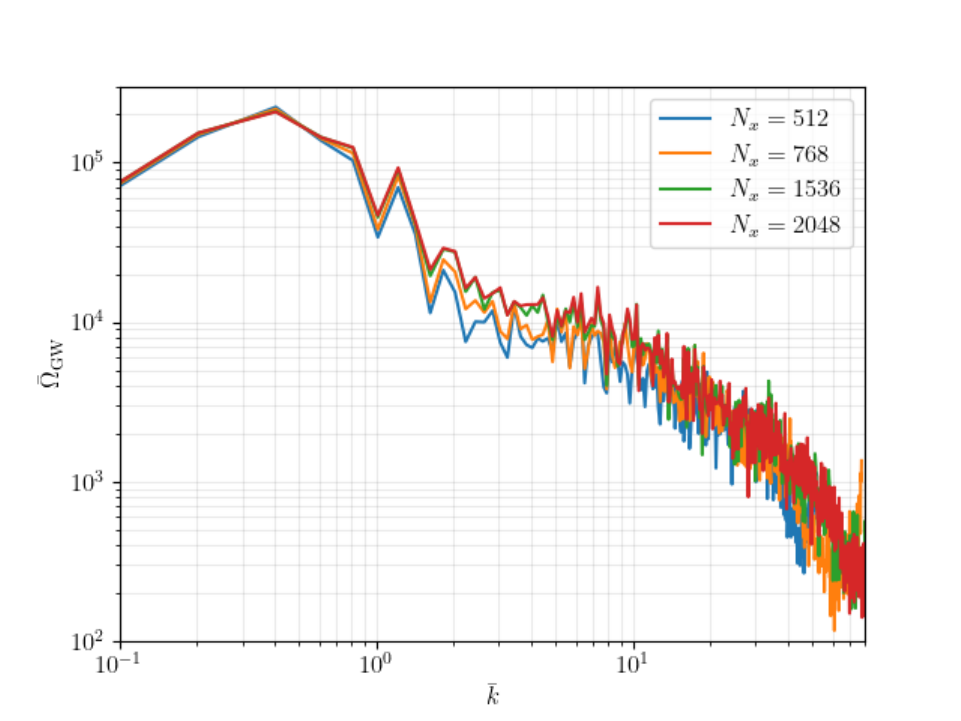}
\hspace{4mm}
\includegraphics[trim={0.5cm 0cm 1.4cm 1.4cm},clip,width=\columnwidth]{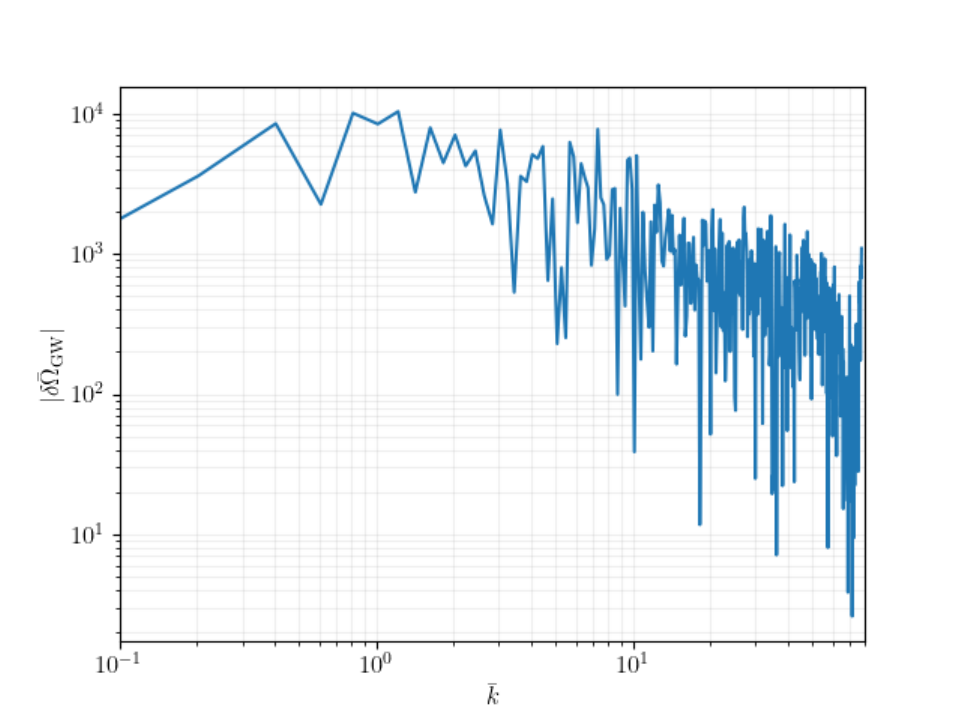}
\caption{Comparisons of the gravitational wave spectrum (averaged across the three "axes" directions) for simulations with different resolutions. They are all performed in the same comoving box, with the same initial conditions, but with different numbers of lattice sites in each direction, $N_x$, and therefore different lattice spacings. On the left we show how the spectrum changes  for four different simulations, while on the right we show the magnitude of the residuals between the $N_x=768$ (which has $p=1.5$, the value that we use for our main results) and $N_x=2048$ ($p=4$). }
\label{Fig:Omega_GW convergence test}
\end{figure*}

In Figure \ref{Fig:Omega_GW convergence test} we show both how the spectrum of gravitational waves changes with the resolution of the simulation and the magnitude of the differences between the $N_x=768$ and $N_x=2048$ simulations. These two examples were chosen because the former has $p=1.5$, which is the choice that we make for our main results, and the latter simply because it is our highest resolution simulation. These plots make it clear that there is a good agreement around the peak, which is the part of the spectrum that we are most interested in, but also that the residuals do not decrease with $k$ as quickly as the spectrum itself --- meaning that the fractional errors increase.
%
%

\begin{figure*}
    \centering
    \includegraphics[trim={0.5cm 0cm 1.4cm 1.4cm},clip,width=\columnwidth]{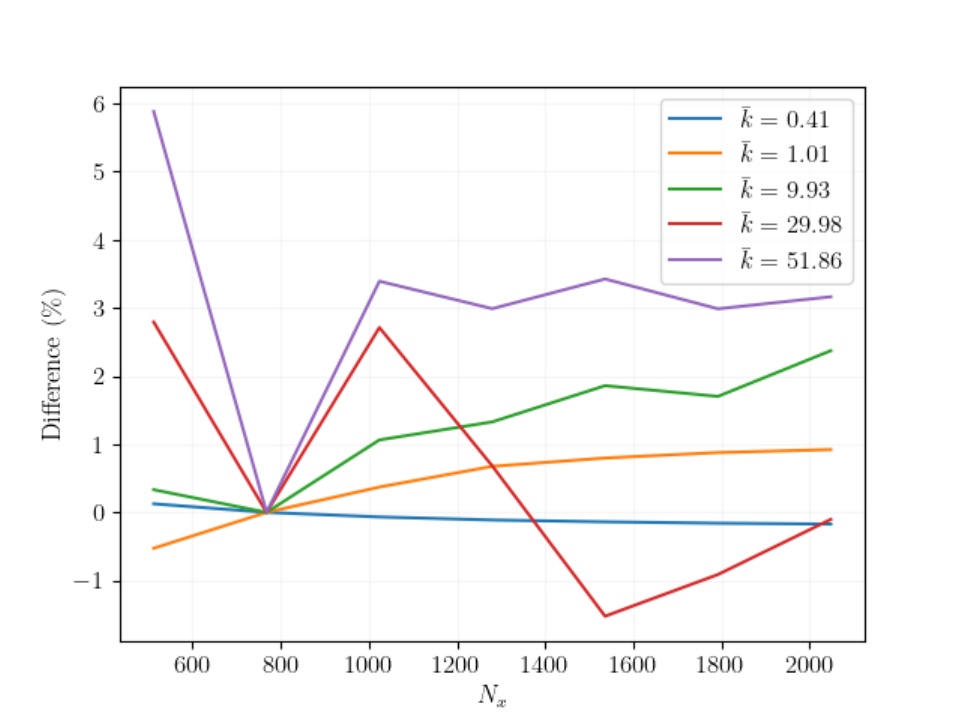}
    \hspace{4mm}
    \includegraphics[trim={0.5cm 0cm 1.4cm 1.3cm},clip,width=\columnwidth]{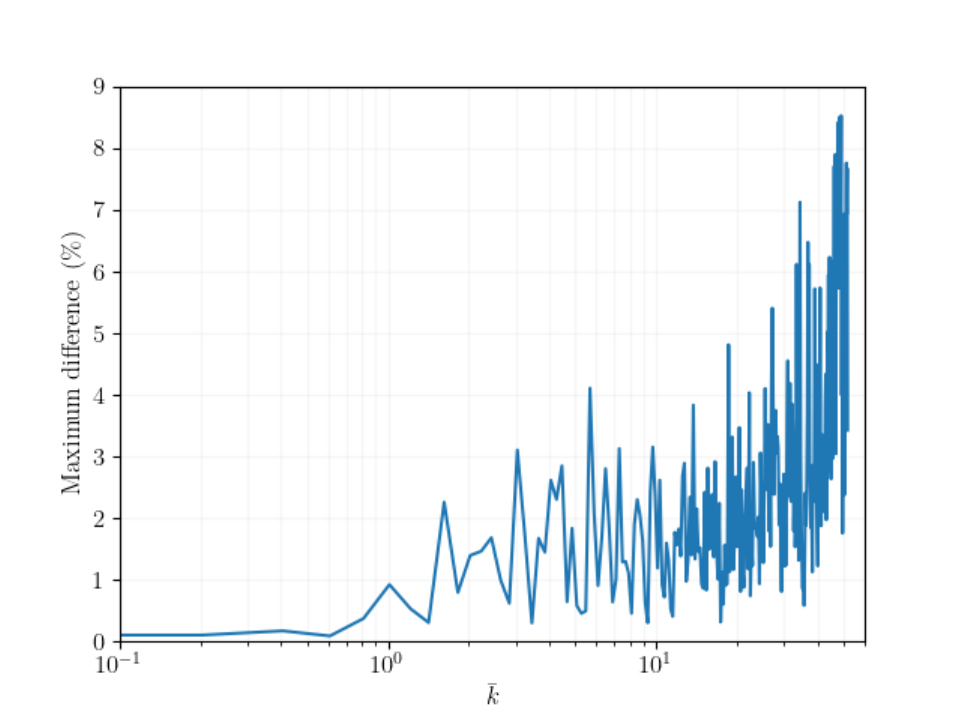}
    \caption{The percentage difference between $\log(\Bar{\Omega}_{\rm GW}(\bar{k}))$ for various values of $N_x$ compared to the $N_x = 768$ case. On the left, we show how a few sample values of $\Bar{k}$ change with the number of grid points. The magnitudes of the differences are much lower for small values of $\bar{k}$ and the curves converge smoothly, whereas for large $\bar{k}$ the convergence is less certain. On the right, we show the maximum percentage difference for the $N_x>768$ simulations, as a function of $\Bar{k}$. The errors grow rapidly as $\bar{k}$ becomes large, rising above $5\%$ consistently for $\Bar{k} \gtrsim 30$. }
    \label{Fig:Percentage difference convergence test}
\end{figure*}

In Figure \ref{Fig:Percentage difference convergence test}, we make this evident by making some comparisons of the percentage difference between the $N_x=768$ case and the other simulations. Note that we are calculating the percentage difference for $\log(\bar{\Omega}_{\rm GW})$, rather than for $\bar{\Omega}_{\rm GW}$, as this is what matters for the fitting procedures that we perform and the other results that are extracted throughout this paper, but it should be kept in mind that this reduces the numbers.

The left plot shows how how the spectrum changes at a few sample values of $\bar{k}$, as a function of the number of grid points. It clearly demonstrates the point that we can have more confidence in the spectrum at small values of $\bar{k}$ as the differences are both smaller and have a more well-behaved convergent behaviour. The right-hand plot supports this argument as it shows the maximum difference between the $N_x=768$ case and the simulations with a finer resolution, which grows with $\bar{k}$. In this work we predominantly concern ourselves with the GW signal in the near-peak region, so we focus on the larger wavelength part of the spectrum which has a sufficiently low level of numerical error.
%

\begin{figure*}
\centering 
\includegraphics[width=\columnwidth]{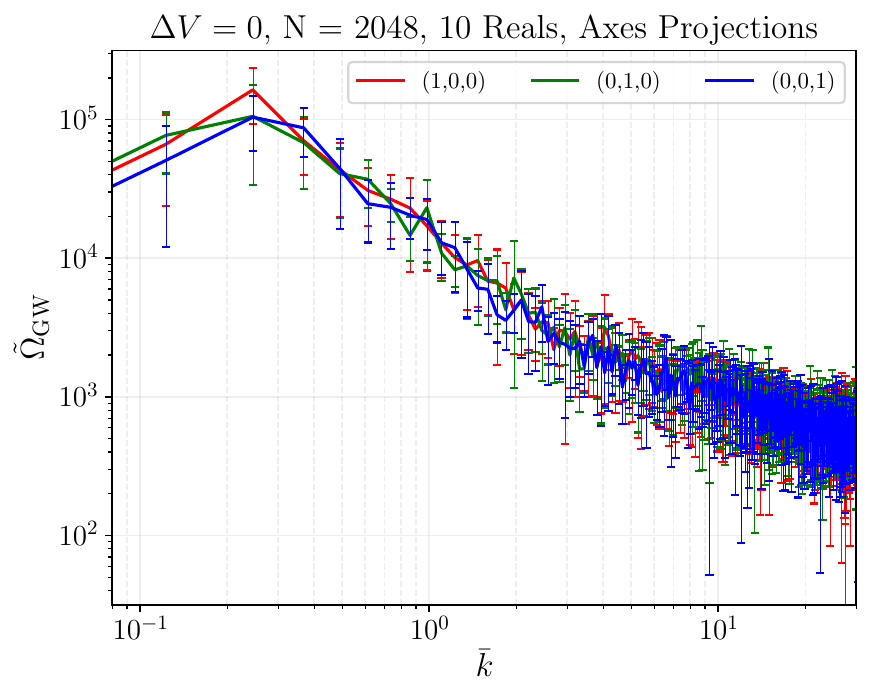}
\hspace{4mm}
\includegraphics[width=\columnwidth]{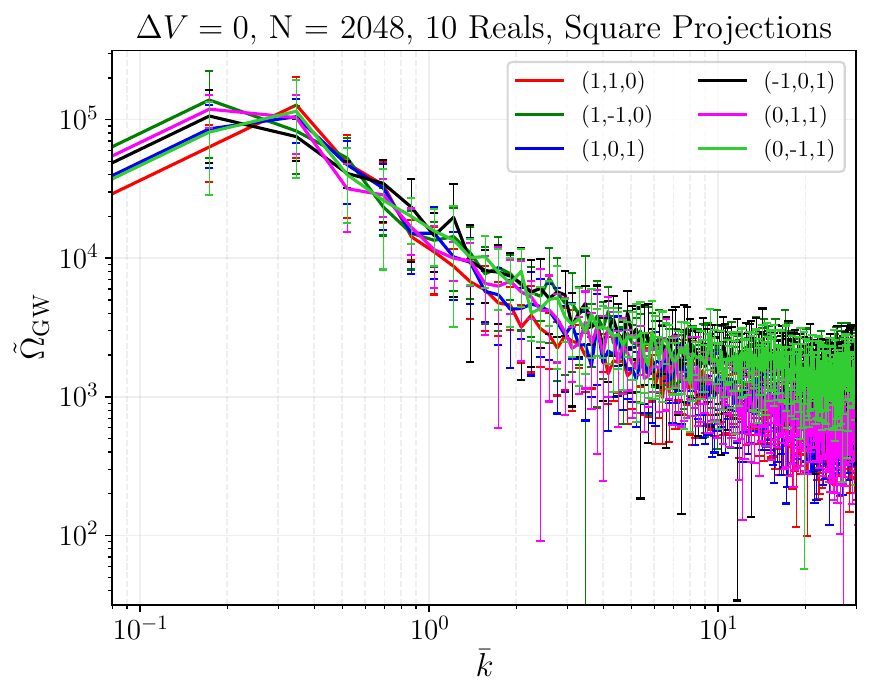}\\
\includegraphics[width=\columnwidth]{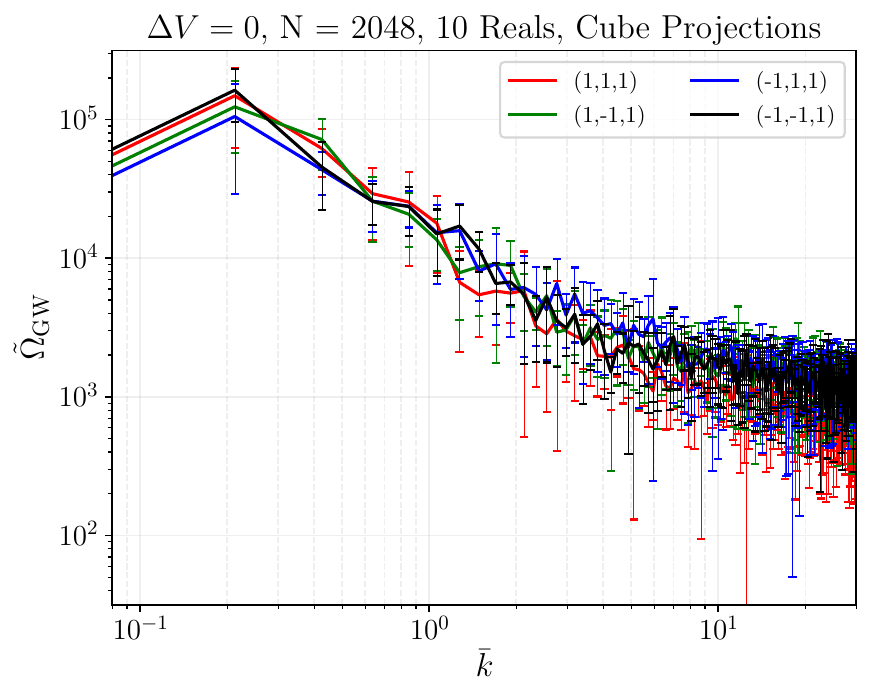}
\hspace{4mm}
\includegraphics[width=\columnwidth]{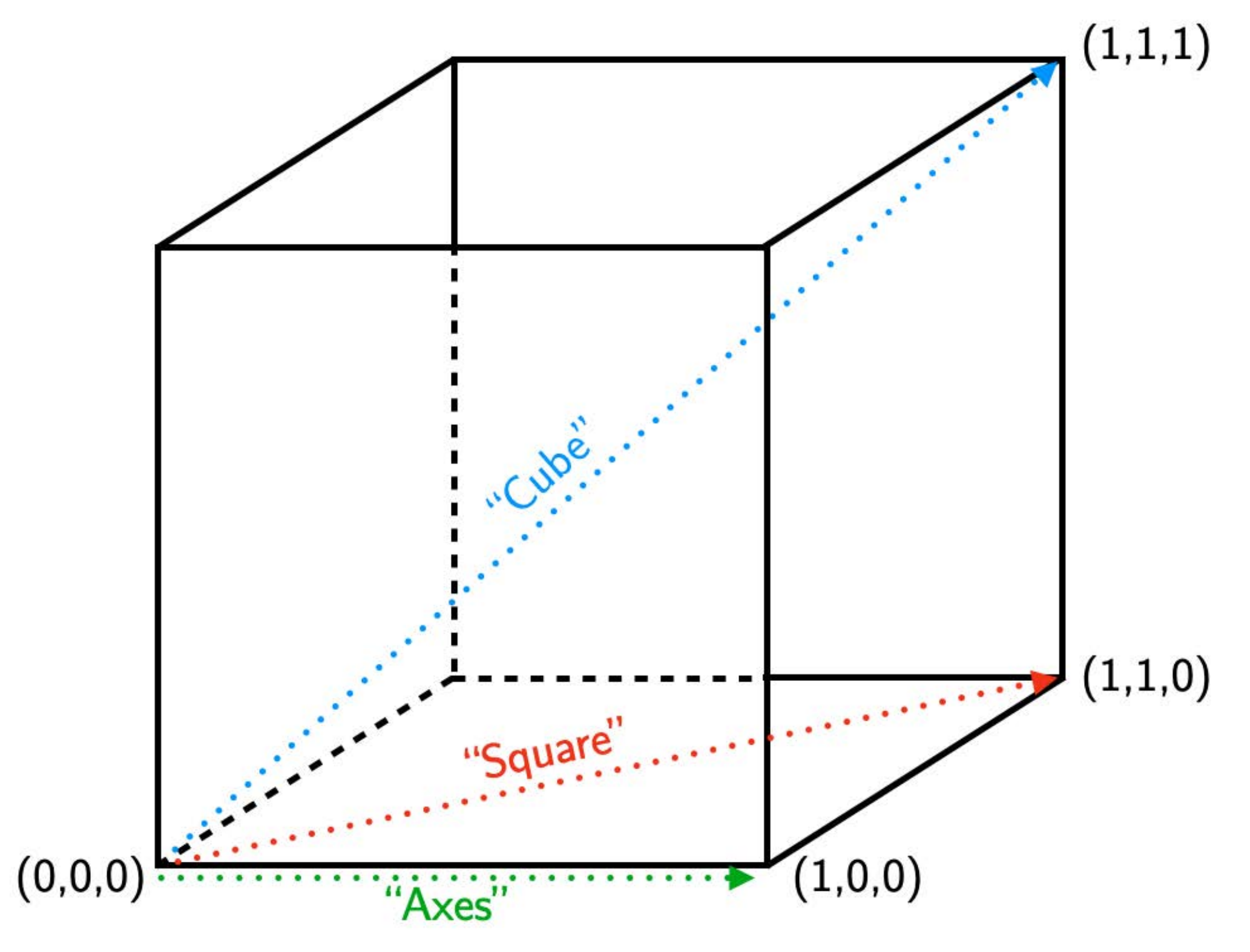}
\caption{$\bar{\Omega}_{\rm GW}$ computed along 13 different line of sight projections for the unbiased scenario, and averaged over ten realizations. All results presented here are at the final timestep, corresponding to $\tau_{\rm f} \simeq 26$. Error bars indicate one standard deviation over the average of realizations. We interpret the fact that the error bars overlap along each line of sight as a justification of the isotropy of the gravitational wave signature. Note that the axes, square, and cube projections each probe independent $k$-modes allowing for an altogether denser sampling of the spectrum, as can be seen by our illustration in the bottom right.}
\label{Fig:Anisotropies}
\end{figure*}

We also check one of the main assumptions of this work --- that the GW spectrum is approximately isotropic --- in Figure \ref{Fig:Anisotropies}. In these plots we compare the spectrum along $13$ different directions (3 ``axes", 6 ``square" and 4 ``cube") for $10$ realisations of unbiased simulations. The three categories of directions are hard to compare directly, as the $\bar{k}$ values for the square and cube directions are larger than those for the axes directions by a geometrical factor of $\sqrt{2}$ and $\sqrt{3}$ respectively, so we compare each category individually. We find that the spectra along each direction agree with each other within their error bars, which we interpret as a confirmation that our assumption of an isotropic signal is appropriate.

%
%

\subsection{Numerical Dictionary} \label{sec:A-numerical-dictionary}
As was touched upon above, here we explicitly list relationships between rescaled numerical quantities (which we express with an overbar), and their physical counterparts. 
\begin{itemize}
    \item Field values are expressed in units of $\eta$, i.e. as $\bar{\phi} = \phi/\eta$. 
    
    \item Comoving coordinates are given as $\bar{x}_{\mu} = (\bar{\tau},\bar{\textbf{x}})$, where $\bar{x}_\mu =\sqrt{\lambda}\eta \, x_\mu$. The numerical conjugate momentum is thus fixed as $\bar{\textbf{k}} = \textbf{k}/\sqrt{\lambda}\eta$. In the paper we often assume isotropy, in which case $\bar{k} = |\bar{\textbf{k}}|$.
    
    \item For completeness, $\bar{\epsilon} = \epsilon/\lambda$, though because this is a ratio of dimensionless quantities, we often choose not to use the barred notation.
    
    \item We operate only in radiation domination, where we set the scale factor to be $a = \bar{\tau}$. In this work, $a(\bar{\tau_{\rm i}}) = 1$ marks the beginning of the simulation. Within this framework, the conformal Hubble scale is given by $\mathcal{H} = \sqrt{\lambda}\eta/\bar{\tau}$, which can be further cast into $H = \sqrt{\lambda}\eta /\bar{\tau}^2$.

    \item In the previous subsection, the quantity $S_{\rm k}$ was defined, which is related to the the gravitational wave spectrum through
    \begin{align}
        \Omega_{\rm GW} = \frac{1}{\rho_{\rm c}}\frac{G}{2\pi^2 V a^4} S_{\rm k},
    \end{align}
    where $V$ is the comoving volume of the box. By dimensional analysis, one can see that $S_{\rm k}$ has mass dimension three. In terms of numerical quantities, one can find that the rescaled (numerical) version of this function is $\bar{S}_{\rm k} = \sqrt{\lambda} \, S_{\rm k}/\eta^3$. This is the quantity computed directly during the simulation. By also noting that $\bar{V} = V \, (\lambda^{1/2} \eta)^3 $ and $\rho_{\rm c} = 3H^2/8\pi G$ we can rewrite the gravitational spectrum as
    \begin{align}
        \Omega_{\rm GW} = \frac{4G^2 \eta^6}{3\pi H^2} \left[ \frac{\bar{S}_{\rm k}}{\bar{a}^4 \bar{V}} \right],
    \end{align}
    where purely numerical quantities to be extracted from simulations have been relegated to the square brackets\footnote{Note that $a = \bar{a}$ since the scale factor is de-facto dimensionless, and $\bar{V} = (N_x\Delta \bar{x})^3$ with $\Delta \bar{x}$ defined above.}. Now, using $H = \sqrt{\lambda}\eta /\bar{\tau}^2$ and $\bar{a} = \bar{\tau}$, we find the satisfyingly simple expression
    \begin{align}
        \Omega_{\rm GW} &= \frac{4}{3\pi}\left(\frac{\eta}{m_{\rm pl}} \right)^4 \left[ \frac{\bar{S}_{\rm k}}{\bar{V}}\right]\\
        &= \tilde{\Omega}_{\rm GW} \left(\frac{\eta}{m_{\rm pl}} \right)^4.
    \end{align}
    The usefulness of this form is immediately evident. Our simulations can be used to compute a generic form for $\tilde{\Omega}_{\rm GW}$, which can then be rescaled to any physical scenario where domain walls are formed at a symmetry breaking scale $\eta$ $(\simeq T_{\rm SSB}^2/m_{\rm pl}$ where $T_{\rm SSB}$ is the temperature of the symmetry breaking phase transition).
\end{itemize}

\section{\label{sec:levelA2}Matching Conditions and Numerical Redshifting}
The main aim of performing numerical simulations is to precisely compute the dynamics of a system during some period of the Universe's history. In order to do this, one needs to understand what physical temperature that the simulation began and ended at, in order to perform matching calculations and properly redshift observable quantities. As discussed above, the physical time that our simulations begin at depend on the free parameters of the model, $H_{\rm i} = \sqrt{\lambda} \eta$. We remain in radiation domination throughout the simulation, meaning the following expression can be used to determine the temperature at the beginning of the simulation. 
%
\begin{align}
   \frac{\pi^2}{30} g(T_{\rm i})  T_{\rm i}^4 = \frac{3 H_{\rm i}^2}{8\pi G}
\end{align}
%
where $g(T_{\rm i})$ is the number of (energetic) degrees of freedom present in the plasma at $T_{\rm i}$. For the high scale temperatures we will often consider, this is given by $g(T \gtrsim {\rm GeV}) = 106.75$. This means that the initial temperature is 
%
\begin{align}
   T_{\rm i} = \left(\frac{90 \lambda}{8\pi^3 g(T_{\rm i})}\right)^{1/4} \sqrt{\eta\,\, m_{\rm pl}}.
\end{align}
%
This initial time is also roughly the temperature of spontaneous symmetry breaking for this vacuum phase transition\footnote{The phase transition takes place in vacuum because we do not consider any thermal corrections to the $V(\phi)$ potential. In this sense symmetry breaking happens when the field is released from Hubble friction, at which point it evolves to one of its two vacua.} The symmetry breaking scale $\eta$ is therefore related to the temperature of the phase transition through $\eta \simeq T_{\rm i}^2/m_{\rm pl}$ where $T_{\rm i} \simeq T_{\rm SSB}$. The endpoint of the simulation takes place at $H_{\rm f} = H_{\rm i}/\bar{\tau}_{\rm f}^2$, which can easily be recast into 
%
\begin{align} \label{eq:Tfinal}
    T_{\rm f} = T_{\rm i} \left(\frac{\bar{\tau}_{\rm i}}{\bar{\tau}_{\rm f}}\right) \left(\frac{g(T_{\rm i})}{g(T_{\rm f})} \right)^{1/4},
\end{align}
%
where we remind the reader that we set $\bar{\tau}_{\rm i} = 1$. Thus, once one makes a choice for $\lambda$ and $\eta$, the start and end points of the simulation become fixed, and one can extract physically relevant quantities. 

\subsection{Gravitational wave amplitude}
Given the form of the gravitational wave amplitude at the end of the simulation ($\bar{\tau}_{\rm f}$), we can now redshift its value to the present day. Ultimately, we are aiming to compute $\Omega_{\rm GW,0}$, given by
%
\begin{align}
    \Omega_{\rm GW,0} = \frac{1}{\rho_{\rm c,0}} \frac{\id \rho_{\rm GW,0}}{\id \, {\rm ln}  \, k_0}.
\end{align}
%
First, note that $\id \, {\rm ln} \, k_0 = \id \, \ln \, k$ since $\id \ln k = \id k/k$, thus any redshifting cancels out. Next, we note that gravitational waves are radiation, which means they redshift like $\rho_{\rm GW,0} = \rho_{\rm GW}(\bar{\tau}_{\rm f}) \cdot (a(\bar{\tau}_{\rm f})/a_0)^4$ from the end of the simulation until today. Finally, the critical frequency today can be related to early times using $\rho_{\rm c,0} = \rho_{\rm c} \cdot  (H_0^2/H^2)$. Throwing this all in, we find
%
\begin{align}
    \Omega_{\rm GW,0} &= \Omega_{\rm GW}(\bar{\tau}_{\rm f}) \left( \frac{g_{0,s}} {g_{\bar{\tau}_{\rm f},s}}\right)^{4/3} \left( \frac{H(\bar{\tau}_{\rm f})}{H_0}\right)^2 \left( \frac{T_0}{ T({\bar{\tau}_{\rm f})}}\right)^4,
\end{align}
%
where in the above expression, $g_{0,s} = g_{\rm s}(T_0) \simeq 3.94$ is the effective number of entropic degrees of freedom today ($g_{\bar{\tau}_{\rm f},s} = g_{\rm s}(T_{\rm f})$ is the same but at time/temperature $\bar{\tau}_{\rm f}/T_{\rm f}$). Now, we make use of the following two relations
%
\begin{align}
    \Omega_{\rm r,0} &= g_0 \frac{8\pi^3 G}{90} \frac{T_0^4}{H_0^2},\\
    \frac{\pi^2}{30} g(T_{\rm f}) T^4_{\rm f} &= \frac{3 H^2_{\rm f}}{8\pi G}.
\end{align}
%
Note that $g_0 \simeq 3.38$ is the effective number of (energetic) degrees of freedom today. Inserting these two expressions allows us to fully eliminate the $H$ and $T$ factors above, leading to the simple result
%
\begin{align}
    \Omega_{\rm GW,0} h^2 &= \Omega_{\rm GW}(\bar{\tau}_{\rm f}) \cdot \Omega_{\rm r,0} h^2 \left( \frac{g(T_{\rm f})}{g_0} \right)\left( \frac{g_{0,s}}{g_s(T_{\rm f})} \right)^{4/3}\\
    &= 6.78 \times 10^{-6} \Omega_{\rm GW}(\bar{\tau}_{\rm f}) \nonumber \\ &\hspace{10mm}\times\left(\frac{\Omega_{\rm r,0} h^2}{2.472 \times 10^{-5}} \right) \left( \frac{g(T_{\rm f})}{106.75}\right)^{-1/3}\,,
\end{align}
where in the last line we have assumed that $g_s \simeq g$. This expression allows us to match the spectrum computed at the end of the simulation with what one would expect today.

\subsection{Redshifting the peak frequency}
The (physical) peak frequency of a gravitational wave spectrum, as observed today, can be related to its value at the end of the simulation by the usual redshifting factor \cite{Maggiore2018},
%
\begin{align}
    f_{\rm m,0} = f_{\rm m,f} \frac{a_{\rm f}}{a_0}.
\end{align}
%
The subscript ``m" refers to the peak frequency. In order to relate this to easily extractable quantities from our simulations, first note that $k_{\rm m,f} = 2\pi f_{\rm m,f}$, with relationship between physical and comoving $k$-modes $k_{\rm m,c} = a_{\rm f} k_{\rm m,f}$. In our simulations, we compute the numerical quantity $\bar{k}_{\rm m} = k_{\rm m,c}/\sqrt{\lambda} \eta$. Putting this all together, we find
%
\begin{align}
    f_0 = \frac{\sqrt{\lambda} \eta}{2\pi} \frac{1}{a_0} \bar{k}_{\rm m}
\end{align}
%
In our conventions, we have set $a(\bar{\tau}_i) = 1$ (that is, $a_0 \neq 1$). Conservation of entropy allows us to rewrite $a_0$ in terms of some earlier time $a_{\rm f}$ through $g_{\rm s}(T_{\rm f}) a_{\rm f}^3 T_{\rm f}^3 = g_{\rm 0, s} a_0^3 T_0^3$. Doing this yields
%
\begin{align}
    f_0 = \frac{\sqrt{\lambda} \eta}{2\pi} \frac{\bar{k}_{\rm m}}{a_{\rm f}}\left(\frac{g_{\rm 0,s}}{g_{\rm s}(T_{\rm f})} \right)^{1/3} \left(\frac{T_0}{T_{\rm f}} \right) 
\end{align}
%
Now we make a replacement for $T_{\rm f}$ through $\rho_{\rm rad} = \rho_{\rm c}$, valid deep in radiation domination. Namely, we use Eq.~\eqref{eq:Tfinal} and insert to find
%
\begin{align}
    f_0 = 22.5\,{\rm Hz}\,\left( \frac{g(T_{\rm f})}{106.75}\right)^{-1/12} \left( \frac{H_{\rm i}}{{\rm GeV}}\right)^{1/2} \bar{k}_{\rm m} 
\end{align}
%
Recall that $\sqrt{\lambda} \eta = H_{\rm i}$ and we again approximate $g_{\rm s}(T_{\rm f}) = g(T_{\rm f})$ in deriving this result. It is also implied that $\bar{k}_{\rm p}$ is extracted at $\bar{\tau}_{\rm f}$. One can instead rephrase this in terms of the symmetry breaking scale ($T_{\rm i} \simeq T_{\rm SSB}$), finding instead
%
\begin{align}
    f_0 = 2.68 \times 10^{-8}\,{\rm Hz}\, &\left(\frac{g_f}{106.75} \right)^{-1/12}  \left( \frac{T_{\rm SSB}}{{\rm GeV}}\right) \, \bar{k}_{\rm m}\,.
\end{align}
%
Keep in mind that all of the information relating to the decay phase is encoded in $\bar{k}_{\rm m} = \bar{k}_{\rm m}(\epsilon/\lambda)$ Heuristically speaking, as $\epsilon/\lambda \rightarrow 0$, the value of the peak frequency extracted from simulation will also follow $\bar{k}_{\rm m} \rightarrow 0$ as can be seen even with our limited dynamic range in Fig.~\ref{Fig:Vac-bias-timeslice-fits-comparison}. If one instead knows the temperature that the network decays, one can then use the fact that $H/k_{\rm m} = 0.91$ to find
%
\begin{align}
    f_0 = 2.94 \times 10^{-8}\,{\rm Hz}\,&\left(\frac{g_f}{106.75} \right)^{1/6}  \left( \frac{T_{\rm dec}}{{\rm GeV}}\right)\,.
\end{align}
%
This commonly used form can be found elsewhere \cite{Maggiore2018}.

\section{\label{sec:levelA3}Gravitational wave spectrum modeling}

The gravitational wave spectrum produced from networks of domain walls has been modelled in other work \cite{Gruber2024}, which we will attempt to improve upon here and additionally infer some of the parameters of the model from our data. We will start from the assumption that the network is in the scaling regime and that all of the energy lost from the network goes into gravitational waves. In principle, this second assumption is easy to relax by simply adding an additional variable to represent the fraction of this energy that goes into gravitational waves, but for the sake of simplicity we will not consider that case. We are also not attempting to model network decay, so the scenario under consideration is one where the domain wall network is present and in the scaling regime from some initial time $t_i$, up until the time of observation at $t_0$.

In the scaling regime, energy is lost by the network at a rate
%
\begin{align}
    \frac{\id\rho_{\rm loss}}{\id t}(t_e) = A\left( \frac{\Tilde{t}}{t_e} \right)^2 \,,
\end{align}
%
and the energy density of gravitational waves emitted in a time interval $t_e$ to $t_e + \id t_e$, with frequencies between $k_e$ and $k_e+dk_e$, is
%
\begin{align}
    \id \rho_{\rm GW} = \id t_e\id k_e \frac{\id\rho_{\rm loss}}{\id t}(t_e)\mathcal{P}(k_e, t_e) \,.
\end{align}
%
The spectral function $\mathcal{P}(k_e,t_e)$ describes how the energy is distributed and it is normalised such that
%
\begin{align}
    \int^\infty_0 \mathcal{P}(k_e, t_e) \id k_e = 1 \,.
\end{align}
%
We consider an instantaneous emission spectrum of the form
%
\begin{align}
    \mathcal{P}(k_e, t_e) = \frac{Bt_e}{\Tilde{t}\Tilde{k}}\frac{x_e^2}{1+x_e^{q+2}}
\end{align}
%
which is expressed in terms of the dimensionless variable, $x_e = k_et_e/\Tilde{k}\Tilde{t}$ and $B$ is a normalisation constant (note that we require $q>1$ for it to be finite). This function was chosen heuristically, with the intention of creating a spectrum which behaves like a power law of the form $k^3$ at low frequencies and another power law with a variable spectral index at high frequencies.

Additionally accounting for the redshifting of the frequencies, using $a(t_e)k_e = a(t_o)k_0$, and the dilution of the energy density due to the expansion of the Universe allows us to deduce that the spectral density at the time of observation is
%
\begin{align}
    \frac{\id\rho_{\rm GW}}{\id k_0} &= \int^{t_0}_{t_i} \id t_e \left( \frac{a(t_e)}{a(t_0)} \right)^3 A\left( \frac{\Tilde{t}}{t_e} \right)^2 \mathcal{P} \,, \nonumber \\
    &= \frac{AB\Tilde{t}}{\Tilde{k}}x_0^2\int^1_{y_i} \id y \frac{y^{1+\gamma}}{1 + x_0^{q+2}y^{(1-\gamma)(q+2)}} \,,
    \label{eq: model GW energy density}
\end{align}
%
where we have set $a \propto t^\gamma$, $x_o = k_ot_o/\Tilde{k}\Tilde{t}$ and $y = t_e/t_o$. In order to compare this with the rest of our results, we can express it as
%
\begin{align}
    \Omega_{\rm GW} = \frac{1}{t_0}8\pi G\Tilde{t}^2ABx_0^3 I_y \,,
\end{align}
%
where $I_y$ is the integral over $y$ from equation (\ref{eq: model GW energy density}) and the combination $x_0^3 I_y$ contains all of the information about the shape of the spectrum, with the other factors only influencing the amplitude. 

\begin{figure}
\centering 
\includegraphics[trim={0.3cm 0cm 1.4cm 1.4cm},clip,width=\columnwidth]{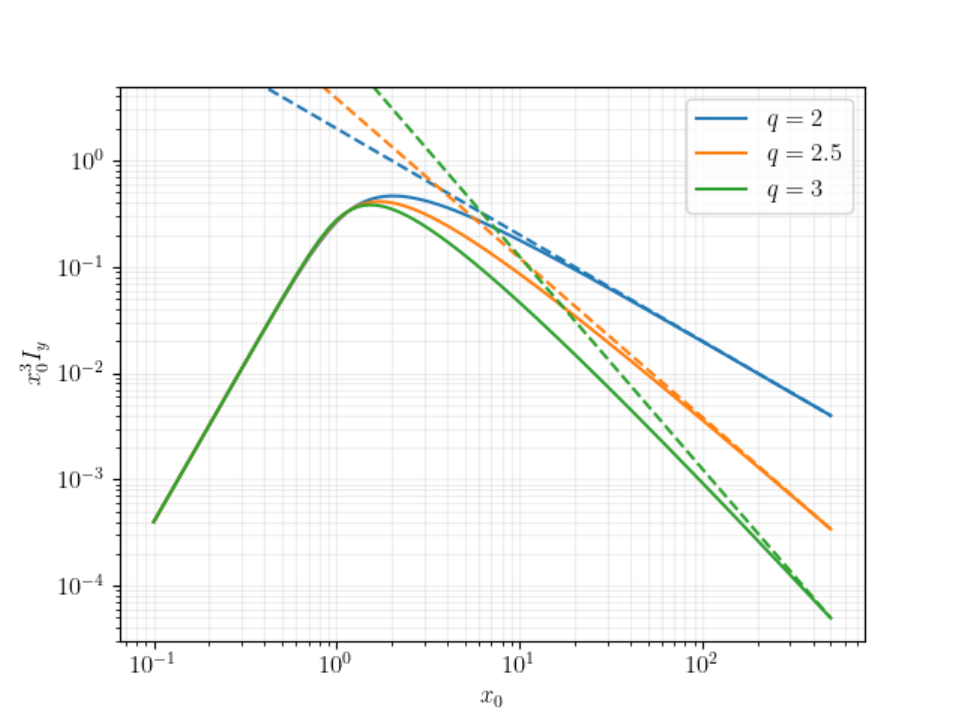}
\caption{The shape of the GW spectrum predicted by our modeling, for a few different values of $q$. The low frequency part of the spectrum looks like the power law $x_o^3$ by design, while the high frequency tail asymptotes to $x_o^{1-q}$. Using our results for the case where $\bar{\epsilon} = 0$, we find that the $q=2.5$ case is the most representative of the spectrum produced in our simulations.}
\label{fig: model integral}
\end{figure}

In Figure \ref{fig: model integral} we show what the spectrum looks like in this model, with the dotted lines representing $x_o^{q-1}$, which is clearly a good fit for the high frequency tail of the spectrum. Our fits to the results from simulations with $\bar{\epsilon} = 0$ have a high frequency spectral index of $\gamma_h\approx -1.5$, which implies that $q=2.5$.
%
%

\section{\label{sec:levelA4}Supplemental Data}
\begin{figure*}
\centering 
\includegraphics[width=\columnwidth]{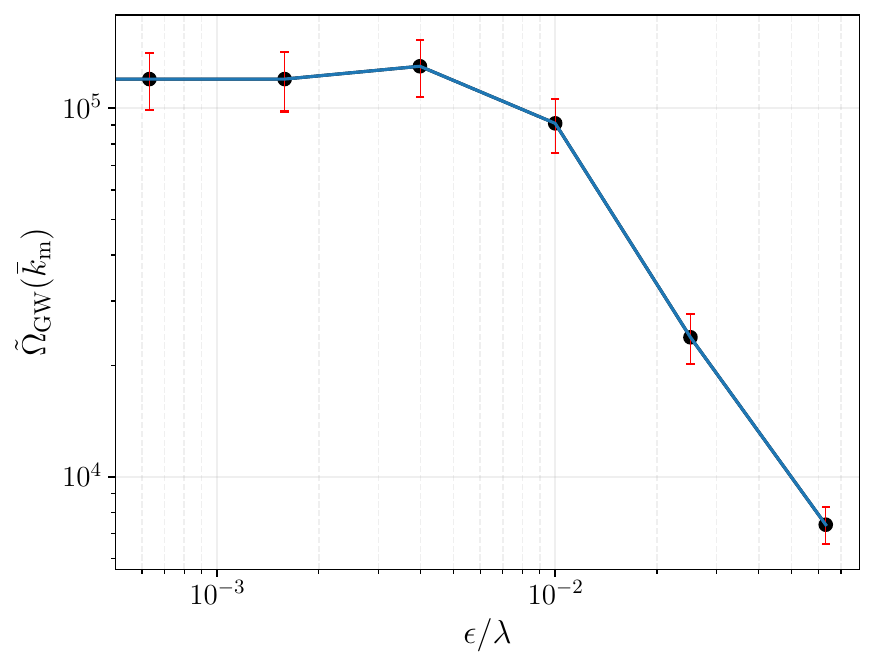}
\hspace{4mm}
\includegraphics[width=\columnwidth]{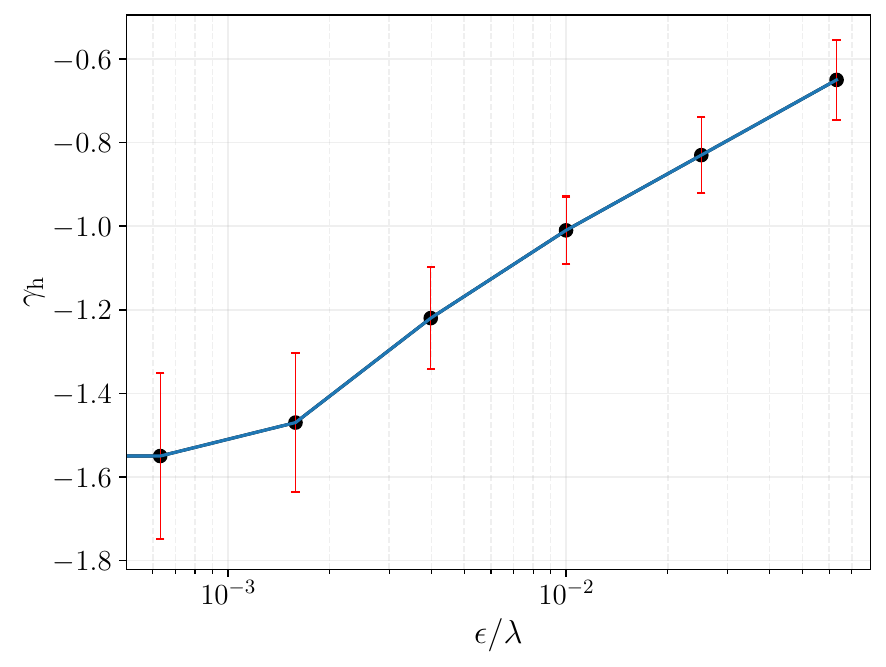}\\
\includegraphics[width=\columnwidth]{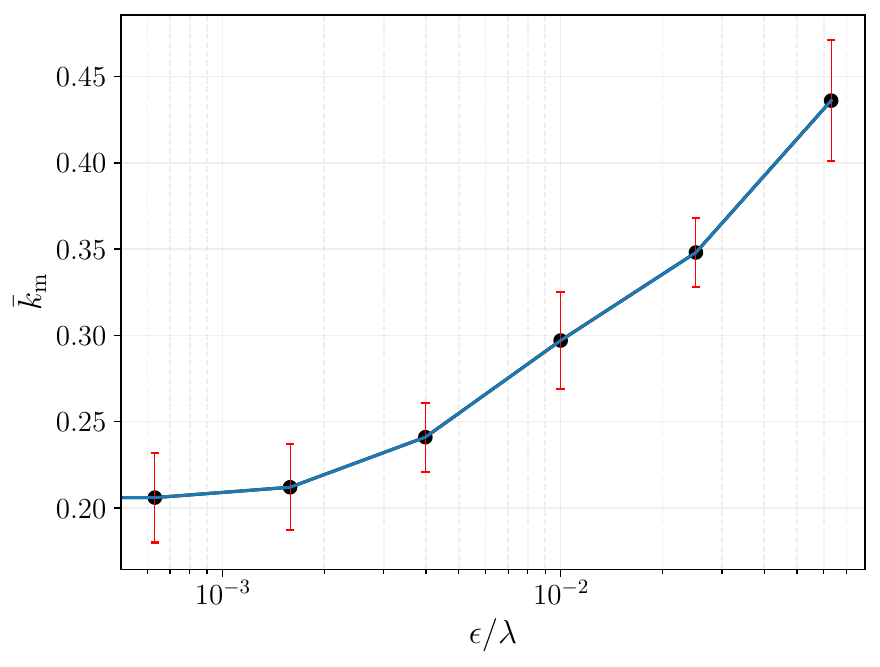}
\hspace{4mm}
\includegraphics[width=\columnwidth]{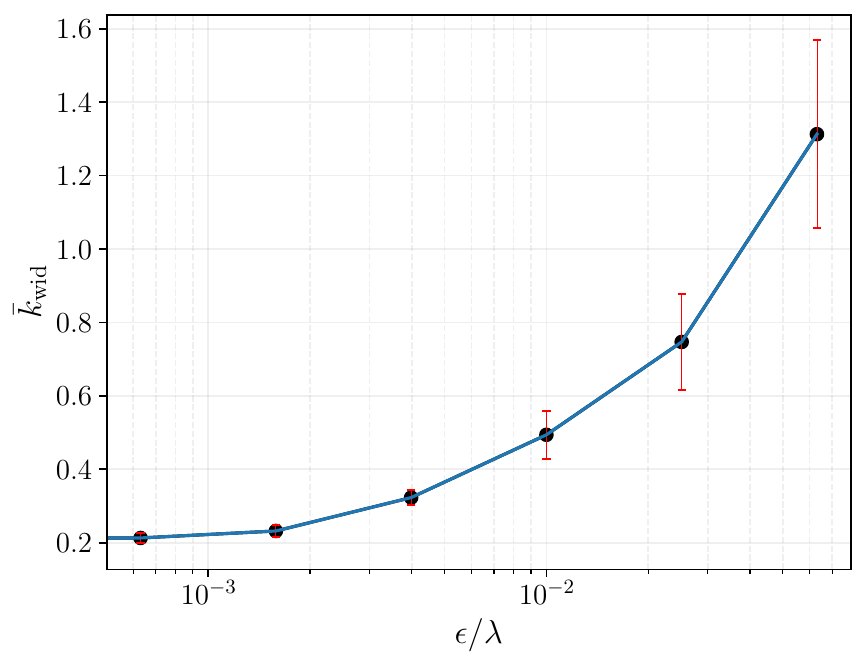}
\caption{Gravitational wave parameters as a function of $\epsilon/\lambda$ for the final ($\bar{\tau} \simeq 26$) timeslice in our simulations. Note that the $\epsilon/\lambda = 10^{-1.6},10^{-2.0},10^{-2.4}$ cases are the most informative, as they are the only scenarios in which the damping, scaling, and decay stages of network evolution are unambiguously resolved.}
\label{Fig:final_slice_summary}
\end{figure*}

Here we provide data for the fits used for each of the timesteps in Figs.~\ref{Fig:No-bias-timeslice-fits} and~\ref{Fig:Vac-bias-timeslice-fits-comparison} which can be found in Tables~\ref{tab:supp-GW-NB}-\ref{tab:supp-GW-VB-12}. As in the main text, error bars represent the standard deviation inferred by fitting to ten different realizations of a particular bias amplitude. When $\bar{\tau}_{\rm dec}$ displays N/A, no significant departure from scaling was found by the end of our simulation. When $\bar{\tau}_{\rm dec}$ is marked with an asterisk, the decay time takes place at $\bar{\tau}_{\rm dec} > \bar{\tau}_{\rm f} \simeq 26$. We show our main observables ($\bar{\Omega}_{\rm GW}(\bar{k}_{\rm m}), \gamma_{\rm h}, \bar{k}_{\rm m}, \bar{k}_{\rm wid}$) at the final timeslice in Fig.~\ref{Fig:final_slice_summary}.

\setlength{\tabcolsep}{10pt}
%
\begin{table*}[t]
    \centering
    \renewcommand{\arraystretch}{1.5} 
    \begin{tabular}{|c|c|c|c|c|c|c|c|}
        \hline
        $\epsilon/\lambda$ & $\bar{k}_{\rm cut}$ & $\bar{\tau}$ & $\bar{k}_{\rm m}$ & $\tilde{\Omega}(\bar{k}_{\rm m})$ & $\bar{k}_{\rm width}$ & $\gamma_{\rm h}$ & $\bar{\tau}_{\rm dec}$ \rule{0pt}{12pt} \\
        \hline
        $0$ & $3$ & $11.3$ & $0.504\pm0.040$ & $(3.82\pm 0.60)\times 10^{3}$ & $0.556\pm0.079$ & $-1.48 \pm 0.240$ & N/A\\
        $0$ & $3$ & $13.8$ & $0.418\pm0.037$ & $(8.86\pm 1.56)\times 10^{3}$ & $0.441\pm0.054$ & $-1.54 \pm 0.216$ & N/A\\
        $0$ & $3$ & $16.4$ & $0.357\pm0.027$ & $(1.76\pm 0.30)\times 10^{4}$ & $0.371\pm0.033$ & $-1.56 \pm 0.167$ & N/A\\
        $0$ & $3$ & $18.9$ & $0.309\pm0.021$ & $(3.18\pm 0.44)\times 10^{4}$ & $0.312\pm0.028$ & $-1.61 \pm 0.158$ & N/A\\
        $0$ & $3$ & $21.5$ & $0.269\pm0.024$ & $(5.24\pm 0.73)\times 10^{4}$ & $0.276\pm0.018$ & $-1.59 \pm 0.173$ & N/A\\
        $0$ & $3$ & $24.1$ & $0.234\pm0.025$ & $(7.95\pm 1.20)\times 10^{4}$ & $0.245\pm0.015$ & $-1.54 \pm 0.177$ & N/A\\
        $0$ & $3$ & $26.6$ & $0.205\pm0.026$ & $(1.20\pm 0.21)\times 10^{5}$ & $0.211\pm0.015$ & $-1.56 \pm 0.201$ & N/A\\
        \hline
    \end{tabular}
    \caption{Timeslice data for the no bias ($\Delta V = 0$) case.}
    \label{tab:supp-GW-NB}
\end{table*}
%
\setlength{\tabcolsep}{6pt}
\setlength{\tabcolsep}{10pt}
%
\begin{table*}[t]
    \centering
    \renewcommand{\arraystretch}{1.5} 
    \begin{tabular}{|c|c|c|c|c|c|c|c|}
        \hline
        $\epsilon/\lambda$ & $\bar{k}_{\rm cut}$ & $\bar{\tau}$ & $\bar{k}_{\rm m}$ & $\tilde{\Omega}(\bar{k}_{\rm m})$ & $\bar{k}_{\rm width}$ & $\gamma_{\rm h}$ & $\bar{\tau}_{\rm dec}$ \rule{0pt}{12pt} \\
        \hline
        $10^{-3.2}$ & $3$ & $11.3$ & $0.504\pm0.040$ & $(3.82\pm 0.59)\times 10^{3}$ & $0.567\pm0.076$ & $-1.48 \pm 0.237$ & N/A\\
        $10^{-3.2}$ & $3$ & $13.8$ & $0.417\pm0.036$ & $(8.85\pm 1.55)\times 10^{3}$ & $0.374\pm0.192$ & $-1.53 \pm 0.210$ & N/A\\
        $10^{-3.2}$ & $3$ & $16.4$ & $0.357\pm0.027$ & $(1.75\pm 0.30)\times 10^{4}$ & $0.367\pm0.012$ & $-1.56 \pm 0.170$ & N/A\\
        $10^{-3.2}$ & $3$ & $18.9$ & $0.310\pm0.020$ & $(3.15\pm 0.43)\times 10^{4}$ & $0.313\pm0.012$ & $-1.58 \pm 0.149$ & N/A\\
        $10^{-3.2}$ & $3$ & $21.5$ & $0.269\pm0.022$ & $(5.11\pm 0.63)\times 10^{4}$ & $0.281\pm0.014$ & $-1.53 \pm 0.152$ & N/A\\
        $10^{-3.2}$ & $3$ & $24.1$ & $0.235\pm0.025$ & $(7.89\pm 1.24)\times 10^{4}$ & $0.246\pm0.017$ & $-1.53 \pm 0.188$ & N/A\\
        $10^{-3.2}$ & $3$ & $26.6$ & $0.206\pm0.026$ & $(1.20\pm 0.21)\times 10^{5}$ & $0.213\pm0.014$ & $-1.55 \pm 0.198$ & N/A\\
        \hline
    \end{tabular}
    \caption{Timeslice data for the vacuum bias $\epsilon/\lambda = 10^{-3.2}$ case.}
    \label{tab:supp-GW-VB-32}
\end{table*}
%
\setlength{\tabcolsep}{6pt}

\setlength{\tabcolsep}{10pt}
%
\begin{table*}[t]
    \centering
    \renewcommand{\arraystretch}{1.5} 
    \begin{tabular}{|c|c|c|c|c|c|c|c|}
        \hline
        $\epsilon/\lambda$ & $\bar{k}_{\rm cut}$ & $\bar{\tau}$ & $\bar{k}_{\rm m}$ & $\tilde{\Omega}(\bar{k}_{\rm m})$ & $\bar{k}_{\rm width}$ & $\gamma_{\rm h}$ & $\bar{\tau}_{\rm dec}$ \rule{0pt}{12pt} \\
        \hline
        $10^{-2.8}$ & $3$ & $11.3$ & $0.503\pm0.041$ & $(3.83\pm 0.59)\times 10^{3}$ & $0.556\pm0.074$ & $-1.48 \pm 0.232$ & N/A\\
        $10^{-2.8}$ & $3$ & $13.8$ & $0.418\pm0.035$ & $(8.84\pm 1.54)\times 10^{3}$ & $0.444\pm0.050$ & $-1.52 \pm 0.203$ & N/A\\
        $10^{-2.8}$ & $3$ & $16.4$ & $0.357\pm0.026$ & $(1.74\pm 0.30)\times 10^{4}$ & $0.378\pm0.034$ & $-1.54 \pm 0.162$ & N/A\\
        $10^{-2.8}$ & $3$ & $18.9$ & $0.312\pm0.020$ & $(3.14\pm 0.39)\times 10^{4}$ & $0.323\pm0.013$ & $-1.54 \pm 0.119$ & N/A\\
        $10^{-2.8}$ & $3$ & $21.5$ & $0.273\pm0.018$ & $(5.08\pm 0.64)\times 10^{4}$ & $0.294\pm0.015$ & $-1.49 \pm 0.135$ & N/A\\
        $10^{-2.8}$ & $3$ & $24.1$ & $0.239\pm0.022$ & $(7.87\pm 1.28)\times 10^{4}$ & $0.259\pm0.018$ & $-1.48 \pm 0.159$ & N/A\\
        $10^{-2.8}$ & $3$ & $26.6$ & $0.212\pm0.025$ & $(1.20\pm 0.22)\times 10^{5}$ & $0.232\pm0.016$ & $-1.47 \pm 0.167$ & N/A\\
        \hline
    \end{tabular}
    \caption{Timeslice data for the vacuum bias $\epsilon/\lambda = 10^{-2.8}$ case.}
    \label{tab:supp-GW-VB-28}
\end{table*}
%
\setlength{\tabcolsep}{6pt}

\setlength{\tabcolsep}{10pt}
%
\begin{table*}[t]
    \centering
    \renewcommand{\arraystretch}{1.5} 
    \begin{tabular}{|c|c|c|c|c|c|c|c|}
        \hline
        $\epsilon/\lambda$ & $\bar{k}_{\rm cut}$ & $\bar{\tau}$ & $\bar{k}_{\rm m}$ & $\tilde{\Omega}(\bar{k}_{\rm m})$ & $\bar{k}_{\rm width}$ & $\gamma_{\rm h}$ & $\bar{\tau}_{\rm dec}$ \rule{0pt}{12pt} \\
        \hline
        $10^{-2.4}$ & $3$ & $11.3$ & $0.504\pm0.040$ & $(3.84\pm 0.58)\times 10^{3}$ & $0.565\pm0.076$ & $-1.46 \pm 0.220$ & $31.71^* \pm 5.26$\\
        $10^{-2.4}$ & $3$ & $13.8$ & $0.418\pm0.032$ & $(8.85\pm 1.48)\times 10^{3}$ & $0.463\pm0.048$ & $-1.47 \pm 0.186$ & $31.71^* \pm 5.26$\\
        $10^{-2.4}$ & $4$ & $16.4$ & $0.361\pm0.024$ & $(1.73\pm 0.30)\times 10^{4}$ & $0.421\pm0.030$ & $-1.38 \pm 0.147$ & $31.71^* \pm 5.26$\\
        $10^{-2.4}$ & $5$ & $18.9$ & $0.320\pm0.018$ & $(3.18\pm 0.46)\times 10^{4}$ & $0.389\pm0.015$ & $-1.33 \pm 0.076$ & $31.71^* \pm 5.26$\\
        $10^{-2.4}$ & $7$ & $21.5$ & $0.286\pm0.014$ & $(5.28\pm 0.71)\times 10^{4}$ & $0.362\pm0.018$ & $-1.28 \pm 0.083$ & $31.71^* \pm 5.26$\\
        $10^{-2.4}$ & $10$ & $24.1$ & $0.260\pm0.018$ & $(8.41\pm 1.47)\times 10^{4}$ & $0.344\pm0.022$ & $-1.23 \pm 0.101$ & $31.71^* \pm 5.26$\\
        $10^{-2.4}$ & $10$ & $26.6$ & $0.241\pm0.020$ & $(1.30\pm 0.23)\times 10^{5}$ & $0.323\pm0.021$ & $-1.22 \pm 0.122$ & $31.71^* \pm 5.26$\\
        \hline
    \end{tabular}
    \caption{Timeslice data for the vacuum bias $\epsilon/\lambda = 10^{-2.4}$ case.}
    \label{tab:supp-GW-VB-24}
\end{table*}
%
\setlength{\tabcolsep}{6pt}

\setlength{\tabcolsep}{10pt}
%
\begin{table*}[t]
    \centering
    \renewcommand{\arraystretch}{1.5} 
    \begin{tabular}{|c|c|c|c|c|c|c|c|}
        \hline
        $\epsilon/\lambda$ & $\bar{k}_{\rm cut}$ & $\bar{\tau}$ & $\bar{k}_{\rm m}$ & $\tilde{\Omega}(\bar{k}_{\rm m})$ & $\bar{k}_{\rm width}$ & $\gamma_{\rm h}$ & $\bar{\tau}_{\rm dec}$ \rule{0pt}{12pt} \\
        \hline
        $10^{-2.0}$ & $3$ & $11.3$ & $0.510\pm0.039$ & $(3.93\pm 0.54)\times 10^{3}$ & $0.614\pm0.081$ & $-1.36 \pm 0.194$ & $20.82 \pm 0.97$\\
        $10^{-2.0}$ & $3$ & $13.8$ & $0.431\pm0.026$ & $(9.28\pm 1.29)\times 10^{3}$ & $0.563\pm0.050$ & $-1.25 \pm 0.134$ & $20.82 \pm 0.97$\\
        $10^{-2.0}$ & $4$ & $16.4$ & $0.387\pm0.020$ & $(1.94\pm 0.29)\times 10^{4}$ & $0.573\pm0.046$ & $-1.11 \pm 0.119$ & $20.82 \pm 0.97$\\
        $10^{-2.0}$ & $5$ & $18.9$ & $0.357\pm0.018$ & $(3.64\pm 0.47)\times 10^{4}$ & $0.570\pm0.059$ & $-1.05 \pm 0.117$ & $20.82 \pm 0.97$\\
        $10^{-2.0}$ & $7$ & $21.5$ & $0.338\pm0.024$ & $(5.99\pm 0.74)\times 10^{4}$ & $0.553\pm0.058$ & $-1.02 \pm 0.102$ & $20.82 \pm 0.97$\\
        $10^{-2.0}$ & $10$ & $24.1$ & $0.318\pm0.028$ & $(8.10\pm 1.00)\times 10^{4}$ & $0.515\pm0.061$ & $-1.03 \pm 0.087$ & $20.82 \pm 0.97$\\
        $10^{-2.0}$ & $10$ & $26.6$ & $0.297\pm0.028$ & $(9.10\pm 1.52)\times 10^{4}$ & $0.494\pm0.066$ & $-1.01 \pm 0.081$ & $20.82 \pm 0.97$\\
        \hline
    \end{tabular}
    \caption{Timeslice data for the vacuum bias $\epsilon/\lambda = 10^{-2.0}$ case.}
    \label{tab:supp-GW-VB-20}
\end{table*}
%
\setlength{\tabcolsep}{6pt}

\setlength{\tabcolsep}{10pt}
%
\begin{table*}[t]
    \centering
    \renewcommand{\arraystretch}{1.5} 
    \begin{tabular}{|c|c|c|c|c|c|c|c|}
        \hline
        $\epsilon/\lambda$ & $\bar{k}_{\rm cut}$ & $\bar{\tau}$ & $\bar{k}_{\rm m}$ & $\tilde{\Omega}(\bar{k}_{\rm m})$ & $\bar{k}_{\rm width}$ & $\gamma_{\rm h}$ & $\bar{\tau}_{\rm dec}$ \rule{0pt}{12pt} \\
        \hline
        $10^{-1.6}$ & $10$ & $11.3$ & $0.551\pm0.039$ & $(4.38\pm 0.43)\times 10^{3}$ & $0.999\pm0.104$ & $-0.94 \pm 0.104$ & $15.29 \pm 0.31$\\
        $10^{-1.6}$ & $13$ & $13.8$ & $0.488\pm0.028$ & $(1.12\pm 0.13)\times 10^{4}$ & $0.994\pm0.103$ & $-0.85 \pm 0.081$ & $15.29 \pm 0.31$\\
        $10^{-1.6}$ & $7$  & $16.4$ & $0.446\pm0.026$ & $(2.16\pm 0.27)\times 10^{4}$ & $0.862\pm0.108$ & $-0.90 \pm 0.098$ & $15.29 \pm 0.31$\\
        $10^{-1.6}$ & $7$  & $18.9$ & $0.419\pm0.023$ & $(2.70\pm 0.31)\times 10^{4}$ & $0.705\pm0.315$ & $-0.90 \pm 0.088$ & $15.29 \pm 0.31$\\
        $10^{-1.6}$ & $7$  & $21.5$ & $0.389\pm0.021$ & $(2.70\pm 0.36)\times 10^{4}$ & $0.739\pm0.087$ & $-0.91 \pm 0.084$ & $15.29 \pm 0.31$\\
        $10^{-1.6}$ & $7$  & $24.1$ & $0.365\pm0.020$ & $(2.56\pm 0.39)\times 10^{4}$ & $0.716\pm0.104$ & $-0.88 \pm 0.088$ & $15.29 \pm 0.31$\\
        $10^{-1.6}$ & $7$  & $26.6$ & $0.348\pm0.020$ & $(2.39\pm 0.37)\times 10^{4}$ & $0.747\pm0.130$ & $-0.83 \pm 0.091$ & $15.29 \pm 0.31$\\
        \hline
    \end{tabular}
    \caption{Timeslice data for the vacuum bias $\epsilon/\lambda = 10^{-1.6}$ case.}
    \label{tab:supp-GW-VB-16}
\end{table*}
%
\setlength{\tabcolsep}{6pt}

\setlength{\tabcolsep}{10pt}
%
\begin{table*}[t]
    \centering
    \renewcommand{\arraystretch}{1.5} 
    \begin{tabular}{|c|c|c|c|c|c|c|c|}
        \hline
        $\epsilon/\lambda$ & $\bar{k}_{\rm cut}$ & $\bar{\tau}$ & $\bar{k}_{\rm m}$ & $\tilde{\Omega}(\bar{k}_{\rm m})$ & $\bar{k}_{\rm width}$ & $\gamma_{\rm h}$ & $\bar{\tau}_{\rm dec}$ \rule{0pt}{12pt} \\
        \hline
        $10^{-1.2}$ & $13$ & $11.3$ & $0.622\pm0.055$ & $(6.22\pm 0.74)\times 10^{3}$ & $1.436\pm0.222$ & $-0.79 \pm 0.129$ & $12.08 \pm 0.15$\\
        $10^{-1.2}$ & $10$ & $13.8$ & $0.553\pm0.044$ & $(9.90\pm 1.04)\times 10^{3}$ & $1.164\pm0.106$ & $-0.84 \pm 0.102$ & $12.08 \pm 0.15$\\
        $10^{-1.2}$ & $8$  & $16.4$ & $0.491\pm0.037$ & $(9.78\pm 1.11)\times 10^{3}$ & $0.969\pm0.061$ & $-0.88 \pm 0.079$ & $12.08 \pm 0.15$\\
        $10^{-1.2}$ & $7$  & $18.9$ & $0.461\pm0.035$ & $(8.80\pm 0.96)\times 10^{3}$ & $0.999\pm0.096$ & $-0.81 \pm 0.077$ & $12.08 \pm 0.15$\\
        $10^{-1.2}$ & $8$  & $21.5$ & $0.444\pm0.034$ & $(8.10\pm 0.86)\times 10^{3}$ & $1.117\pm0.155$ & $-0.74 \pm 0.084$ & $12.08 \pm 0.15$\\
        $10^{-1.2}$ & $9$  & $24.1$ & $0.437\pm0.035$ & $(7.68\pm 0.86)\times 10^{3}$ & $1.063\pm0.492$ & $-0.69 \pm 0.087$ & $12.08 \pm 0.15$\\
        $10^{-1.2}$ & $7$  & $26.6$ & $0.436\pm0.035$ & $(7.41\pm 0.85)\times 10^{3}$ & $1.313\pm0.256$ & $-0.65 \pm 0.095$ & $12.08 \pm 0.15$\\
        \hline
    \end{tabular}
    \caption{Timeslice data for the vacuum bias $\epsilon/\lambda = 10^{-1.2}$ case.}
    \label{tab:supp-GW-VB-12}
\end{table*}
%
\setlength{\tabcolsep}{6pt}
%
%
\clearpage
\bibliographystyle{apsrev4-1}

\begin{thebibliography}{58}%
\makeatletter
\providecommand \@ifxundefined [1]{%
 \@ifx{#1\undefined}
}%
\providecommand \@ifnum [1]{%
 \ifnum #1\expandafter \@firstoftwo
 \else \expandafter \@secondoftwo
 \fi
}%
\providecommand \@ifx [1]{%
 \ifx #1\expandafter \@firstoftwo
 \else \expandafter \@secondoftwo
 \fi
}%
\providecommand \natexlab [1]{#1}%
\providecommand \enquote  [1]{``#1''}%
\providecommand \bibnamefont  [1]{#1}%
\providecommand \bibfnamefont [1]{#1}%
\providecommand \citenamefont [1]{#1}%
\providecommand \href@noop [0]{\@secondoftwo}%
\providecommand \href [0]{\begingroup \@sanitize@url \@href}%
\providecommand \@href[1]{\@@startlink{#1}\@@href}%
\providecommand \@@href[1]{\endgroup#1\@@endlink}%
\providecommand \@sanitize@url [0]{\catcode `\\12\catcode `\$12\catcode `\&12\catcode `\#12\catcode `\^12\catcode `\_12\catcode `\%12\relax}%
\providecommand \@@startlink[1]{}%
\providecommand \@@endlink[0]{}%
\providecommand \url  [0]{\begingroup\@sanitize@url \@url }%
\providecommand \@url [1]{\endgroup\@href {#1}{\urlprefix }}%
\providecommand \urlprefix  [0]{URL }%
\providecommand \Eprint [0]{\href }%
\providecommand \doibase [0]{http://dx.doi.org/}%
\providecommand \selectlanguage [0]{\@gobble}%
\providecommand \bibinfo  [0]{\@secondoftwo}%
\providecommand \bibfield  [0]{\@secondoftwo}%
\providecommand \translation [1]{[#1]}%
\providecommand \BibitemOpen [0]{}%
\providecommand \bibitemStop [0]{}%
\providecommand \bibitemNoStop [0]{.\EOS\space}%
\providecommand \EOS [0]{\spacefactor3000\relax}%
\providecommand \BibitemShut  [1]{\csname bibitem#1\endcsname}%
\let\auto@bib@innerbib\@empty
\bibitem [{\citenamefont {Agazie}\ \emph {et~al.}(2023)\citenamefont {Agazie} \emph {et~al.}}]{NANOGravDetection2023}%
  \BibitemOpen
  \bibfield  {author} {\bibinfo {author} {\bibfnamefont {G.}~\bibnamefont {Agazie}} \emph {et~al.} (\bibinfo {collaboration} {NANOGrav}),\ }\href {\doibase 10.3847/2041-8213/acdac6} {\bibfield  {journal} {\bibinfo  {journal} {Astrophys. J. Lett.}\ }\textbf {\bibinfo {volume} {951}},\ \bibinfo {pages} {L8} (\bibinfo {year} {2023})},\ \Eprint {http://arxiv.org/abs/2306.16213} {arXiv:2306.16213 [astro-ph.HE]} \BibitemShut {NoStop}%
\bibitem [{\citenamefont {Antoniadis}\ \emph {et~al.}(2023)\citenamefont {Antoniadis} \emph {et~al.}}]{EPTADetection2023}%
  \BibitemOpen
  \bibfield  {author} {\bibinfo {author} {\bibfnamefont {J.}~\bibnamefont {Antoniadis}} \emph {et~al.} (\bibinfo {collaboration} {EPTA}),\ }\href {\doibase 10.48550/arXiv.2306.16214} {\bibfield  {journal} {\bibinfo  {journal} {arXiv e-prints}\ ,\ \bibinfo {eid} {arXiv:2306.16214}} (\bibinfo {year} {2023})},\ \Eprint {http://arxiv.org/abs/2306.16214} {arXiv:2306.16214 [astro-ph.HE]} \BibitemShut {NoStop}%
\bibitem [{\citenamefont {Reardon}\ \emph {et~al.}(2023)\citenamefont {Reardon} \emph {et~al.}}]{PPTADetection2023}%
  \BibitemOpen
  \bibfield  {author} {\bibinfo {author} {\bibfnamefont {D.~J.}\ \bibnamefont {Reardon}} \emph {et~al.},\ }\href {\doibase 10.3847/2041-8213/acdd02} {\bibfield  {journal} {\bibinfo  {journal} {Astrophys. J. Lett.}\ }\textbf {\bibinfo {volume} {951}},\ \bibinfo {pages} {L6} (\bibinfo {year} {2023})},\ \Eprint {http://arxiv.org/abs/2306.16215} {arXiv:2306.16215 [astro-ph.HE]} \BibitemShut {NoStop}%
\bibitem [{\citenamefont {Xu}\ \emph {et~al.}(2023)\citenamefont {Xu} \emph {et~al.}}]{CPTADetection2023}%
  \BibitemOpen
  \bibfield  {author} {\bibinfo {author} {\bibfnamefont {H.}~\bibnamefont {Xu}} \emph {et~al.},\ }\href {\doibase 10.1088/1674-4527/acdfa5} {\bibfield  {journal} {\bibinfo  {journal} {Res. Astron. Astrophys.}\ }\textbf {\bibinfo {volume} {23}},\ \bibinfo {pages} {075024} (\bibinfo {year} {2023})},\ \Eprint {http://arxiv.org/abs/2306.16216} {arXiv:2306.16216 [astro-ph.HE]} \BibitemShut {NoStop}%
\bibitem [{\citenamefont {Hellings}\ and\ \citenamefont {Downs}(1983)}]{Hellings1983}%
  \BibitemOpen
  \bibfield  {author} {\bibinfo {author} {\bibfnamefont {R.~w.}\ \bibnamefont {Hellings}}\ and\ \bibinfo {author} {\bibfnamefont {G.~s.}\ \bibnamefont {Downs}},\ }\href {\doibase 10.1086/183954} {\bibfield  {journal} {\bibinfo  {journal} {Astrophys. J. Lett.}\ }\textbf {\bibinfo {volume} {265}},\ \bibinfo {pages} {L39} (\bibinfo {year} {1983})}\BibitemShut {NoStop}%
\bibitem [{\citenamefont {Arzoumanian}\ \emph {et~al.}(2020)\citenamefont {Arzoumanian} \emph {et~al.}}]{NANOGrav2020}%
  \BibitemOpen
  \bibfield  {author} {\bibinfo {author} {\bibfnamefont {Z.}~\bibnamefont {Arzoumanian}} \emph {et~al.} (\bibinfo {collaboration} {NANOGrav}),\ }\href {\doibase 10.3847/2041-8213/abd401} {\bibfield  {journal} {\bibinfo  {journal} {Astrophys. J. Lett.}\ }\textbf {\bibinfo {volume} {905}},\ \bibinfo {pages} {L34} (\bibinfo {year} {2020})},\ \Eprint {http://arxiv.org/abs/2009.04496} {arXiv:2009.04496 [astro-ph.HE]} \BibitemShut {NoStop}%
\bibitem [{\citenamefont {Afzal}\ \emph {et~al.}(2023)\citenamefont {Afzal} \emph {et~al.}}]{NANOGrav2023Exotic}%
  \BibitemOpen
  \bibfield  {author} {\bibinfo {author} {\bibfnamefont {A.}~\bibnamefont {Afzal}} \emph {et~al.} (\bibinfo {collaboration} {NANOGrav}),\ }\href {\doibase 10.3847/2041-8213/acdc91} {\bibfield  {journal} {\bibinfo  {journal} {Astrophys. J. Lett.}\ }\textbf {\bibinfo {volume} {951}},\ \bibinfo {pages} {L11} (\bibinfo {year} {2023})},\ \Eprint {http://arxiv.org/abs/2306.16219} {arXiv:2306.16219 [astro-ph.HE]} \BibitemShut {NoStop}%
\bibitem [{\citenamefont {Begelman}\ \emph {et~al.}(1980)\citenamefont {Begelman}, \citenamefont {Blandford},\ and\ \citenamefont {Rees}}]{Begelman1980}%
  \BibitemOpen
  \bibfield  {author} {\bibinfo {author} {\bibfnamefont {M.~C.}\ \bibnamefont {Begelman}}, \bibinfo {author} {\bibfnamefont {R.~D.}\ \bibnamefont {Blandford}}, \ and\ \bibinfo {author} {\bibfnamefont {M.~J.}\ \bibnamefont {Rees}},\ }\href {\doibase 10.1038/287307a0} {\bibfield  {journal} {\bibinfo  {journal} {Nature}\ }\textbf {\bibinfo {volume} {287}},\ \bibinfo {pages} {307} (\bibinfo {year} {1980})}\BibitemShut {NoStop}%
\bibitem [{\citenamefont {{Phinney}}(2001)}]{Phinney2001}%
  \BibitemOpen
  \bibfield  {author} {\bibinfo {author} {\bibfnamefont {E.~S.}\ \bibnamefont {{Phinney}}},\ }\href {\doibase 10.48550/arXiv.astro-ph/0108028} {\bibfield  {journal} {\bibinfo  {journal} {arXiv e-prints}\ ,\ \bibinfo {eid} {astro-ph/0108028}} (\bibinfo {year} {2001})},\ \Eprint {http://arxiv.org/abs/astro-ph/0108028} {arXiv:astro-ph/0108028 [astro-ph]} \BibitemShut {NoStop}%
\bibitem [{\citenamefont {Vilenkin}\ and\ \citenamefont {Shellard}(2000)}]{Vilenkin2000}%
  \BibitemOpen
  \bibfield  {author} {\bibinfo {author} {\bibfnamefont {A.}~\bibnamefont {Vilenkin}}\ and\ \bibinfo {author} {\bibfnamefont {E.~P.~S.}\ \bibnamefont {Shellard}},\ }\href@noop {} {\emph {\bibinfo {title} {{Cosmic Strings and Other Topological Defects}}}}\ (\bibinfo  {publisher} {Cambridge University Press},\ \bibinfo {year} {2000})\BibitemShut {NoStop}%
\bibitem [{\citenamefont {Kibble}(1976)}]{Kibble1976}%
  \BibitemOpen
  \bibfield  {author} {\bibinfo {author} {\bibfnamefont {T.~W.~B.}\ \bibnamefont {Kibble}},\ }\href {\doibase 10.1088/0305-4470/9/8/029} {\bibfield  {journal} {\bibinfo  {journal} {J. Phys. A}\ }\textbf {\bibinfo {volume} {9}},\ \bibinfo {pages} {1387} (\bibinfo {year} {1976})}\BibitemShut {NoStop}%
\bibitem [{\citenamefont {Kibble}(1980)}]{Kibble1980}%
  \BibitemOpen
  \bibfield  {author} {\bibinfo {author} {\bibfnamefont {T.~W.~B.}\ \bibnamefont {Kibble}},\ }\href {\doibase 10.1016/0370-1573(80)90091-5} {\bibfield  {journal} {\bibinfo  {journal} {Phys. Rept.}\ }\textbf {\bibinfo {volume} {67}},\ \bibinfo {pages} {183} (\bibinfo {year} {1980})}\BibitemShut {NoStop}%
\bibitem [{\citenamefont {Press}\ \emph {et~al.}(1989)\citenamefont {Press}, \citenamefont {Ryden},\ and\ \citenamefont {Spergel}}]{Press1989}%
  \BibitemOpen
  \bibfield  {author} {\bibinfo {author} {\bibfnamefont {W.~H.}\ \bibnamefont {Press}}, \bibinfo {author} {\bibfnamefont {B.~S.}\ \bibnamefont {Ryden}}, \ and\ \bibinfo {author} {\bibfnamefont {D.~N.}\ \bibnamefont {Spergel}},\ }\href {\doibase 10.1086/168151} {\bibfield  {journal} {\bibinfo  {journal} {Astrophys. J.}\ }\textbf {\bibinfo {volume} {347}},\ \bibinfo {pages} {590} (\bibinfo {year} {1989})}\BibitemShut {NoStop}%
\bibitem [{\citenamefont {Hindmarsh}(1996)}]{Hindmarsh1996}%
  \BibitemOpen
  \bibfield  {author} {\bibinfo {author} {\bibfnamefont {M.}~\bibnamefont {Hindmarsh}},\ }\href {\doibase 10.1103/PhysRevLett.77.4495} {\bibfield  {journal} {\bibinfo  {journal} {Phys. Rev. Lett.}\ }\textbf {\bibinfo {volume} {77}},\ \bibinfo {pages} {4495} (\bibinfo {year} {1996})},\ \Eprint {http://arxiv.org/abs/hep-ph/9605332} {arXiv:hep-ph/9605332} \BibitemShut {NoStop}%
\bibitem [{\citenamefont {Oliveira}\ \emph {et~al.}(2005)\citenamefont {Oliveira}, \citenamefont {Martins},\ and\ \citenamefont {Avelino}}]{Oliveira2004}%
  \BibitemOpen
  \bibfield  {author} {\bibinfo {author} {\bibfnamefont {J.~C. R.~E.}\ \bibnamefont {Oliveira}}, \bibinfo {author} {\bibfnamefont {C.~J. A.~P.}\ \bibnamefont {Martins}}, \ and\ \bibinfo {author} {\bibfnamefont {P.~P.}\ \bibnamefont {Avelino}},\ }\href {\doibase 10.1103/PhysRevD.71.083509} {\bibfield  {journal} {\bibinfo  {journal} {Phys. Rev. D}\ }\textbf {\bibinfo {volume} {71}},\ \bibinfo {pages} {083509} (\bibinfo {year} {2005})},\ \Eprint {http://arxiv.org/abs/hep-ph/0410356} {arXiv:hep-ph/0410356} \BibitemShut {NoStop}%
\bibitem [{\citenamefont {Avelino}\ \emph {et~al.}(2005)\citenamefont {Avelino}, \citenamefont {Martins},\ and\ \citenamefont {Oliveira}}]{Avelino2005}%
  \BibitemOpen
  \bibfield  {author} {\bibinfo {author} {\bibfnamefont {P.~P.}\ \bibnamefont {Avelino}}, \bibinfo {author} {\bibfnamefont {C.~J. A.~P.}\ \bibnamefont {Martins}}, \ and\ \bibinfo {author} {\bibfnamefont {J.~C. R.~E.}\ \bibnamefont {Oliveira}},\ }\href {\doibase 10.1103/PhysRevD.72.083506} {\bibfield  {journal} {\bibinfo  {journal} {Phys. Rev. D}\ }\textbf {\bibinfo {volume} {72}},\ \bibinfo {pages} {083506} (\bibinfo {year} {2005})},\ \Eprint {http://arxiv.org/abs/hep-ph/0507272} {arXiv:hep-ph/0507272} \BibitemShut {NoStop}%
\bibitem [{\citenamefont {Leite}\ and\ \citenamefont {Martins}(2011)}]{Leite2011}%
  \BibitemOpen
  \bibfield  {author} {\bibinfo {author} {\bibfnamefont {A.~M.~M.}\ \bibnamefont {Leite}}\ and\ \bibinfo {author} {\bibfnamefont {C.~J. A.~P.}\ \bibnamefont {Martins}},\ }\href {\doibase 10.1103/PhysRevD.84.103523} {\bibfield  {journal} {\bibinfo  {journal} {Phys. Rev. D}\ }\textbf {\bibinfo {volume} {84}},\ \bibinfo {pages} {103523} (\bibinfo {year} {2011})},\ \Eprint {http://arxiv.org/abs/1110.3486} {arXiv:1110.3486 [hep-ph]} \BibitemShut {NoStop}%
\bibitem [{\citenamefont {Martins}\ \emph {et~al.}(2016)\citenamefont {Martins}, \citenamefont {Rybak}, \citenamefont {Avgoustidis},\ and\ \citenamefont {Shellard}}]{Martins2016}%
  \BibitemOpen
  \bibfield  {author} {\bibinfo {author} {\bibfnamefont {C.~J. A.~P.}\ \bibnamefont {Martins}}, \bibinfo {author} {\bibfnamefont {I.~Y.}\ \bibnamefont {Rybak}}, \bibinfo {author} {\bibfnamefont {A.}~\bibnamefont {Avgoustidis}}, \ and\ \bibinfo {author} {\bibfnamefont {E.~P.~S.}\ \bibnamefont {Shellard}},\ }\href {\doibase 10.1103/PhysRevD.93.043534} {\bibfield  {journal} {\bibinfo  {journal} {Phys. Rev. D}\ }\textbf {\bibinfo {volume} {93}},\ \bibinfo {pages} {043534} (\bibinfo {year} {2016})},\ \Eprint {http://arxiv.org/abs/1602.01322} {arXiv:1602.01322 [hep-ph]} \BibitemShut {NoStop}%
\bibitem [{\citenamefont {Zeldovich}\ \emph {et~al.}(1974)\citenamefont {Zeldovich}, \citenamefont {Kobzarev},\ and\ \citenamefont {Okun}}]{Zeldovich1974}%
  \BibitemOpen
  \bibfield  {author} {\bibinfo {author} {\bibfnamefont {Y.~B.}\ \bibnamefont {Zeldovich}}, \bibinfo {author} {\bibfnamefont {I.~Y.}\ \bibnamefont {Kobzarev}}, \ and\ \bibinfo {author} {\bibfnamefont {L.~B.}\ \bibnamefont {Okun}},\ }\href@noop {} {\bibfield  {journal} {\bibinfo  {journal} {Zh. Eksp. Teor. Fiz.}\ }\textbf {\bibinfo {volume} {67}},\ \bibinfo {pages} {3} (\bibinfo {year} {1974})}\BibitemShut {NoStop}%
\bibitem [{\citenamefont {Vilenkin}(1981)}]{Vilenkin1981}%
  \BibitemOpen
  \bibfield  {author} {\bibinfo {author} {\bibfnamefont {A.}~\bibnamefont {Vilenkin}},\ }\href {\doibase 10.1103/PhysRevD.23.852} {\bibfield  {journal} {\bibinfo  {journal} {Phys. Rev. D}\ }\textbf {\bibinfo {volume} {23}},\ \bibinfo {pages} {852} (\bibinfo {year} {1981})}\BibitemShut {NoStop}%
\bibitem [{\citenamefont {Widrow}(1989)}]{Widrow1989}%
  \BibitemOpen
  \bibfield  {author} {\bibinfo {author} {\bibfnamefont {L.~M.}\ \bibnamefont {Widrow}},\ }\href {\doibase 10.1103/PhysRevD.40.1002} {\bibfield  {journal} {\bibinfo  {journal} {Phys. Rev. D}\ }\textbf {\bibinfo {volume} {40}},\ \bibinfo {pages} {1002} (\bibinfo {year} {1989})}\BibitemShut {NoStop}%
\bibitem [{\citenamefont {Larsson}\ \emph {et~al.}(1997)\citenamefont {Larsson}, \citenamefont {Sarkar},\ and\ \citenamefont {White}}]{Larsson1996}%
  \BibitemOpen
  \bibfield  {author} {\bibinfo {author} {\bibfnamefont {S.~E.}\ \bibnamefont {Larsson}}, \bibinfo {author} {\bibfnamefont {S.}~\bibnamefont {Sarkar}}, \ and\ \bibinfo {author} {\bibfnamefont {P.~L.}\ \bibnamefont {White}},\ }\href {\doibase 10.1103/PhysRevD.55.5129} {\bibfield  {journal} {\bibinfo  {journal} {Phys. Rev. D}\ }\textbf {\bibinfo {volume} {55}},\ \bibinfo {pages} {5129} (\bibinfo {year} {1997})},\ \Eprint {http://arxiv.org/abs/hep-ph/9608319} {arXiv:hep-ph/9608319} \BibitemShut {NoStop}%
\bibitem [{\citenamefont {Gelmini}\ \emph {et~al.}(1989)\citenamefont {Gelmini}, \citenamefont {Gleiser},\ and\ \citenamefont {Kolb}}]{Gelmini1988}%
  \BibitemOpen
  \bibfield  {author} {\bibinfo {author} {\bibfnamefont {G.~B.}\ \bibnamefont {Gelmini}}, \bibinfo {author} {\bibfnamefont {M.}~\bibnamefont {Gleiser}}, \ and\ \bibinfo {author} {\bibfnamefont {E.~W.}\ \bibnamefont {Kolb}},\ }\href {\doibase 10.1103/PhysRevD.39.1558} {\bibfield  {journal} {\bibinfo  {journal} {Phys. Rev. D}\ }\textbf {\bibinfo {volume} {39}},\ \bibinfo {pages} {1558} (\bibinfo {year} {1989})}\BibitemShut {NoStop}%
\bibitem [{\citenamefont {Correia}\ \emph {et~al.}(2014)\citenamefont {Correia}, \citenamefont {Leite},\ and\ \citenamefont {Martins}}]{Correia2014}%
  \BibitemOpen
  \bibfield  {author} {\bibinfo {author} {\bibfnamefont {J.~R. C. C.~C.}\ \bibnamefont {Correia}}, \bibinfo {author} {\bibfnamefont {I.~S. C.~R.}\ \bibnamefont {Leite}}, \ and\ \bibinfo {author} {\bibfnamefont {C.~J. A.~P.}\ \bibnamefont {Martins}},\ }\href {\doibase 10.1103/PhysRevD.90.023521} {\bibfield  {journal} {\bibinfo  {journal} {Phys. Rev. D}\ }\textbf {\bibinfo {volume} {90}},\ \bibinfo {pages} {023521} (\bibinfo {year} {2014})},\ \Eprint {http://arxiv.org/abs/1407.3905} {arXiv:1407.3905 [hep-ph]} \BibitemShut {NoStop}%
\bibitem [{\citenamefont {Correia}\ \emph {et~al.}(2018)\citenamefont {Correia}, \citenamefont {Leite},\ and\ \citenamefont {Martins}}]{Correia2018}%
  \BibitemOpen
  \bibfield  {author} {\bibinfo {author} {\bibfnamefont {J.~R. C. C.~C.}\ \bibnamefont {Correia}}, \bibinfo {author} {\bibfnamefont {I.~S. C.~R.}\ \bibnamefont {Leite}}, \ and\ \bibinfo {author} {\bibfnamefont {C.~J. A.~P.}\ \bibnamefont {Martins}},\ }\href {\doibase 10.1103/PhysRevD.97.083521} {\bibfield  {journal} {\bibinfo  {journal} {Phys. Rev. D}\ }\textbf {\bibinfo {volume} {97}},\ \bibinfo {pages} {083521} (\bibinfo {year} {2018})},\ \Eprint {http://arxiv.org/abs/1804.10761} {arXiv:1804.10761 [astro-ph.CO]} \BibitemShut {NoStop}%
\bibitem [{\citenamefont {Hiramatsu}\ \emph {et~al.}(2010)\citenamefont {Hiramatsu}, \citenamefont {Kawasaki},\ and\ \citenamefont {Saikawa}}]{Hiramatsu2010}%
  \BibitemOpen
  \bibfield  {author} {\bibinfo {author} {\bibfnamefont {T.}~\bibnamefont {Hiramatsu}}, \bibinfo {author} {\bibfnamefont {M.}~\bibnamefont {Kawasaki}}, \ and\ \bibinfo {author} {\bibfnamefont {K.}~\bibnamefont {Saikawa}},\ }\href {\doibase 10.1088/1475-7516/2010/05/032} {\bibfield  {journal} {\bibinfo  {journal} {JCAP}\ }\textbf {\bibinfo {volume} {05}},\ \bibinfo {pages} {032} (\bibinfo {year} {2010})},\ \Eprint {http://arxiv.org/abs/1002.1555} {arXiv:1002.1555 [astro-ph.CO]} \BibitemShut {NoStop}%
\bibitem [{\citenamefont {Kawasaki}\ and\ \citenamefont {Saikawa}(2011)}]{Kawasaki2011}%
  \BibitemOpen
  \bibfield  {author} {\bibinfo {author} {\bibfnamefont {M.}~\bibnamefont {Kawasaki}}\ and\ \bibinfo {author} {\bibfnamefont {K.}~\bibnamefont {Saikawa}},\ }\href {\doibase 10.1088/1475-7516/2011/09/008} {\bibfield  {journal} {\bibinfo  {journal} {JCAP}\ }\textbf {\bibinfo {volume} {09}},\ \bibinfo {pages} {008} (\bibinfo {year} {2011})},\ \Eprint {http://arxiv.org/abs/1102.5628} {arXiv:1102.5628 [astro-ph.CO]} \BibitemShut {NoStop}%
\bibitem [{\citenamefont {Hiramatsu}\ \emph {et~al.}(2014)\citenamefont {Hiramatsu}, \citenamefont {Kawasaki},\ and\ \citenamefont {Saikawa}}]{Hiramatsu2013}%
  \BibitemOpen
  \bibfield  {author} {\bibinfo {author} {\bibfnamefont {T.}~\bibnamefont {Hiramatsu}}, \bibinfo {author} {\bibfnamefont {M.}~\bibnamefont {Kawasaki}}, \ and\ \bibinfo {author} {\bibfnamefont {K.}~\bibnamefont {Saikawa}},\ }\href {\doibase 10.1088/1475-7516/2014/02/031} {\bibfield  {journal} {\bibinfo  {journal} {JCAP}\ }\textbf {\bibinfo {volume} {02}},\ \bibinfo {pages} {031} (\bibinfo {year} {2014})},\ \Eprint {http://arxiv.org/abs/1309.5001} {arXiv:1309.5001 [astro-ph.CO]} \BibitemShut {NoStop}%
\bibitem [{\citenamefont {Saikawa}(2017)}]{Saikawa2017}%
  \BibitemOpen
  \bibfield  {author} {\bibinfo {author} {\bibfnamefont {K.}~\bibnamefont {Saikawa}},\ }\href {\doibase 10.3390/universe3020040} {\bibfield  {journal} {\bibinfo  {journal} {Universe}\ }\textbf {\bibinfo {volume} {3}},\ \bibinfo {pages} {40} (\bibinfo {year} {2017})},\ \Eprint {http://arxiv.org/abs/1703.02576} {arXiv:1703.02576 [hep-ph]} \BibitemShut {NoStop}%
\bibitem [{\citenamefont {Kitajima}\ \emph {et~al.}(2024)\citenamefont {Kitajima}, \citenamefont {Lee}, \citenamefont {Murai}, \citenamefont {Takahashi},\ and\ \citenamefont {Yin}}]{Kitajima2023}%
  \BibitemOpen
  \bibfield  {author} {\bibinfo {author} {\bibfnamefont {N.}~\bibnamefont {Kitajima}}, \bibinfo {author} {\bibfnamefont {J.}~\bibnamefont {Lee}}, \bibinfo {author} {\bibfnamefont {K.}~\bibnamefont {Murai}}, \bibinfo {author} {\bibfnamefont {F.}~\bibnamefont {Takahashi}}, \ and\ \bibinfo {author} {\bibfnamefont {W.}~\bibnamefont {Yin}},\ }\href {\doibase 10.1016/j.physletb.2024.138586} {\bibfield  {journal} {\bibinfo  {journal} {Phys. Lett. B}\ }\textbf {\bibinfo {volume} {851}},\ \bibinfo {pages} {138586} (\bibinfo {year} {2024})},\ \Eprint {http://arxiv.org/abs/2306.17146} {arXiv:2306.17146 [hep-ph]} \BibitemShut {NoStop}%
\bibitem [{\citenamefont {Ferreira}\ \emph {et~al.}(2024)\citenamefont {Ferreira}, \citenamefont {Notari}, \citenamefont {Pujol\`as},\ and\ \citenamefont {Rompineve}}]{Ferreira2024}%
  \BibitemOpen
  \bibfield  {author} {\bibinfo {author} {\bibfnamefont {R.~Z.}\ \bibnamefont {Ferreira}}, \bibinfo {author} {\bibfnamefont {A.}~\bibnamefont {Notari}}, \bibinfo {author} {\bibfnamefont {O.}~\bibnamefont {Pujol\`as}}, \ and\ \bibinfo {author} {\bibfnamefont {F.}~\bibnamefont {Rompineve}},\ }\href {\doibase 10.1088/1475-7516/2024/06/020} {\bibfield  {journal} {\bibinfo  {journal} {JCAP}\ }\textbf {\bibinfo {volume} {06}},\ \bibinfo {pages} {020} (\bibinfo {year} {2024})},\ \Eprint {http://arxiv.org/abs/2401.14331} {arXiv:2401.14331 [astro-ph.CO]} \BibitemShut {NoStop}%
\bibitem [{\citenamefont {Dankovsky}\ \emph {et~al.}(2024)\citenamefont {Dankovsky}, \citenamefont {Babichev}, \citenamefont {Gorbunov}, \citenamefont {Ramazanov},\ and\ \citenamefont {Vikman}}]{Dankovsky2024}%
  \BibitemOpen
  \bibfield  {author} {\bibinfo {author} {\bibfnamefont {I.}~\bibnamefont {Dankovsky}}, \bibinfo {author} {\bibfnamefont {E.}~\bibnamefont {Babichev}}, \bibinfo {author} {\bibfnamefont {D.}~\bibnamefont {Gorbunov}}, \bibinfo {author} {\bibfnamefont {S.}~\bibnamefont {Ramazanov}}, \ and\ \bibinfo {author} {\bibfnamefont {A.}~\bibnamefont {Vikman}},\ }\href {\doibase 10.1088/1475-7516/2024/09/047} {\bibfield  {journal} {\bibinfo  {journal} {JCAP}\ }\textbf {\bibinfo {volume} {09}},\ \bibinfo {pages} {047} (\bibinfo {year} {2024})},\ \Eprint {http://arxiv.org/abs/2406.17053} {arXiv:2406.17053 [astro-ph.CO]} \BibitemShut {NoStop}%
\bibitem [{\citenamefont {Caprini}\ \emph {et~al.}(2009)\citenamefont {Caprini}, \citenamefont {Durrer}, \citenamefont {Konstandin},\ and\ \citenamefont {Servant}}]{Caprini2009}%
  \BibitemOpen
  \bibfield  {author} {\bibinfo {author} {\bibfnamefont {C.}~\bibnamefont {Caprini}}, \bibinfo {author} {\bibfnamefont {R.}~\bibnamefont {Durrer}}, \bibinfo {author} {\bibfnamefont {T.}~\bibnamefont {Konstandin}}, \ and\ \bibinfo {author} {\bibfnamefont {G.}~\bibnamefont {Servant}},\ }\href {\doibase 10.1103/PhysRevD.79.083519} {\bibfield  {journal} {\bibinfo  {journal} {Phys. Rev. D}\ }\textbf {\bibinfo {volume} {79}},\ \bibinfo {pages} {083519} (\bibinfo {year} {2009})},\ \Eprint {http://arxiv.org/abs/0901.1661} {arXiv:0901.1661 [astro-ph.CO]} \BibitemShut {NoStop}%
\bibitem [{\citenamefont {Cai}\ \emph {et~al.}(2020)\citenamefont {Cai}, \citenamefont {Pi},\ and\ \citenamefont {Sasaki}}]{Cai2019}%
  \BibitemOpen
  \bibfield  {author} {\bibinfo {author} {\bibfnamefont {R.-G.}\ \bibnamefont {Cai}}, \bibinfo {author} {\bibfnamefont {S.}~\bibnamefont {Pi}}, \ and\ \bibinfo {author} {\bibfnamefont {M.}~\bibnamefont {Sasaki}},\ }\href {\doibase 10.1103/PhysRevD.102.083528} {\bibfield  {journal} {\bibinfo  {journal} {Phys. Rev. D}\ }\textbf {\bibinfo {volume} {102}},\ \bibinfo {pages} {083528} (\bibinfo {year} {2020})},\ \Eprint {http://arxiv.org/abs/1909.13728} {arXiv:1909.13728 [astro-ph.CO]} \BibitemShut {NoStop}%
\bibitem [{\citenamefont {Cyr}\ \emph {et~al.}(2023)\citenamefont {Cyr}, \citenamefont {Kite}, \citenamefont {Chluba}, \citenamefont {Hill}, \citenamefont {Jeong}, \citenamefont {Acharya}, \citenamefont {Bolliet},\ and\ \citenamefont {Patil}}]{Cyr2023}%
  \BibitemOpen
  \bibfield  {author} {\bibinfo {author} {\bibfnamefont {B.}~\bibnamefont {Cyr}}, \bibinfo {author} {\bibfnamefont {T.}~\bibnamefont {Kite}}, \bibinfo {author} {\bibfnamefont {J.}~\bibnamefont {Chluba}}, \bibinfo {author} {\bibfnamefont {J.~C.}\ \bibnamefont {Hill}}, \bibinfo {author} {\bibfnamefont {D.}~\bibnamefont {Jeong}}, \bibinfo {author} {\bibfnamefont {S.~K.}\ \bibnamefont {Acharya}}, \bibinfo {author} {\bibfnamefont {B.}~\bibnamefont {Bolliet}}, \ and\ \bibinfo {author} {\bibfnamefont {S.~P.}\ \bibnamefont {Patil}},\ }\href@noop {} {\  (\bibinfo {year} {2023})},\ \Eprint {http://arxiv.org/abs/2309.02366} {arXiv:2309.02366 [astro-ph.CO]} \BibitemShut {NoStop}%
\bibitem [{\citenamefont {Garagounis}\ and\ \citenamefont {Hindmarsh}(2003)}]{Garagounis2002}%
  \BibitemOpen
  \bibfield  {author} {\bibinfo {author} {\bibfnamefont {T.}~\bibnamefont {Garagounis}}\ and\ \bibinfo {author} {\bibfnamefont {M.}~\bibnamefont {Hindmarsh}},\ }\href {\doibase 10.1103/PhysRevD.68.103506} {\bibfield  {journal} {\bibinfo  {journal} {Phys. Rev. D}\ }\textbf {\bibinfo {volume} {68}},\ \bibinfo {pages} {103506} (\bibinfo {year} {2003})},\ \Eprint {http://arxiv.org/abs/hep-ph/0212359} {arXiv:hep-ph/0212359} \BibitemShut {NoStop}%
\bibitem [{\citenamefont {Battye}\ \emph {et~al.}(2020)\citenamefont {Battye}, \citenamefont {Pilaftsis},\ and\ \citenamefont {Viatic}}]{Battye2020}%
  \BibitemOpen
  \bibfield  {author} {\bibinfo {author} {\bibfnamefont {R.~A.}\ \bibnamefont {Battye}}, \bibinfo {author} {\bibfnamefont {A.}~\bibnamefont {Pilaftsis}}, \ and\ \bibinfo {author} {\bibfnamefont {D.~G.}\ \bibnamefont {Viatic}},\ }\href {\doibase 10.1103/PhysRevD.102.123536} {\bibfield  {journal} {\bibinfo  {journal} {Phys. Rev. D}\ }\textbf {\bibinfo {volume} {102}},\ \bibinfo {pages} {123536} (\bibinfo {year} {2020})},\ \Eprint {http://arxiv.org/abs/2010.09840} {arXiv:2010.09840 [hep-ph]} \BibitemShut {NoStop}%
\bibitem [{\citenamefont {Stauffer}(1979)}]{Stauffer1978}%
  \BibitemOpen
  \bibfield  {author} {\bibinfo {author} {\bibfnamefont {D.}~\bibnamefont {Stauffer}},\ }\href {\doibase 10.1016/0370-1573(79)90060-7} {\bibfield  {journal} {\bibinfo  {journal} {Phys. Rept.}\ }\textbf {\bibinfo {volume} {54}},\ \bibinfo {pages} {1} (\bibinfo {year} {1979})}\BibitemShut {NoStop}%
\bibitem [{\citenamefont {Vafa}(2005)}]{Vafa2005}%
  \BibitemOpen
  \bibfield  {author} {\bibinfo {author} {\bibfnamefont {C.}~\bibnamefont {Vafa}},\ }\href@noop {} {\  (\bibinfo {year} {2005})},\ \Eprint {http://arxiv.org/abs/hep-th/0509212} {arXiv:hep-th/0509212} \BibitemShut {NoStop}%
\bibitem [{\citenamefont {Banks}\ and\ \citenamefont {Seiberg}(2011)}]{Banks2010}%
  \BibitemOpen
  \bibfield  {author} {\bibinfo {author} {\bibfnamefont {T.}~\bibnamefont {Banks}}\ and\ \bibinfo {author} {\bibfnamefont {N.}~\bibnamefont {Seiberg}},\ }\href {\doibase 10.1103/PhysRevD.83.084019} {\bibfield  {journal} {\bibinfo  {journal} {Phys. Rev. D}\ }\textbf {\bibinfo {volume} {83}},\ \bibinfo {pages} {084019} (\bibinfo {year} {2011})},\ \Eprint {http://arxiv.org/abs/1011.5120} {arXiv:1011.5120 [hep-th]} \BibitemShut {NoStop}%
\bibitem [{\citenamefont {Harlow}\ and\ \citenamefont {Ooguri}(2021)}]{Harlow2018}%
  \BibitemOpen
  \bibfield  {author} {\bibinfo {author} {\bibfnamefont {D.}~\bibnamefont {Harlow}}\ and\ \bibinfo {author} {\bibfnamefont {H.}~\bibnamefont {Ooguri}},\ }\href {\doibase 10.1007/s00220-021-04040-y} {\bibfield  {journal} {\bibinfo  {journal} {Commun. Math. Phys.}\ }\textbf {\bibinfo {volume} {383}},\ \bibinfo {pages} {1669} (\bibinfo {year} {2021})},\ \Eprint {http://arxiv.org/abs/1810.05338} {arXiv:1810.05338 [hep-th]} \BibitemShut {NoStop}%
\bibitem [{\citenamefont {King}\ \emph {et~al.}(2024)\citenamefont {King}, \citenamefont {Roshan}, \citenamefont {Wang}, \citenamefont {White},\ and\ \citenamefont {Yamazaki}}]{King2023}%
  \BibitemOpen
  \bibfield  {author} {\bibinfo {author} {\bibfnamefont {S.~F.}\ \bibnamefont {King}}, \bibinfo {author} {\bibfnamefont {R.}~\bibnamefont {Roshan}}, \bibinfo {author} {\bibfnamefont {X.}~\bibnamefont {Wang}}, \bibinfo {author} {\bibfnamefont {G.}~\bibnamefont {White}}, \ and\ \bibinfo {author} {\bibfnamefont {M.}~\bibnamefont {Yamazaki}},\ }\href {\doibase 10.1088/1475-7516/2024/05/071} {\bibfield  {journal} {\bibinfo  {journal} {JCAP}\ }\textbf {\bibinfo {volume} {05}},\ \bibinfo {pages} {071} (\bibinfo {year} {2024})},\ \Eprint {http://arxiv.org/abs/2311.12487} {arXiv:2311.12487 [hep-ph]} \BibitemShut {NoStop}%
\bibitem [{\citenamefont {Gouttenoire}\ \emph {et~al.}(2025)\citenamefont {Gouttenoire}, \citenamefont {King}, \citenamefont {Roshan}, \citenamefont {Wang}, \citenamefont {White},\ and\ \citenamefont {Yamazaki}}]{Gouttenoire2025}%
  \BibitemOpen
  \bibfield  {author} {\bibinfo {author} {\bibfnamefont {Y.}~\bibnamefont {Gouttenoire}}, \bibinfo {author} {\bibfnamefont {S.~F.}\ \bibnamefont {King}}, \bibinfo {author} {\bibfnamefont {R.}~\bibnamefont {Roshan}}, \bibinfo {author} {\bibfnamefont {X.}~\bibnamefont {Wang}}, \bibinfo {author} {\bibfnamefont {G.}~\bibnamefont {White}}, \ and\ \bibinfo {author} {\bibfnamefont {M.}~\bibnamefont {Yamazaki}},\ }\href@noop {} {\  (\bibinfo {year} {2025})},\ \Eprint {http://arxiv.org/abs/2501.16414} {arXiv:2501.16414 [hep-ph]} \BibitemShut {NoStop}%
\bibitem [{\citenamefont {Krajewski}\ \emph {et~al.}(2021)\citenamefont {Krajewski}, \citenamefont {Kwapisz}, \citenamefont {Lalak},\ and\ \citenamefont {Lewicki}}]{Krajewski2021}%
  \BibitemOpen
  \bibfield  {author} {\bibinfo {author} {\bibfnamefont {T.}~\bibnamefont {Krajewski}}, \bibinfo {author} {\bibfnamefont {J.~H.}\ \bibnamefont {Kwapisz}}, \bibinfo {author} {\bibfnamefont {Z.}~\bibnamefont {Lalak}}, \ and\ \bibinfo {author} {\bibfnamefont {M.}~\bibnamefont {Lewicki}},\ }\href {\doibase 10.1103/PhysRevD.104.123522} {\bibfield  {journal} {\bibinfo  {journal} {Phys. Rev. D}\ }\textbf {\bibinfo {volume} {104}},\ \bibinfo {pages} {123522} (\bibinfo {year} {2021})},\ \Eprint {http://arxiv.org/abs/2103.03225} {arXiv:2103.03225 [astro-ph.CO]} \BibitemShut {NoStop}%
\bibitem [{\citenamefont {Kitajima}\ \emph {et~al.}(2023)\citenamefont {Kitajima}, \citenamefont {Lee}, \citenamefont {Takahashi},\ and\ \citenamefont {Yin}}]{Kitajima2023b}%
  \BibitemOpen
  \bibfield  {author} {\bibinfo {author} {\bibfnamefont {N.}~\bibnamefont {Kitajima}}, \bibinfo {author} {\bibfnamefont {J.}~\bibnamefont {Lee}}, \bibinfo {author} {\bibfnamefont {F.}~\bibnamefont {Takahashi}}, \ and\ \bibinfo {author} {\bibfnamefont {W.}~\bibnamefont {Yin}},\ }\href@noop {} {\  (\bibinfo {year} {2023})},\ \Eprint {http://arxiv.org/abs/2311.14590} {arXiv:2311.14590 [hep-ph]} \BibitemShut {NoStop}%
\bibitem [{\citenamefont {Roshan}\ and\ \citenamefont {White}(2025)}]{Roshan2024}%
  \BibitemOpen
  \bibfield  {author} {\bibinfo {author} {\bibfnamefont {R.}~\bibnamefont {Roshan}}\ and\ \bibinfo {author} {\bibfnamefont {G.}~\bibnamefont {White}},\ }\href {\doibase 10.1103/RevModPhys.97.015001} {\bibfield  {journal} {\bibinfo  {journal} {Rev. Mod. Phys.}\ }\textbf {\bibinfo {volume} {97}},\ \bibinfo {pages} {015001} (\bibinfo {year} {2025})},\ \Eprint {http://arxiv.org/abs/2401.04388} {arXiv:2401.04388 [hep-ph]} \BibitemShut {NoStop}%
\bibitem [{\citenamefont {Ferreira}\ \emph {et~al.}(2023)\citenamefont {Ferreira}, \citenamefont {Notari}, \citenamefont {Pujolas},\ and\ \citenamefont {Rompineve}}]{Ferreira2022}%
  \BibitemOpen
  \bibfield  {author} {\bibinfo {author} {\bibfnamefont {R.~Z.}\ \bibnamefont {Ferreira}}, \bibinfo {author} {\bibfnamefont {A.}~\bibnamefont {Notari}}, \bibinfo {author} {\bibfnamefont {O.}~\bibnamefont {Pujolas}}, \ and\ \bibinfo {author} {\bibfnamefont {F.}~\bibnamefont {Rompineve}},\ }\href {\doibase 10.1088/1475-7516/2023/02/001} {\bibfield  {journal} {\bibinfo  {journal} {JCAP}\ }\textbf {\bibinfo {volume} {02}},\ \bibinfo {pages} {001} (\bibinfo {year} {2023})},\ \Eprint {http://arxiv.org/abs/2204.04228} {arXiv:2204.04228 [astro-ph.CO]} \BibitemShut {NoStop}%
\bibitem [{\citenamefont {Ade}\ \emph {et~al.}(2019)\citenamefont {Ade} \emph {et~al.}}]{SimonsObservatory2018}%
  \BibitemOpen
  \bibfield  {author} {\bibinfo {author} {\bibfnamefont {P.}~\bibnamefont {Ade}} \emph {et~al.} (\bibinfo {collaboration} {Simons Observatory}),\ }\href {\doibase 10.1088/1475-7516/2019/02/056} {\bibfield  {journal} {\bibinfo  {journal} {JCAP}\ }\textbf {\bibinfo {volume} {02}},\ \bibinfo {pages} {056} (\bibinfo {year} {2019})},\ \Eprint {http://arxiv.org/abs/1808.07445} {arXiv:1808.07445 [astro-ph.CO]} \BibitemShut {NoStop}%
\bibitem [{\citenamefont {Carroll}(2019)}]{Carroll2004}%
  \BibitemOpen
  \bibfield  {author} {\bibinfo {author} {\bibfnamefont {S.~M.}\ \bibnamefont {Carroll}},\ }\href {\doibase 10.1017/9781108770385} {\emph {\bibinfo {title} {{Spacetime and Geometry}: {An Introduction to General Relativity}}}}\ (\bibinfo  {publisher} {Cambridge University Press},\ \bibinfo {year} {2019})\BibitemShut {NoStop}%
\bibitem [{\citenamefont {Gr\"uber}\ \emph {et~al.}(2024)\citenamefont {Gr\"uber}, \citenamefont {Sousa},\ and\ \citenamefont {Avelino}}]{Gruber2024}%
  \BibitemOpen
  \bibfield  {author} {\bibinfo {author} {\bibfnamefont {D.}~\bibnamefont {Gr\"uber}}, \bibinfo {author} {\bibfnamefont {L.}~\bibnamefont {Sousa}}, \ and\ \bibinfo {author} {\bibfnamefont {P.~P.}\ \bibnamefont {Avelino}},\ }\href {\doibase 10.1103/PhysRevD.110.023505} {\bibfield  {journal} {\bibinfo  {journal} {Phys. Rev. D}\ }\textbf {\bibinfo {volume} {110}},\ \bibinfo {pages} {023505} (\bibinfo {year} {2024})},\ \Eprint {http://arxiv.org/abs/2403.09816} {arXiv:2403.09816 [gr-qc]} \BibitemShut {NoStop}%
\bibitem [{\citenamefont {Dufaux}\ \emph {et~al.}(2007)\citenamefont {Dufaux}, \citenamefont {Bergman}, \citenamefont {Felder}, \citenamefont {Kofman},\ and\ \citenamefont {Uzan}}]{Dufaux2007}%
  \BibitemOpen
  \bibfield  {author} {\bibinfo {author} {\bibfnamefont {J.~F.}\ \bibnamefont {Dufaux}}, \bibinfo {author} {\bibfnamefont {A.}~\bibnamefont {Bergman}}, \bibinfo {author} {\bibfnamefont {G.~N.}\ \bibnamefont {Felder}}, \bibinfo {author} {\bibfnamefont {L.}~\bibnamefont {Kofman}}, \ and\ \bibinfo {author} {\bibfnamefont {J.-P.}\ \bibnamefont {Uzan}},\ }\href {\doibase 10.1103/PhysRevD.76.123517} {\bibfield  {journal} {\bibinfo  {journal} {Phys. Rev. D}\ }\textbf {\bibinfo {volume} {76}},\ \bibinfo {pages} {123517} (\bibinfo {year} {2007})},\ \Eprint {http://arxiv.org/abs/0707.0875} {arXiv:0707.0875 [astro-ph]} \BibitemShut {NoStop}%
\bibitem [{\citenamefont {Pujolas}\ and\ \citenamefont {Zahariade}(2023)}]{Pujolas2022}%
  \BibitemOpen
  \bibfield  {author} {\bibinfo {author} {\bibfnamefont {O.}~\bibnamefont {Pujolas}}\ and\ \bibinfo {author} {\bibfnamefont {G.}~\bibnamefont {Zahariade}},\ }\href {\doibase 10.1103/PhysRevD.107.123527} {\bibfield  {journal} {\bibinfo  {journal} {Phys. Rev. D}\ }\textbf {\bibinfo {volume} {107}},\ \bibinfo {pages} {123527} (\bibinfo {year} {2023})},\ \Eprint {http://arxiv.org/abs/2212.11204} {arXiv:2212.11204 [hep-th]} \BibitemShut {NoStop}%
\bibitem [{\citenamefont {Campeti}\ \emph {et~al.}(2021)\citenamefont {Campeti}, \citenamefont {Komatsu}, \citenamefont {Poletti},\ and\ \citenamefont {Baccigalupi}}]{Campeti2020}%
  \BibitemOpen
  \bibfield  {author} {\bibinfo {author} {\bibfnamefont {P.}~\bibnamefont {Campeti}}, \bibinfo {author} {\bibfnamefont {E.}~\bibnamefont {Komatsu}}, \bibinfo {author} {\bibfnamefont {D.}~\bibnamefont {Poletti}}, \ and\ \bibinfo {author} {\bibfnamefont {C.}~\bibnamefont {Baccigalupi}},\ }\href {\doibase 10.1088/1475-7516/2021/01/012} {\bibfield  {journal} {\bibinfo  {journal} {JCAP}\ }\textbf {\bibinfo {volume} {01}},\ \bibinfo {pages} {012} (\bibinfo {year} {2021})},\ \Eprint {http://arxiv.org/abs/2007.04241} {arXiv:2007.04241 [astro-ph.CO]} \BibitemShut {NoStop}%
\bibitem [{\citenamefont {Ellis}\ \emph {et~al.}(2024)\citenamefont {Ellis}, \citenamefont {Fairbairn}, \citenamefont {Franciolini}, \citenamefont {H\"utsi}, \citenamefont {Iovino}, \citenamefont {Lewicki}, \citenamefont {Raidal}, \citenamefont {Urrutia}, \citenamefont {Vaskonen},\ and\ \citenamefont {Veerm\"ae}}]{Ellis2023}%
  \BibitemOpen
  \bibfield  {author} {\bibinfo {author} {\bibfnamefont {J.}~\bibnamefont {Ellis}}, \bibinfo {author} {\bibfnamefont {M.}~\bibnamefont {Fairbairn}}, \bibinfo {author} {\bibfnamefont {G.}~\bibnamefont {Franciolini}}, \bibinfo {author} {\bibfnamefont {G.}~\bibnamefont {H\"utsi}}, \bibinfo {author} {\bibfnamefont {A.}~\bibnamefont {Iovino}}, \bibinfo {author} {\bibfnamefont {M.}~\bibnamefont {Lewicki}}, \bibinfo {author} {\bibfnamefont {M.}~\bibnamefont {Raidal}}, \bibinfo {author} {\bibfnamefont {J.}~\bibnamefont {Urrutia}}, \bibinfo {author} {\bibfnamefont {V.}~\bibnamefont {Vaskonen}}, \ and\ \bibinfo {author} {\bibfnamefont {H.}~\bibnamefont {Veerm\"ae}},\ }\href {\doibase 10.1103/PhysRevD.109.023522} {\bibfield  {journal} {\bibinfo  {journal} {Phys. Rev. D}\ }\textbf {\bibinfo {volume} {109}},\ \bibinfo {pages} {023522} (\bibinfo {year} {2024})},\ \Eprint {http://arxiv.org/abs/2308.08546} {arXiv:2308.08546 [astro-ph.CO]} \BibitemShut {NoStop}%
\bibitem [{\citenamefont {Aggarwal}\ \emph {et~al.}(2025)\citenamefont {Aggarwal} \emph {et~al.}}]{Aggarwal2025}%
  \BibitemOpen
  \bibfield  {author} {\bibinfo {author} {\bibfnamefont {N.}~\bibnamefont {Aggarwal}} \emph {et~al.},\ }\href@noop {} {\  (\bibinfo {year} {2025})},\ \Eprint {http://arxiv.org/abs/2501.11723} {arXiv:2501.11723 [gr-qc]} \BibitemShut {NoStop}%
\bibitem [{\citenamefont {Kite}\ \emph {et~al.}(2021)\citenamefont {Kite}, \citenamefont {Ravenni}, \citenamefont {Patil},\ and\ \citenamefont {Chluba}}]{Kite2020}%
  \BibitemOpen
  \bibfield  {author} {\bibinfo {author} {\bibfnamefont {T.}~\bibnamefont {Kite}}, \bibinfo {author} {\bibfnamefont {A.}~\bibnamefont {Ravenni}}, \bibinfo {author} {\bibfnamefont {S.~P.}\ \bibnamefont {Patil}}, \ and\ \bibinfo {author} {\bibfnamefont {J.}~\bibnamefont {Chluba}},\ }\href {\doibase 10.1093/mnras/stab1558} {\bibfield  {journal} {\bibinfo  {journal} {Mon. Not. Roy. Astron. Soc.}\ }\textbf {\bibinfo {volume} {505}},\ \bibinfo {pages} {4396} (\bibinfo {year} {2021})},\ \Eprint {http://arxiv.org/abs/2010.00040} {arXiv:2010.00040 [astro-ph.CO]} \BibitemShut {NoStop}%
\bibitem [{\citenamefont {Yeh}\ \emph {et~al.}(2022)\citenamefont {Yeh}, \citenamefont {Shelton}, \citenamefont {Olive},\ and\ \citenamefont {Fields}}]{Yeh2022}%
  \BibitemOpen
  \bibfield  {author} {\bibinfo {author} {\bibfnamefont {T.-H.}\ \bibnamefont {Yeh}}, \bibinfo {author} {\bibfnamefont {J.}~\bibnamefont {Shelton}}, \bibinfo {author} {\bibfnamefont {K.~A.}\ \bibnamefont {Olive}}, \ and\ \bibinfo {author} {\bibfnamefont {B.~D.}\ \bibnamefont {Fields}},\ }\href {\doibase 10.1088/1475-7516/2022/10/046} {\bibfield  {journal} {\bibinfo  {journal} {JCAP}\ }\textbf {\bibinfo {volume} {10}},\ \bibinfo {pages} {046} (\bibinfo {year} {2022})},\ \Eprint {http://arxiv.org/abs/2207.13133} {arXiv:2207.13133 [astro-ph.CO]} \BibitemShut {NoStop}%
\bibitem [{\citenamefont {Maggiore}(2018)}]{Maggiore2018}%
  \BibitemOpen
  \bibfield  {author} {\bibinfo {author} {\bibfnamefont {M.}~\bibnamefont {Maggiore}},\ }\href@noop {} {\emph {\bibinfo {title} {{Gravitational Waves. Vol. 2: Astrophysics and Cosmology}}}}\ (\bibinfo  {publisher} {Oxford University Press},\ \bibinfo {year} {2018})\BibitemShut {NoStop}%
\end{thebibliography}
%

\end{document}